%% file: pasj.tex
\documentclass[proof]{pasj01}

\usepackage{natbib,bm}
\usepackage[pdftex]{graphicx,color}  
\usepackage{lineno}
\usepackage[T1]{fontenc} 

\Received{$\langle$reception date$\rangle$}
\Accepted{$\langle$acception date$\rangle$}
\Published{$\langle$publication date$\rangle$}

\usepackage[varg]{txfonts}

\graphicspath{{fig}{fig/fig_lcs}{fig/complex_lc}}
\begin{document}

\title{Fast optical flares from M dwarfs detected by \\ a one-second-cadence survey with Tomo-e Gozen} 
\author{Masataka \textsc{Aizawa}\altaffilmark{1*}, 
        Kojiro \textsc{Kawana}\altaffilmark{2}, 
        Kazumi \textsc{Kashiyama}\altaffilmark{2,3,4,5}, 
        Ryou \textsc{Ohsawa}\altaffilmark{6,7}, 
        Hajime \textsc{Kawahara}\altaffilmark{3,8}, 
        Fumihiro \textsc{Naokawa}\altaffilmark{2,3}, 
        Tomoyuki \textsc{Tajiri}\altaffilmark{8}, 
        Noriaki \textsc{Arima}\altaffilmark{6,9},
        Hanchun \textsc{Jiang}\altaffilmark{10},
        Tilman \textsc{Hartwig}\altaffilmark{2,4,11},
        Kotaro \textsc{Fujisawa}\altaffilmark{2}, 
        Toshikazu \textsc{Shigeyama} \altaffilmark{3,9}, 
        Ko \textsc{Arimatsu}  \altaffilmark{12},
        Mamoru \textsc{Doi} \altaffilmark{3,6,13,14}, 
        Toshihiro \textsc{Kasuga} \altaffilmark{15}, 
        Naoto \textsc{Kobayashi} \altaffilmark{6,7, 13,14}, 
        Sohei \textsc{Kondo} \altaffilmark{7},
        Yuki \textsc{Mori} \altaffilmark{7},
        Shin-ichiro \textsc{Okumura} \altaffilmark{16},
        Satoshi \textsc{Takita} \altaffilmark{7}, 
        Shigeyuki \textsc{Sako}\altaffilmark{6,13,14}
        }

\altaffiltext{1}{Tsung-Dao Lee Institute, Shanghai Jiao Tong University, Shengrong Road 520, 201210 Shanghai, P. R. China}
\altaffiltext{2}{Department of Physics, Graduate School of Science, The University of Tokyo, Bunkyo-ku, Tokyo 113-0033, Japan}
\altaffiltext{3}{Research Center for the Early Universe, Graduate School of Science, The University of Tokyo, Bunkyo-ku, Tokyo 113-0033, Japan}
\altaffiltext{4}{Kavli Institute for the Physics and Mathematics of the Universe (Kavli IPMU,WPI), The University of Tokyo, Chiba 277-8582, Japan}
\altaffiltext{5}{Astronomical Institute, Graduate School of Science, Tohoku University, Aoba, Sendai 980-8578, Japan}
\altaffiltext{6}{Institute of Astronomy, Graduate School of Science, The University of Tokyo, 2-21-1 Osawa, Mitaka, Tokyo 181-0015, Japan}
\altaffiltext{7}{Kiso Observatory,Institute of Astronomy,Graduate School of Science, The University of Tokyo, 10762-30, Mitake, Kiso-machi, Kiso-gun, Nagano 397-0101, Japan}
\altaffiltext{8}{Department of Earth and Planetary Science, The University of Tokyo, 7-3-1, Hongo, Tokyo, Japan}
\altaffiltext{9}{Department of Astronomy, Graduate School of Science, The University of Tokyo, 7-3-1 Hongo, Bunkyo-ku, Tokyo 113-0033, Japan}
\altaffiltext{10}{School of Astronomy and Space Science, Nanjing University, Nanjing 210093, China}
\altaffiltext{11}{Institute for Physics of Intelligence, School of Science, The University of Tokyo, Bunkyo, Tokyo 113-0033, Japan}
\altaffiltext{12}{The Hakubi Center/Astronomical Observatory, Graduate School of Science, Kyoto University Kitashirakawa-oiwake-cho, Sakyo-ku, Kyoto 606-8502, Japan}
\altaffiltext{13}{UTokyo Organization for Planetary Space Science, The University of Tokyo, 7-3-1 Hongo, Tokyo 113-0033, Japan}
\altaffiltext{14}{Collaborative Research Organization for Space Science and Technology, The University of Tokyo, 7-3-1 Hongo, Tokyo 113-0033, Japan}
\altaffiltext{15}{National Astronomical Observatory of Japan, 2-21-1 Osawa, Mitaka, Tokyo 181-8588, Japan}
\altaffiltext{16}{Japan Spaceguard Association, Bisei Spaceguard Center, 1716-3 Okura, Bisei-cho, Ibara, Okayama 714-1411, Japan}

\email{aizw\underline{{ }{ }}ms@sjtu.edu.cn}

\maketitle
\begin{abstract}
We report a one-second-cadence wide-field survey for M-dwarf flares using the Tomo-e Gozen camera mounted on the Kiso Schmidt telescope. We detect 22 flares from M3-M5 dwarfs with rise times and amplitudes ranging from $5\, \mathrm{sec} \lesssim t_\mathrm{rise} \lesssim 100\,\mathrm{sec}$ and $0.5 \lesssim \Delta F/F_{\star} \lesssim 20$, respectively. The flare light curves mostly show steeper rises and shallower decays than those obtained from the Kepler one-minute cadence data and tend to have flat peak structures. Assuming a blackbody spectrum with temperatures of $9,000-15,000\,\mathrm{K}$, the peak luminosities and bolometric energies are estimated to be $10^{29}\,\mathrm{erg\,sec^{-1}} \lesssim L_\mathrm{peak} \lesssim 10^{31}\,\mathrm{erg\,sec^{-1}}$ and $10^{31}\,\mathrm{erg} \lesssim E_{\rm bol} \lesssim 10^{34}\,\mathrm{erg}$, which constitutes the bright end of fast optical flares for M dwarfs. We confirm that more than 90\% of the host stars of the detected flares are magnetically active based on their H$\alpha$ emission line intensities obtained by LAMOST. The estimated occurrence rate of the detected flares is $\sim 0.7$ per day per an active star, indicating they are common in magnetically active M dwarfs. We argue that the flare light curves can be explained by the chromospheric compression model; the rise time is broadly consistent with the Alfv\'en transit time of a magnetic loop with a length scale of $l_\mathrm{loop} \sim 10^4\,\mathrm{km}$ and a field strength of $1,000\,\mathrm{G}$, while the decay time is likely determined by the radiative cooling of the compressed chromosphere down to near the photosphere with a temperature of $\gtrsim 10,000\,\mathrm{K}$. These flares from M dwarfs could be a major contamination source for a future search of fast optical transients of unknown types.
\end{abstract}
\section{Introduction}
Stellar flares are triggered by magnetic reconnection in the corona~\citep{2011LRSP....8....6S}. The magnetic energy released in the reconnection is transferred to high energy particles that can produce a variety of abrupt brightening in a wide wavelength range depending on the strength and geometrical configuration of the reconnecting fields. The high-energy particles and photons could have a great impact on the atmospheric and terrestrial environment of the surrounding planets (if any) and thus important for the habitability \citep{2020IJAsB..19..136A}. 

Flares in visible bands are called white light flares (WLFs), which were firstly identified in a Solar flare \citep{1859MNRAS..20...13C}. Solar WLFs are commonly associated with X-ray flares, which are attributed to bremsstrahlung of accelerated electrons with $>$ 10 keV injected downward towards the chromosphere~\citep[e.g.,][]{1971SoPh...18..489B}.  
The dissipation of the kinetic energy of the high-energy electron beam heats up the upper chromosphere and the enhanced Balmer-continuum radiation can back-warm the underlying transition region, where the WLFs can be produced. This scenario has been supported both by spectroscopic observations and radiative hydrodynamic simulations for moderate to strong flares~\citep[e.g.,][]{1990ApJ...365..391M,2005ApJ...630..573A}. 
However, the direct observation of solar WLF is still challenging, and chances to study rare, energetic events are limited.  

Late type stars are known to be magnetically active, and thus more frequently produce flares including energetic ones. A strong WLF from an M dwarf is generally associated with blackbody radiation with a temperature larger than $10^4\,\mathrm{K}$~\citep[e.g.][]{1992ApJS...78..565H,2003ApJ...597..535H,2013ApJS..207...15K}. 
Based on the radiative hydrodynamic simulations~\citep{2015SoPh..290.3487K}, 
such a hot continuum component can only be produced by injecting a more intense electron beam than those inferred for mild to strong solar WLFs, which would significantly compress the chromosphere. 
The mechanisms of formation, propagation, and energy dissipation of the electron beam in an energetic flare are still uncertain, and M dwarfs are unique targets for studying this subject; flares can be detected with a large signal-to-noise ratio due to its relatively low stellar luminosity.

Flare light curves also include key information to unveil the physical properties of the flares. High-cadence wide-field photometric surveys with, e.g., the Kepler space telescope and the Transiting Exoplanet Survey Satellite (TESS), have identified many optical flares in their high-precision data, which enable to conduct an unbiased statistical analysis on the flares, e.g., calculating correlations among the flare parameters and stellar properties, and constructing the template shape of the flare light curves~\citep[e.g.][]{2012Natur.485..478M,2013ApJ...771..127N,2013ApJS..209....5S,2014ApJ...797..121H, 2014ApJ...797..122D,2016ApJ...829...23D,2020AJ....159...60G, 2021arXiv211013155H}. The observational cadences are as short as 20 seconds for TESS and one minute for Kepler, so these observations have revealed mainly a population of flares with durations longer than one minute.


Stellar flares, however, can have timescales comparable to several seconds, and such fast flares can be missed or unresolved by the previous surveys.  For instance, the Alfv\'en transit time of a magnetic loop responsible for a typical M-dwarf flare can be described as~\citep[e.g.][]{2011LRSP....8....6S}; 
\begin{equation}
    t_{\rm A} = 5 \,\mathrm{sec}\,\left(\frac{B}{1,000\,\mathrm{G}}\right)^{-1} \left(\frac{n}{10^{10}\,\mathrm{cm^{-3}}}\right)^{1/2}   \left(\frac{l_{\rm loop}}{10^{4}\,\mathrm{km}}\right) \left(\frac{M_{A}}{0.1}\right)^{-1}, \label{eq:time_A}
\end{equation}
where $B$ is the magnetic field strength, $n$ is the coronal density, $l_{\rm loop}$ is the loop size, and $M_{A} \sim 0.1\mbox{-}0.01$ is the dimensionless reconnection rate. Thus, $O(1)$ sec cadence observations are desired to resolve the energy injection process of the flare. In fact, recently \cite{2021ApJ...911L..25M} identified a very fast flare lasting a few seconds across radio and UV bands from Proxima Centauri. The flare was simultaneously observed by TESS, but the cadence was two minutes, which was not enough to resolve the flare light curve in the optical band. On the other hand, \cite{2016ApJ...820...95K} monitored 5 active M dwarfs using 3 narrow band filters on ULTRACAM at 1-second cadence. They detected 20 large flares in the total observation time of 40 hrs, and confirmed that some optical flare light curves rise within $O(10)$ sec as expected from equation (\ref{eq:time_A}).

Previous optical flare searches with very high cadence ($\leq$ 1 sec) are limited to target observations. A wide-field survey with a $O(1)$ sec cadence could uncover a new aspect of optical flares, e.g., by detecting relatively rare bright events, which will be useful to  understand the flare mechanism in general and its association with the host star properties. For this purpose, the Tomo-e Gozen camera mounted on the 1.05\,m Kiso Schmidt telescope is a unique instrument since it can simultaneously realize a wide field of view of 20.7 deg$^{2}$ and a high cadence as short as 0.5 sec~\citep{2018SPIE10702E..0JS}. We here report newly detected 22 of fast optical flares from M dwarfs with the Tomo-e Gozen camera. 

This paper is organized as follows. We describe our search method in section \ref{sec:method}. We show the observed properties of the detected fast optical flares in section \ref{sec:result}. We summarize the current observations, discuss the observed timescales, and present future prospects in section \ref{sec:discussion}.



\section{A search for fast optical flares from M dwarfs with Tomo-e Gozen}\label{sec:method}

\subsection{The HeSO survey}
We use the Tomo-e Gozen camera mounted on the 1.05\,m Kiso Schmidt telescope~\citep{2018SPIE10702E..0JS}. Tomo-e Gozen can take consecutive images at 1-2 frame-per-second~(fps) with a field-of-view (FOV) of 20.7\,deg$^2$ covered by 84 chips of 2k $\times$ 1k CMOS image sensors,  which are sensitive to 300--1,000 nm with a peak efficiency of 0.67 at 500 nm~\citep{2018SPIE10709E..1TK}. One pixel corresponds to $\sim 1.19\arcsec$, and the typical seeing ranges from a few to several arcsecs. The present study is a part of an on-going survey for sub-minute variabilities of Galactic stellar objects, the Hertz Stellar Object (HeSO) survey. 

In this study, we use the HeSO survey data obtained in 2019-2020 at 1-2 fps. With the sampling rate of 2 fps, the limiting V-band magnitude at $5\sigma$ significance is $\simeq 17$ in good conditions \citep[cf.][]{2020PASJ...72....3R}. The total survey time is $\sim 40 $ hrs, which spreads over 31 different nights. Most of the data were taken in the direction of the Earth's shadow in order to suppress the contamination by artificial satellite reflection of the solar light. 

\subsection{Target selection}\label{sec:target_selection} 
In this study, we prepare a set of 520,519 of M dwarf (candidates) in the all-sky region from the TESS Input Catalog~\citep[TIC;][]{2019AJ....158..138S} based on the following conditions: $M_{\star} \leq 0.4\,M_{\odot}$, $R_{\star} \leq 0.4\,R_{\odot}$,  $T_\mathrm{eff} \leq 4,200\,\mathrm{K}$, and $G_\mathrm{mag}\leq 17$. The former three conditions are for selecting M dwarfs, and the last condition is to detect targets with $5\,\sigma$ significance assuming that the spectral responses of Tomo-e Gozen are similar to that of Gaia. There are typically $200\mbox{-}300$ target M dwarfs in an FOV of Tomo-e Gozen. We also prepare a catalog of reference stars by retrieving objects with $11 < G_\mathrm{mag} < 16$ from the Gaia DR2 catalog. The number of the reference stars is roughly 100 times larger than that of the target M dwarfs. 

\begin{figure*}
\begin{center}
\includegraphics[width=.45 \linewidth]{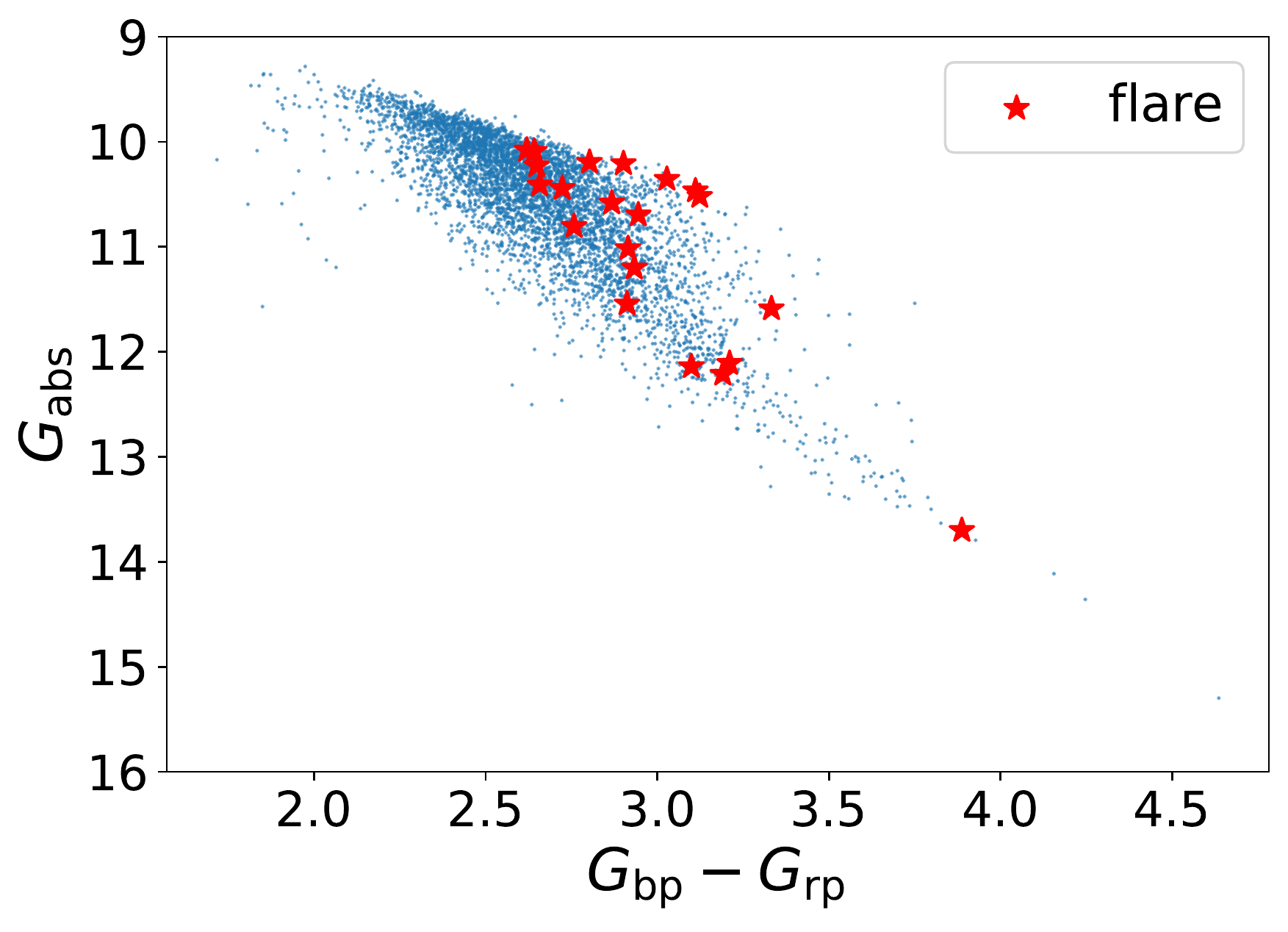}
\includegraphics[width=.48 \linewidth]{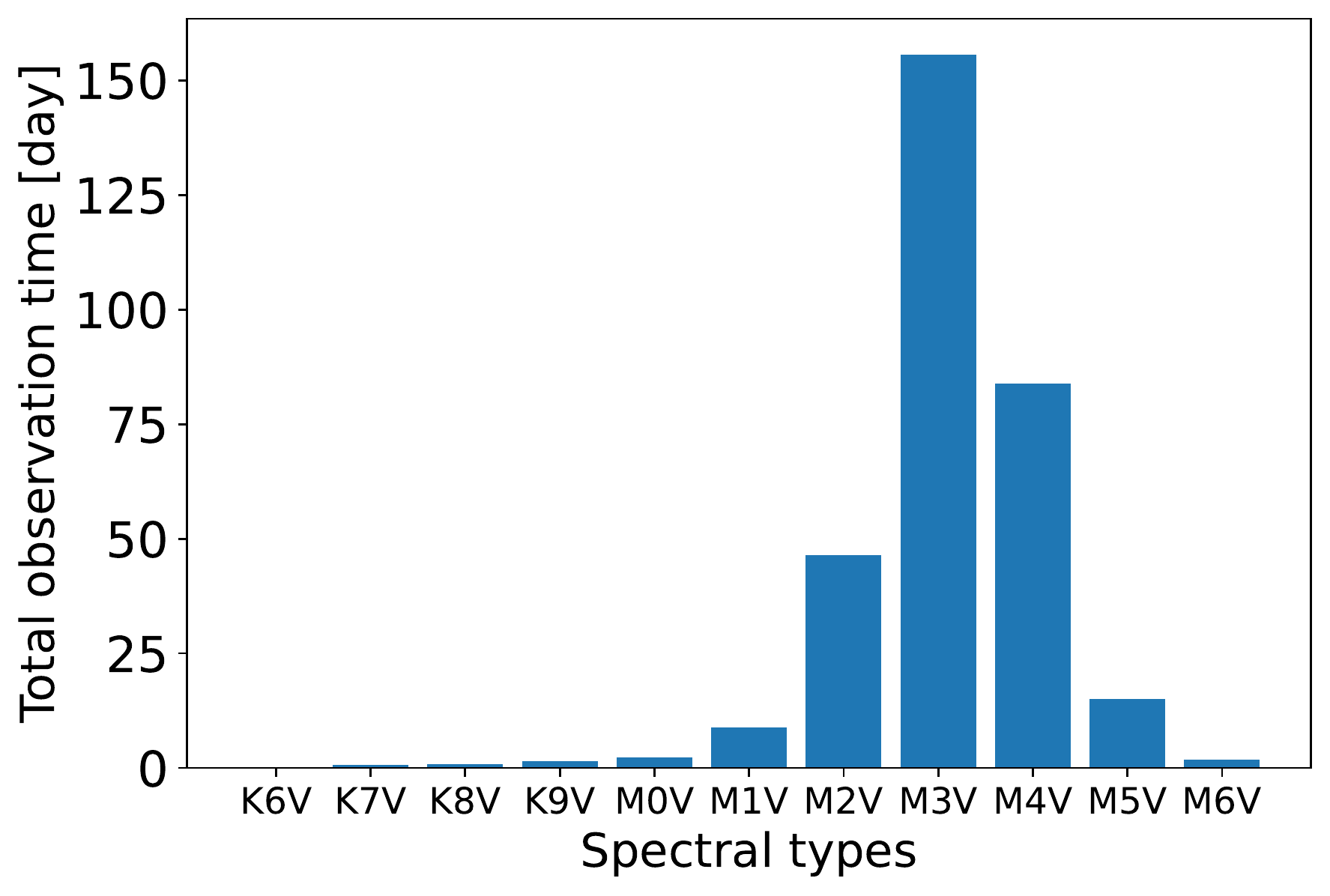}
\end{center}
\caption{(Left) The Gaia's Hertzsprung-Russell diagram \citep{2018A&A...616A...1G} showing the observed M dwarfs in our survey. The red stars indicate the stars with flare detection. (Right) Histogram of the total observation time for each spectral type. }
\label{fig:fit_hr}
\end{figure*}

The left panel in figure~\ref{fig:fit_hr} shows the observed M dwarfs on the Gaia's Hertzsprung-Russell (HR) diagram. The total number of the observed M dwarfs is 5,692, some of which were observed on multiple dates. Note that we only pick up M dwarfs observed for more than 20 minutes. The right panel in figure \ref{fig:fit_hr} shows the integrated observation time for each spectral type; the spectral type is determined from the effective temperature in the TIC using an empirical relation~\citep{2013ApJS..208....9P}. The total observation time is 317.0 M-dwarf days. Given that the observation conditions are bad for 1/3--1/2 of the current data set (e.g. due to cloudy and/or hazy weather), the effective total observation time is 100--150 M-dwarf days. 

\subsection{Data reduction and light curve production \label{sec:lc_mv}}
The raw data of our survey are video data at 1 or 2 fps for a $\sim 20\,\mathrm{deg}^2$ patch of the sky, covered by 84 CMOS image sensors. We first conduct the dark and flat-field correction and assign World Coordinate System (WCS) to each data frame with {astrometry.net}~\citep{2010AJ....139.1782L}. We detect objects in each data frame with extracting the background by utilizing {SExtractor}~\citep{1996A&AS..117..393B}. We identify our targets and reference stars by cross-matching the detected objects with the Gaia DR2 catalog~\citep{2018A&A...616A...1G}. We save the video data, consecutively cutting out 20 $\times$ 20 pixels around their centroids, to check the robustness of the detection and their intrinsic variability if any.


We calculate the light curves of the detected targets and reference stars by performing aperture photometry with several different settings. We adopt two types of variable aperture implemented in SExtractor to estimate \texttt{FLUX\_AUTO} and \texttt{FLUX\_ISO}~\citep{1996A&AS..117..393B}. We also estimate the flux for fixed circle apertures around each object with radii of $(5, 7, 10, 14)$ pixels, defining them as \texttt{FLUX\_r5}, \texttt{FLUX\_r7}, \texttt{FLUX\_r10}, and \texttt{FLUX\_r14}, respectively. In total, we produce 6 types of raw light curves. 

We divide the detected objects into groups with respect to CMOS sensors, where they are on. For each group, we remove common systematic trends from the raw light curves by utilizing the singular value decomposition (SVD); we construct the cotrending basis vectors in each group \citep{2012PASP..124..985S,2012PASP..124.1000S}, and subtract the common trends from the raw light curves. Note that the mean flux for each star is not changed in the subtraction. The detailed implementation is explained in appendix \ref{sec:cot}. The left panel in figure \ref{fig:fig_reduced} shows an example of the light curve detrending. One can see that the ``dip" in the middle of the raw light curve, likely caused by a cloud crossing, is successfully removed, while the flare indicated by the dotted lines remains in the detrended light curve. 

\begin{figure*}
\begin{center}
\includegraphics[width=.44 \linewidth]{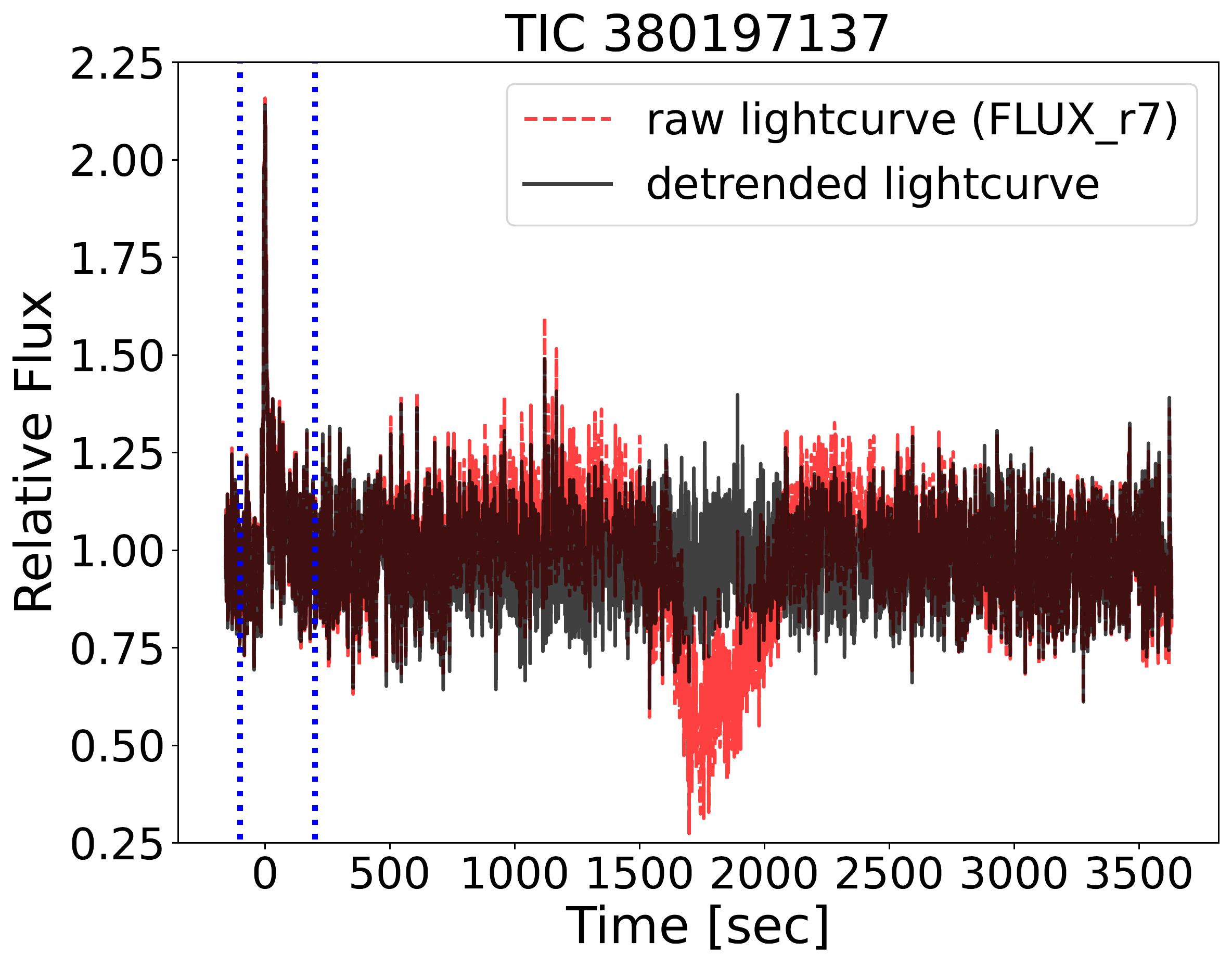}
\includegraphics[width=.47 \linewidth]{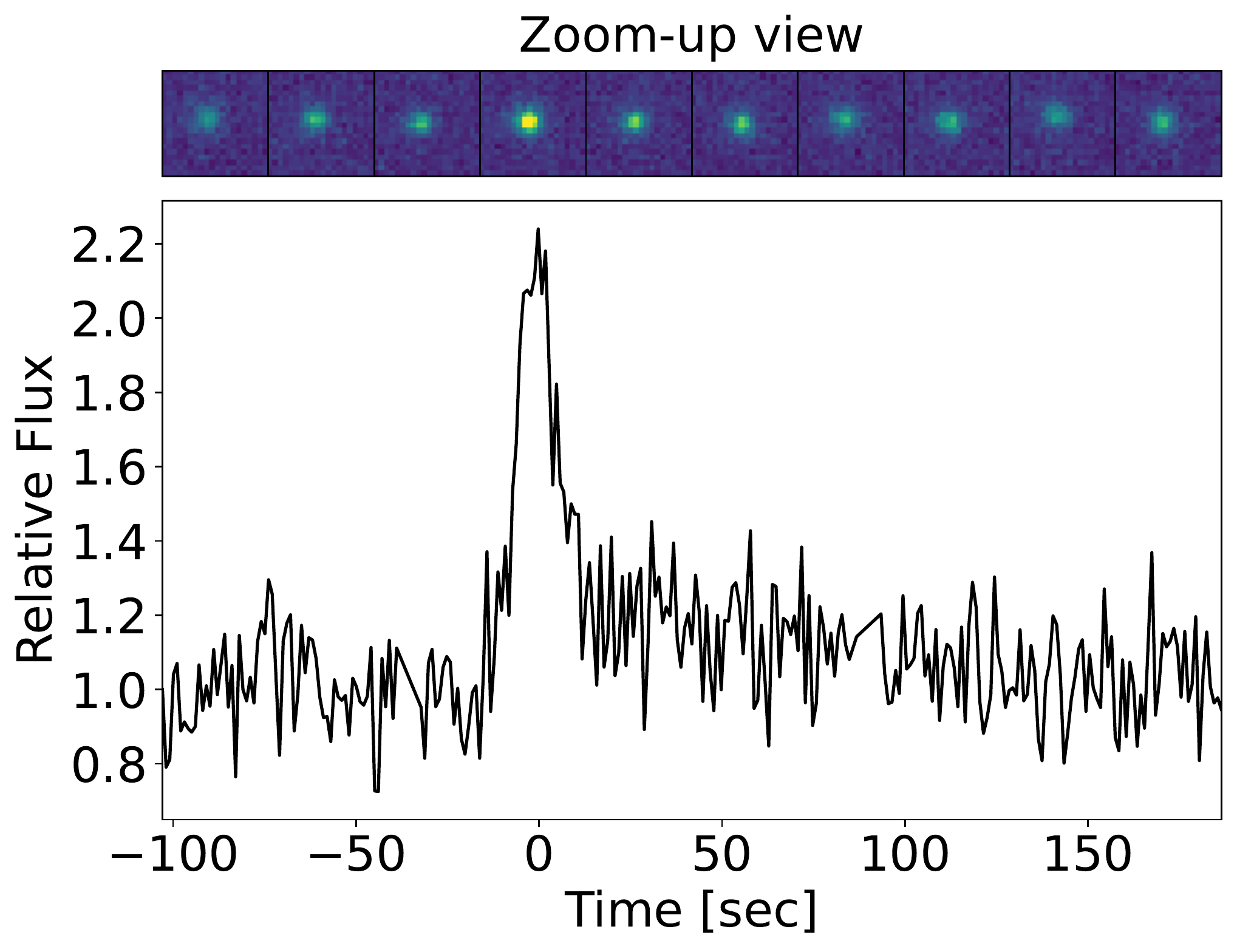}
\end{center}
\caption{(Left) Flare light curve of TIC 380197137 ($G_{\rm mag} = 15.62$). The red dahsed line shows the light curve obtained by the simple aperture photometry on the raw data while the black line shows the light curve after removing the systematic noise. The dotted lines indicate the region zoomed up in the right panel. (Right) Zoom-up view of the light curve in the flaring phase. In the upper panels, we show series of averaged images with a the sizes of 20 $\times$ 20 pixels ($\simeq24\arcsec$ for one side). We show integrated images over 10 seconds with an interval of $\sim28$ seconds, and a clear brightening is seen around the flare peak. }
\label{fig:fig_reduced}
\end{figure*}

\subsection{Visual inspection and flare detection}
We search for M dwarf flares in the calculated light curves. First, three persons visually inspected the raw and detrended light curves independently to make lists of candidate events. Each person chooses a aperture size for each date, mostly \texttt{FLUX\_r5}-\texttt{FLUX\_r10}, and investigate the lightcurves for the fixed circle aperture. Second, the candidate events were merged, and  we investigated the light curves and videos in detail; we checked the light curves of reference stars near the target to confirm whether the candidate event is intrinsic to the target. Finally, by checking the video data together with all kinds of light curves (\texttt{FLUX\_r5}-\texttt{FLUX\_r14}, \texttt{FLUX\_AUTO}, and \texttt{FLUX\_ISO}), we exclude contamination from nearby star, meteor and/or the background noise. We also apply the differential imaging~\citep{2013PASP..125..889B,2019AJ....157..218K,2020ApJS..251...18T}, where we compare the integrated images of the target before and during the flare phase, in order to confirm the flare activity.

\section{Results}\label{sec:result}
We have identified 22 flares from different M dwarfs (table \ref{table:star_para} and figure \ref{fig:fit_hr}). Based on the light curve shapes, the flares can be divided into classical flares and complex flares. The former has one single spike while the latter includes multiple spikes during the flare activity. We show the light curves of 20 classical flares in appendix \ref{sec:all_lcs} and two complex flares in subsection \ref{sec:complex_lc}. The light curves are contaminated from nearby stars or atmospheric noise in three cases (TIC 21286088, TIC 2430176281, TIC 98751507), for which we show the details of the analysis in appendix \ref{sec:all_lcs}. 


In the following, we first analyze the classical flares by fitting the light curves by a simple model to derive the basic parameters, i.e., rise time, decay time, peak luminosity, and total emitted energy, and construct an average profile of the light curves. We compare the properties of the detected flares with those reported in previous observations~\citep[e.g.,][]{2014ApJ...797..121H,2016ApJ...820...95K, 2021arXiv211013155H}. After showing the properties of the complex flares, we investigate other activity indicators of the flare stars, such as rotation periods and H$\alpha$ line intensities, to ensure whether the fast optical flares come from magnetically active M dwarfs. 

\input{table_stellar.tex}

\subsection{Classical flare} 
\subsubsection{Light-curve shape}\label{sec:model fitting}
As previously reported cases~\citep[e.g.,][]{2014ApJ...797..122D}, our classical flare light curves show fast rise and exponential decays. Thanks to the high cadence of our observation, a flat top structure of the flare peak is resolved in the most cases; some show a prolonged flare peak. Based on these observations, we adopt the following model for fitting the flare flux relative to the average flux; 

\footnotesize
\begin{eqnarray}
  \hspace*{-2.5cm}
 \frac{F_{\rm flare}(t)}{\bar{F}} \equiv f(t) = \left\{
  \begin{array}{ll}
   \displaystyle f_{\rm peak}\left[\sum_{i=1}^{4}a_{i} \left(\frac{t-t_{\rm peak}}{t_{\rm rise}}\right)^{i} + 1\right]   &  (t_{\rm start} < t < t_{\rm peak}), \\
   f_{\rm peak}   & (t_{\rm peak} < t <t_{\rm peak}+ \Delta t_{\rm peak}), \\
   \displaystyle {\cal C} f_{\rm peak } \exp\left[-\frac{(t-t_{\rm peak}- \Delta t_{\rm peak})}{\tau_{{\rm fast}}}\right]+(1-{\cal C})f_{\rm peak }  \exp\left[-\frac{(t-t_{\rm peak}- \Delta t_{\rm peak})}{\tau_{\rm slow}}\right] &  (t_{\rm peak}+ \Delta t_{\rm peak} < t). \\
  \end{array} \right.
  \label{eq:lc_fit_model}
\end{eqnarray}
\normalsize

\begin{figure*}
\begin{center}
\includegraphics[width=.63 \linewidth]{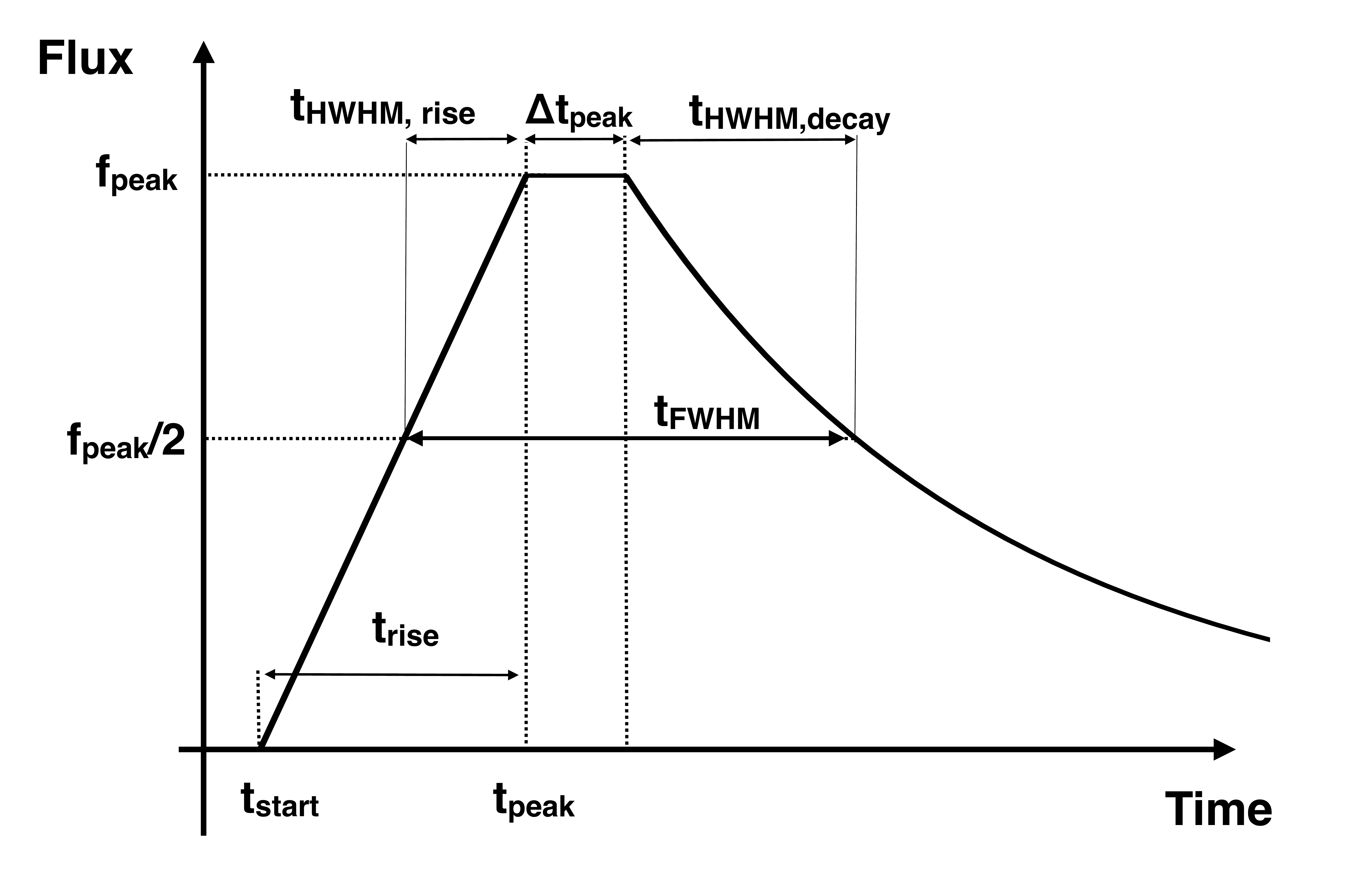}
\end{center}
\caption{Schematic illustration of the flare light-curve model (equation \ref{eq:lc_fit_model}). }
\label{fig:flare_example}
\end{figure*}

Figure \ref{fig:flare_example} illustrates the light-curve model. Here, $\bar{F}$ is the average flux of the star before the flare, $F_{\rm flare}(t)$ is the flare flux, and $f_{\rm peak}$ is the flare peak amplitude. Following \cite{2014ApJ...797..122D}, the rise part is fitted by a 4th-order polynomial function of time normalized by $t_\mathrm{rise}$ with $t_\mathrm{start}$ being the onset of the flare. We have a constraint of  $a_{1} - a_{2} + a_{3} - a_{4} = 1$ on the polynomial coefficients from the continuity condition at $t = t_\mathrm{\rm start}$. The flat peak duration is $\Delta t_\mathrm{peak}$. The decay part consists of two exponential functions with decay times of $\tau_\mathrm{fast}$ and $\tau_\mathrm{slow}$, and ${\cal C}$ represents the ratio of the flare declines between the fast / slow decay phases~\citep[see also][]{2014ApJ...797..122D}. In total, we have 10 free parameters. 

We fit the light curve data shown in appendix \ref{sec:all_lcs} with equation (\ref{eq:lc_fit_model}), performing a Bayesian inference of the model parameters. To sample the posterior distribution, we use Hamiltonian Monte Carlo (HMC e.g. \cite{1987PhLB..195..216D}) with No-U-Turn Sampler (NUTS; \cite{2011arXiv1111.4246H}) implemented on JAX \citep{jax2018github} and NumPyro \citep{2018arXiv181009538B,2019arXiv191211554P}. We take 10,000 samples for the burn-in step, and additional 10,000 samples to construct the posterior distribution. In the sampling, we impose a prior so that $f_\mathrm{flare}(t)$ and its time derivative are positive in the rise part ($ t_{\rm start} < t < t_{\rm peak}$). We confirm that the sampling is converged; the effective number of samples $n_{\rm eff}$ is sufficiently large and a $r$-hat value is close to 1.0 in all the fitting. After the fitting, we also compute the full width at half maximum (FWHM) of the flare peak and the half-width at half maximum (HWHM) for the rise and decay phases,  $t_{\rm FWHM}$, $t_{\rm HWHM, rise}$, and $t_{\rm HWHM, decay}$, respectively.

\input{flare_mcmc.tex}
\input{flare_fwhm.tex}

\begin{figure*}
\begin{center}
\includegraphics[width=0.45 \linewidth]{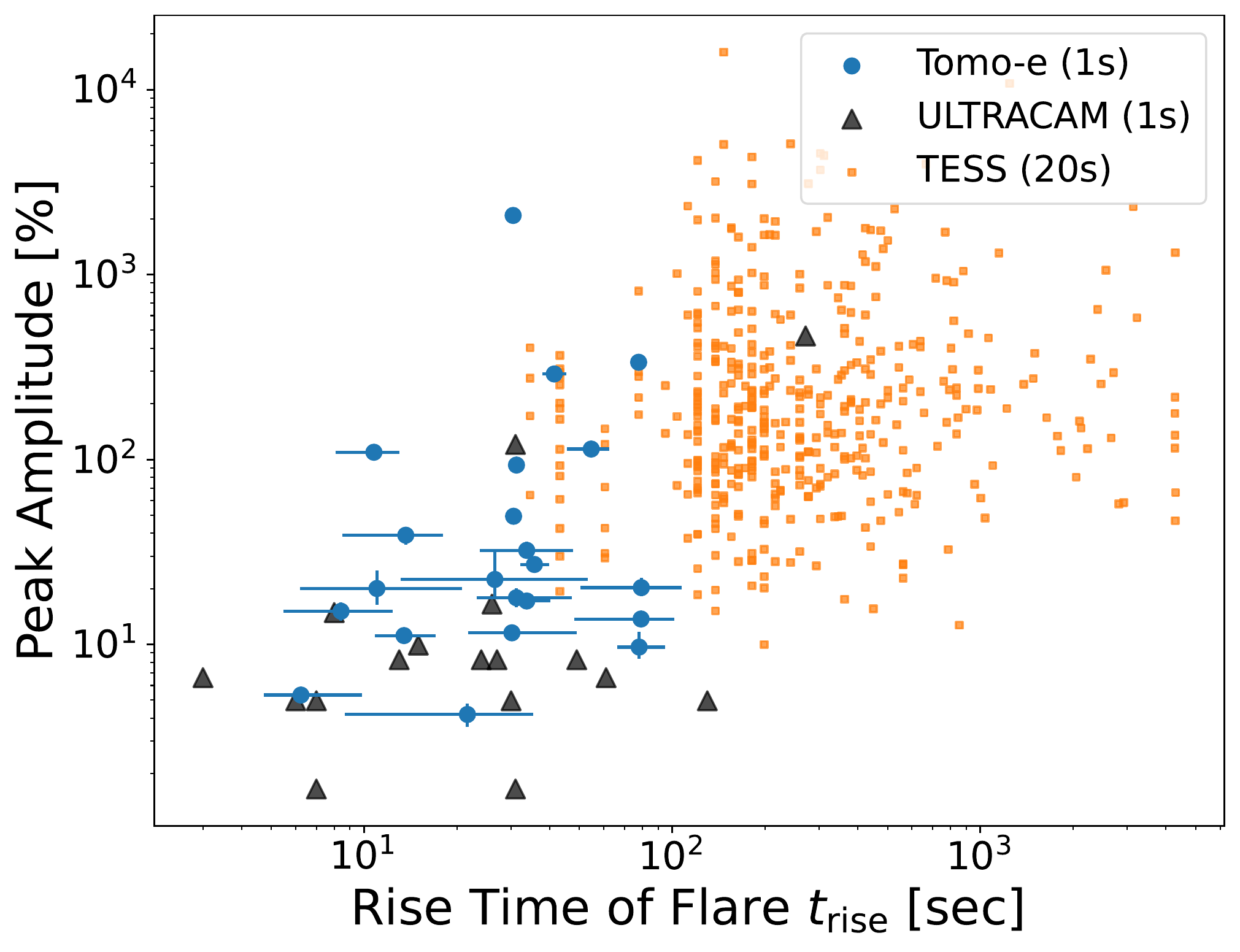}
\includegraphics[width=0.45 \linewidth]{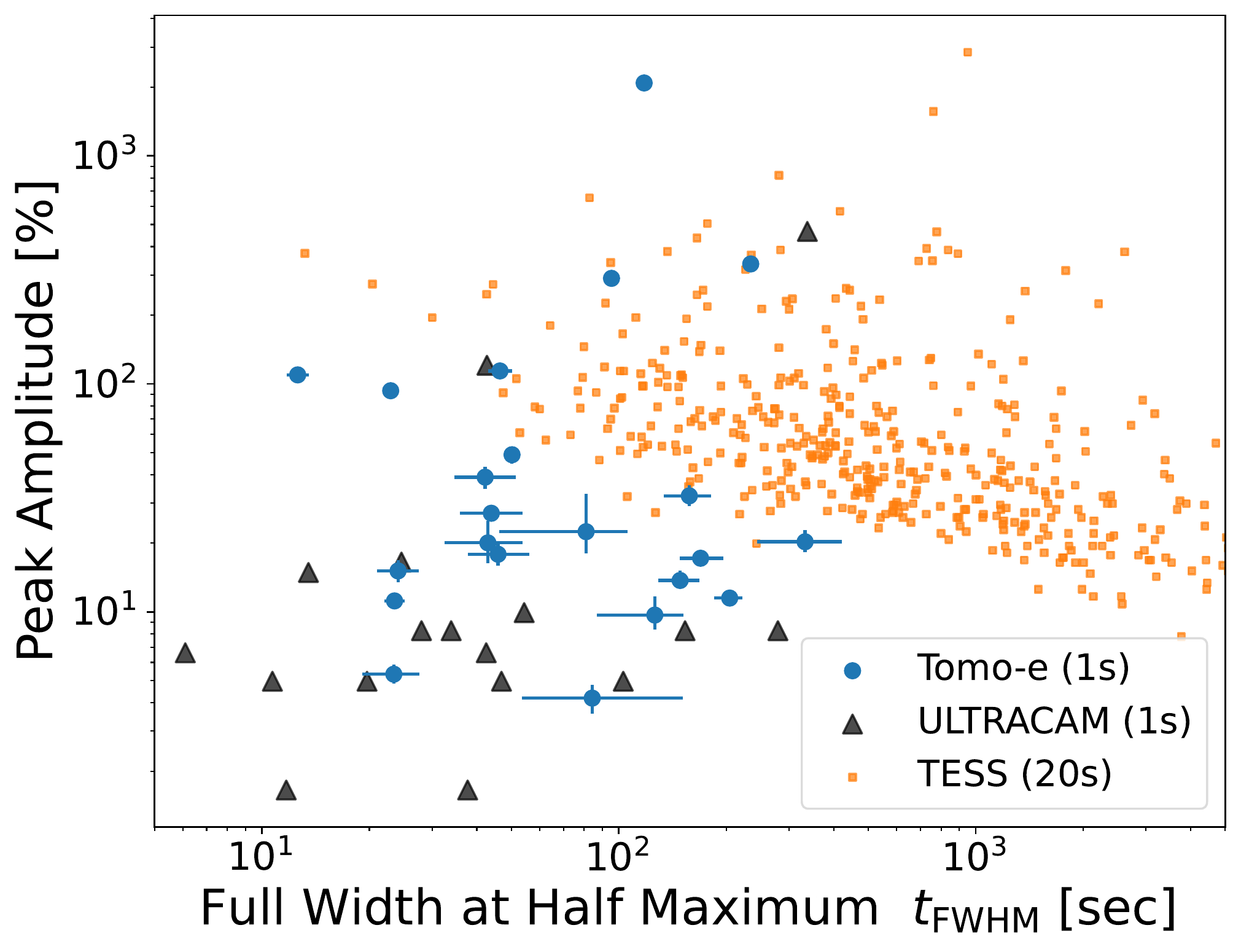}
\end{center}
\caption{(left) Rise time and peak amplitude of optical flares from M dwarfs. The blue circles, black triangles, and orange squares represent the detected flares by Tomo-e Gozen (this work), ULTRACAM~\citep{2016ApJ...820...95K}, and TESS~\citep{2021arXiv211013155H}, respectively. (right) FWHM durations vs peak amplitudes of the same flares.}
\label{fig:time_vs_amp}
\end{figure*}

\begin{figure*}
\begin{center}
\includegraphics[width=0.43 \linewidth]{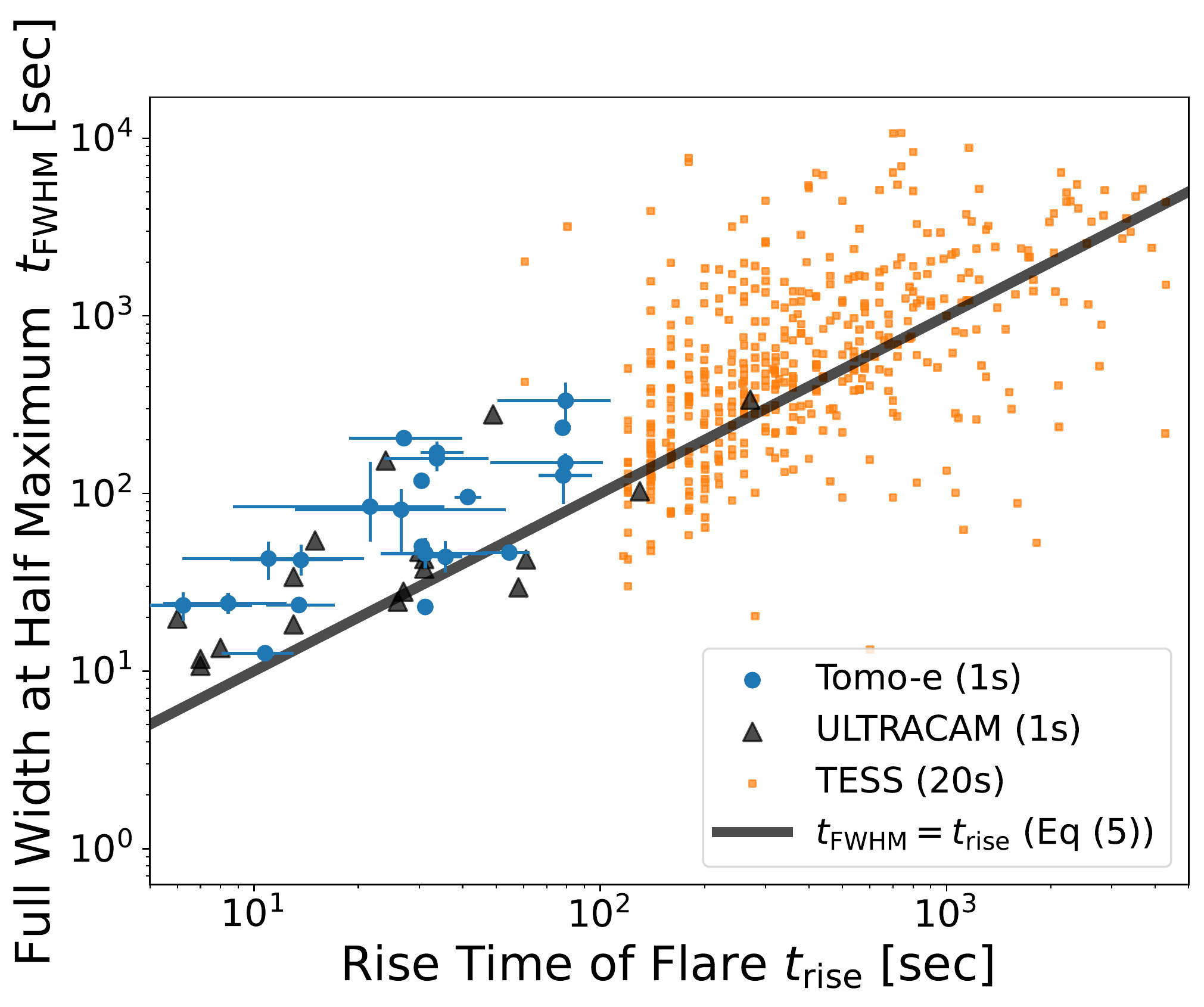}
\includegraphics[width=0.48 \linewidth]{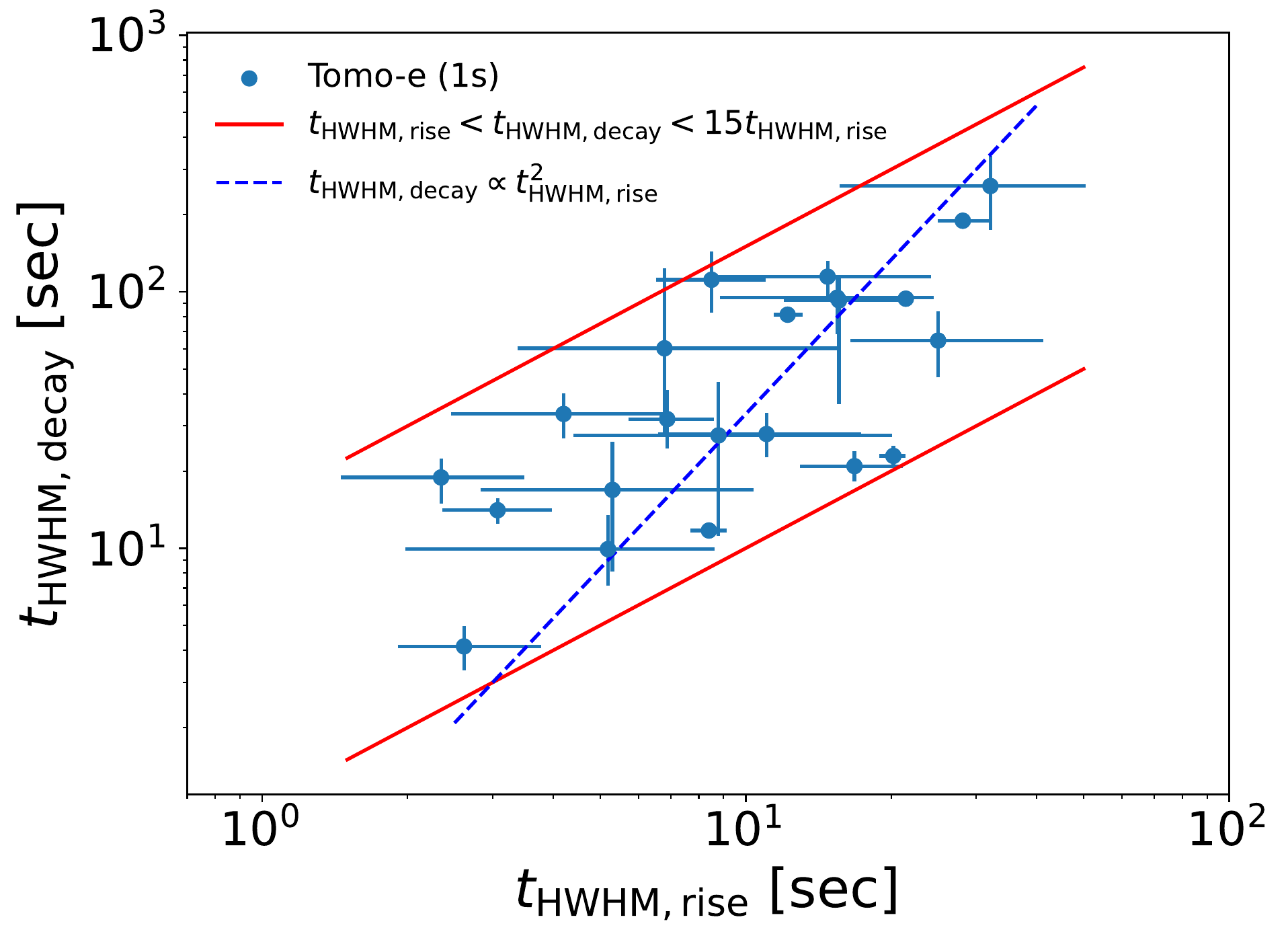}
\end{center}
\caption{(left) Rise time and full width at half maximum (FWHM) of optical flares from M dwarfs. The grey solid line corresponds to $t_{\rm FWHM} =  t_{\rm rise}$. (right) Comparison of half width at half maximum for rise and decay phases. The red solid lines correspond to $t_{\rm HWHM, rise} < t_{\rm HWHM, decay}<15 t_{\rm HWHM, rise}$, and the blue dashed line expresses the relation $t_{\rm HWHM, decay} \propto t_{\rm HWHM, rise} ^{2}$ as discussed in subsection \ref{sec:time}.  }
\label{fig:trise_vs_t_12}
\end{figure*}

The obtained model parameters with $1\,\sigma$ errors are shown in tables \ref{table:flare_paras}--\ref{table:flare_fwhm}. Note that we miss the rise phase of the flare light curve of TIC 21286574. We also plot the flare timescales and amplitudes in figures \ref{fig:time_vs_amp}--\ref{fig:trise_vs_t_12}. The detected flares are all fast; the rise times are all less than $100\,\mathrm{sec}$ and even less than $10\,\mathrm{sec}$ in some cases. Most of the flares have amplitudes ranging from 5\% to 100\% while several cases have amplitudes larger than 100\%, up to $\sim$ 2,000\%. The decay phase is reasonably fitted by the two exponential functions and the extent of the decay is always dominated by the fast decay phase, i.e., ${\cal C} > 0.5$. Note that the uncertainty of $\tau_\mathrm{slow}$ is relatively large in some cases (e.g. TIC 374272810) when the observation is incomplete in the late phase of the flare. There can be a weak correlation between the flare timescale and amplitude (figure \ref{fig:time_vs_amp}); slower flares have larger peak amplitudes. On the other hand, we confirm a correlation between the rise and decay times as $t_\mathrm{HWHM,decay} \propto t_\mathrm{HWHM,rise}{}^{p}$ with $1 \lesssim p \lesssim 2$ (figure \ref{fig:trise_vs_t_12}). We also find that the rise time is always shorter than the decay time by up to $\sim 15$ times.

In figures \ref{fig:time_vs_amp}--\ref{fig:trise_vs_t_12}, we also show previously detected optical flares by TESS and ULTRACAM. The TESS flares are from \cite{2021arXiv211013155H}; 3,792 flares with bolometric emitted energies of $E_\mathrm{bol} > 10^{32}\,\mathrm{erg}$ are detected in the TESS 20-second-cadence data for 226 active M dwarfs (Cycle 3 GI Program 3174) . 
For the TESS flares, $t_\mathrm{rise}$ is defined as the time difference between the flare start time and the flare peak time and $t_\mathrm{FWHM}$ are estimated only for 440 flares with an equivalent duration longer than 50 sec. The ULTRACAM flares are from \cite{2016ApJ...820...95K}, where 22 flares are detected from 5 active M dwarfs by the one-second-cadence observation with the three narrow band filters at 3,500$\AA$, 4,179$\AA$, and 6,010$\AA$. For the flare parameters, we refer to their table 6. As for the flare amplitude, we take into account the difference of the response functions of Tomo-e Gozen, TESS, and ULTRACAM; the observed flare amplitudes by each detector are proportional to $\int d \lambda \; B(\lambda, T_{\rm flare}) S(\lambda)/\int d \lambda \; B(\lambda, T_{\star}) S(\lambda)$ and the ratio of this factor should be multiplied for a fair comparison. Here $B(\lambda, T)$ is the radiation intensity of a blackbody, $T_\mathrm{\star}$ is the stellar temperature, $T_\mathrm{flare}$ is the temperature of the blackbody component of the flare, and $S(\lambda)$ is the response function of the detector (see equation \ref{eq:calc_f} in sub-subsection \ref{sec:flareenergy}). We adopt the official response function~\footnote{https://heasarc.gsfc.nasa.gov/docs/tess/the-tess-space-telescope.html} for the TESS flares and a $\delta$ function centered at $\lambda =$ 6010$\,\AA$ for the narrow band observation by ULTRACAM. The response function of Tomo-e Gozen is from \cite{2018SPIE10709E..1TK} with linearly extrapolating it for $\lambda < 3,700\,\AA$. Assuming $T_{\star}=3,300\,\mathrm{K}$ and $T_{\rm flare}=9,000-15,000\,\mathrm{K}$,  the correction factor is $3.5-4.9$ for the TESS flares and $1.5-1.8$ for the ULTRACAM flares. We adopt the correction factors for $T_{\rm flare}=12,000 \mathrm{K}$ in the following. 

Although there are some overlaps between the distributions of the flare parameters, our flare sample constitutes a population different both from the TESS and ULTRACAM flares. In figure \ref{fig:time_vs_amp}, our sample extends to shorter rise times compared with the TESS flares; by construction, the TESS 20-second-cadence search can only detect or time-resolve flares with $t_\mathrm{rise} \geq 20\,\mathrm{sec}$. Our results clarify that O(1) sec cadence observation is crucial for searching for the fast end of optical flares. Note that the total observation time by \cite{2021arXiv211013155H} is $\sim$ 20,000 active M-dwarf days, and the effective flare search duration is 700--1,000 times longer than our survey given that only 20\% of our targets are active stars (see sub-subsection\ref{sec:event_rate}). Thus, the TESS 20-sec-cadence search is much more complete for flares with $t_\mathrm{rise} \gtrsim 20\,\mathrm{sec}$. 
On the other hand, our sample extends to higher amplitudes compared with the ULTRACAM flares; the total observation time by \cite{2016ApJ...820...95K} is $\sim $ 2 active M-dwarf days, which is one  order-of-magnitude shorter than our survey for active M dwarfs while it has a better sensitivity with a larger telescope aperture size of $\sim 4$ m. Accordingly, our survey is more sensitive to brighter but less frequent flares. In fact, we detect the bright end of fast optical flares (see figure \ref{fig:time_vs_amp}). In figure \ref{fig:trise_vs_t_12}, we find that the correlation between the rise and decay times is broadly consistent between the different observations. 

\begin{figure*}[h]
\begin{center}
\includegraphics[width=0.45 \linewidth]{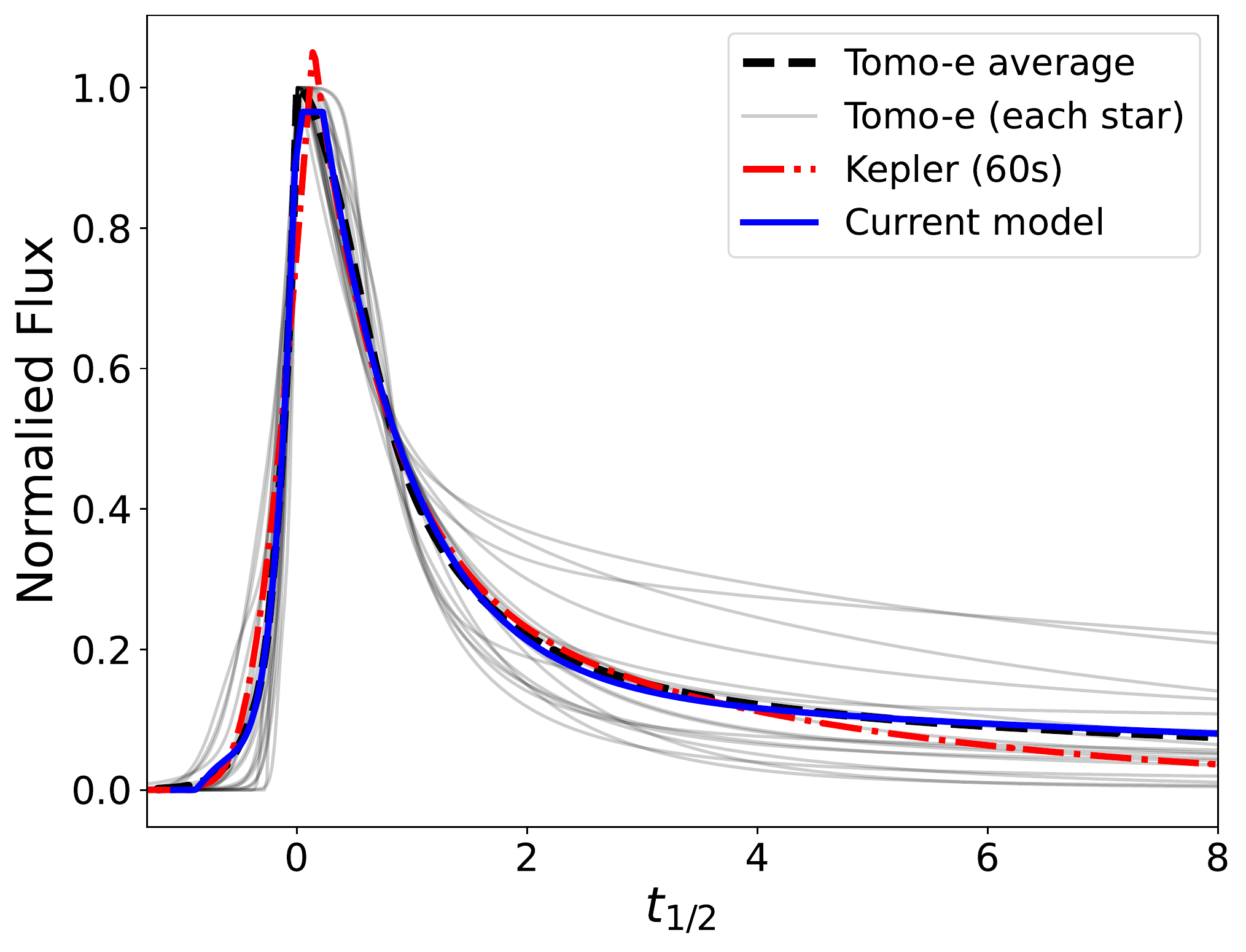}
\includegraphics[width=0.45 \linewidth]{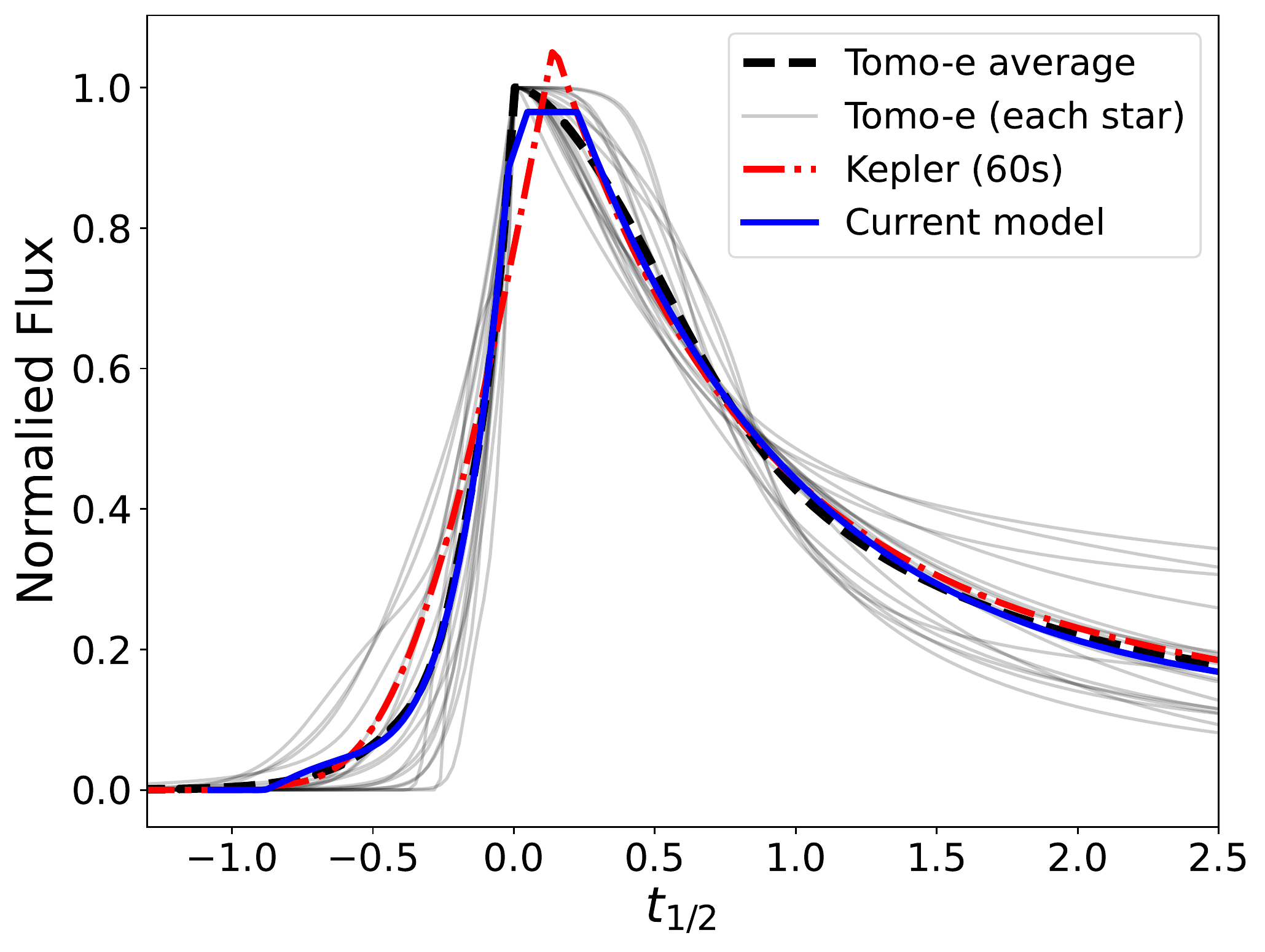}
\end{center}
\caption{(left) Normalized flare light curves of the 18 classical flares (gray lines) and their average profile (black dashed line). The time $t_{1/2}$ is also normalized by the full-width-half-maximum (FWHM) of each flare (equation \ref{eq:t_1_2}). The blue line shows our flare light curve template obtained by fitting the average profile. The red dash-dotted line shows the template obtained by the one-minute-cadence Kepler observation for GJ 1243~\cite{2014ApJ...797..122D}. (right) Zoom-up view of the left panel.  }
\label{fig:average_profile}
\end{figure*}

For an additional comparison between the current flare samples and previously detected ones, we construct an average profile of our flare light curves. To this end, we first rescale each light curve with respect to the flare peak, i.e., $f(t) \rightarrow \bar{f}(\mathrm{t_\mathrm{1/2}}) \equiv f(t_\mathrm{1/2})/f_\mathrm{peak}$, where 
\begin{equation}
t_{\rm 1/2}\equiv \frac{t-t_{\rm peak}}{t_{\rm FWHM}} \label{eq:t_1_2}. 
\end{equation} 
Second, we marginalize each rescaled light curve over the posterior distribution of the model parameters (grey lines in figure \ref{fig:average_profile}). Third, we take the simple mean of the rescaled and marginalized light curves (black dashed line in figure \ref{fig:average_profile}). In this calculation, we exclude complex flares (TIC 15904458 and TIC 358561826), a flare contaminated by atmospheres (TIC 20536892), a flare with no data for the rise part (TIC 21286574). Finally, we re-fit the average profile with the light curve model with rescaling $f_\mathrm{peak}$ and $t_\mathrm{peak}$. The derived template shape is

\footnotesize
\begin{equation} 
  \hspace*{-2.0cm}
     \bar{f}_{\rm Tomo-e}(t_{1/2})=  \left\{ \begin{array}{ll}
  1 + 4.54 (t_{1/2}/t_{\rm rise}) + 8.31 (t_{1/2}/t_{\rm rise})^{2}  + 6.82 (t_{1/2}/t_{\rm rise})^{3} + 2.05 (t_{1/2}/t_{\rm rise})^{4}  &(-t_{\rm rise} < t_{1/2} < 0),\\
  1 & (0 < t_{1/2} < 0.23),  \\ 
    0.85\exp(- 1.28 t_{1/2}) + 0.15 \exp(- 0.080 t_{1/2}), &  (0.23 < t_{1/2}), 
  \end{array} \right.
\label{eq:current_model} 
\end{equation}
\normalsize

where $t_{\rm rise} = 0.88$ and shown in figure \ref{fig:average_profile} (blue solid line). 
For comparison, we also show the flare light curve template obtained by \cite{2014ApJ...797..122D} based on the one-minute Kepler light curves of GJ 1243: 
\footnotesize
\begin{equation}
  \hspace*{-2.0cm}
    \bar{f}_{\rm Kepler}(t_{1/2})=  \left\{ \begin{array}{ll}
    1 + 1.941 (t_{1/2}/t_{\rm rise})- 0.175 (t_{1/2}/t_{\rm rise})^{2}  - 2.246 (t_{1/2}/t_{\rm rise})^{3} - 1.125 (t_{1/2}/t_{\rm rise})^{4} & (-t_{\rm rise} < t_{1/2} < 0),\\
    0.689 \exp(- 1.6t_{1/2}) + 0.303 \exp(-0.2783 t_{1/2}) &  (0 < t_{1/2}), 
  \end{array} \right.
\label{eq:daven}
\end{equation}
\normalsize

where $t_{\rm rise} = 1$ in this case. 
The average profile of our flares and the Kepler flares are roughly consistent but there are some noticeable differences. Our flares show steeper rise as shown in figure \ref{fig:trise_vs_t_12}, and the flare top structure is resolved in our template while $t_\mathrm{rise} = t_\mathrm{FWHM}$ and the flare top structure is not considered in the Kepler template. Indeed, some of the flares show clear flare top features in their lightcurves: TIC 98751507, TIC 119031076, TIC 185243119, TIC 243017627, TIC 374272810, and TIC 380197137. In addition, our flares show shallower decay; the slow decay phase has a longer duration in our template. 


\subsubsection{Peak luminosity and total emitted energy} \label{sec:flareenergy}
In order to estimate the peak bolometric luminosity and total emitted energy of an optical flare, one needs to have spectral information. In general, the spectrum at the flare peak consists of the blackbody component with an temperature of $T_\mathrm{flare} \gtrsim 10^4\,\mathrm{K}$, and the Paschen and Balmer recombination emissions~\citep[e.g.,][]{2016ApJ...820...95K}. The Tomo-e Gozen bandpass filter mainly detects the blackbody component. In this case, the observed flare amplitude can be connected to the flare temperature $T_\mathrm{flare}$ and the effective size of the emission region $A_\mathrm{flare}$ as follows~\citep[e.g.][]{2013ApJS..209....5S,2017ApJ...851...91N}: 
\begin{equation}\label{eq:calc_f}
    f(t) = \frac{A_{\rm flare}(t)}{A_{\rm \star}} \frac{ \int d \lambda \; B(\lambda, T_{\rm flare}) S(\lambda)}{\int d \lambda \; B(\lambda, T_{\star}) S(\lambda)} \label{eq:flare_frac},
\end{equation}
where $A_{\rm \star} = \pi R_{\star}^{2}$ is the projected area of the stellar surface. 
Then, the peak bolometric luminosity of the flare can be given as 
\begin{eqnarray}
L_{\rm peak} = \sigma_{\rm SB} T_{\rm flare}^{4} A_{\rm flare}(t_\mathrm{peak}) =   \pi R_{\star}^{2} \sigma_{\rm SB} T_{\rm flare}^{4}  \frac{\int d \lambda \; B(\lambda, T_{\star}) S(\lambda)} { \int d \lambda \; B(\lambda, T_{\rm flare}) S(\lambda)} f_\mathrm{peak}, 
\end{eqnarray}
where $\sigma_{\rm SB}$ is the Stefan–Boltzmann constant, $S(\lambda)$ is the response function of the instrument. The length scale of the emission region, which would be comparable to the length scale of the reconnecting field, can be also estimated from the peak flux as 
\begin{equation}
l_{\rm peak} =\sqrt{ A_\mathrm{flare} (t_\mathrm{peak})} = \sqrt{ \pi R_{\star}^{2} \frac{\int d \lambda \; B(\lambda, T_{\star}) S(\lambda)} { \int d \lambda \; B(\lambda, T_{\rm flare}) S(\lambda)} f_{\rm peak}}.  \label{eq:length}
\end{equation}
The total emitted energy can be estimated as 
\begin{equation}
E_{\rm bol} = \sigma_{\rm SB} T_{\rm flare}^{4} \int  A_{\rm flare}(t) \; dt = \pi R_{\star}^{2} \sigma_{\rm SB} T_{\rm flare}^{4}  \int \frac{\int d \lambda \; B(\lambda, T_{\star}) S(\lambda)} { \int d \lambda \; B(\lambda, T_{\rm flare}) S(\lambda)} f(t) dt. \label{eq:e_bol}
\end{equation}
When computing the bolometric energy with equation (\ref{eq:e_bol}), we substitute a fitted mean model from the posterior distribution to $f(t)$, and integrate it over the time range shown in the left panels of figure \ref{fig:lc_flares}. In some cases, we miss the later phase of the flare light curve, for which $E_\mathrm{bol}$ is underestimated.

\input{flare_mcmc_energy.tex}

\begin{figure*}
\begin{center}
\includegraphics[width=0.65 \linewidth]{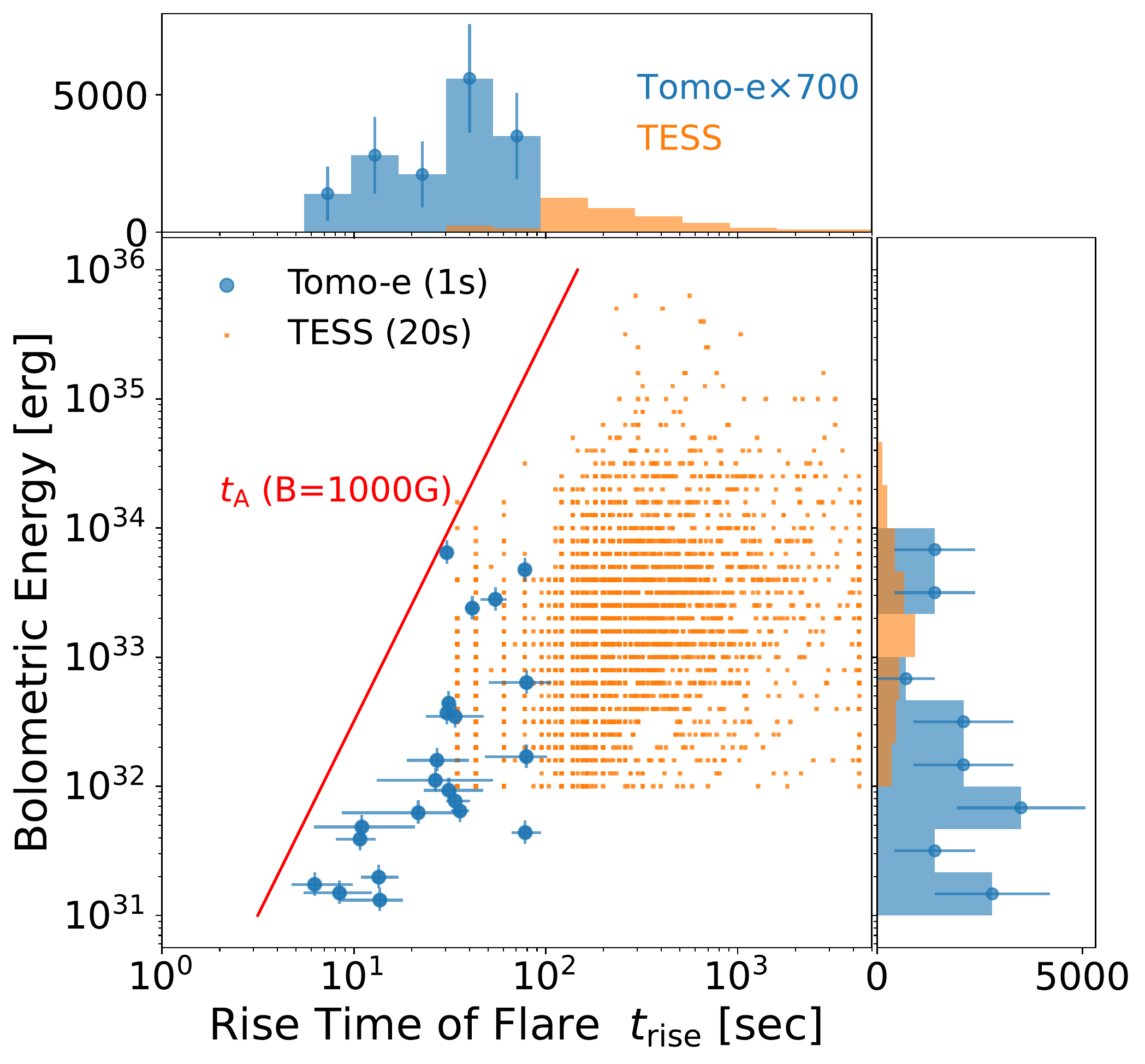}
\end{center}
\caption{Rise time vs bolometric energy of flares detected by Tomo-e Gozen (blue circles). Orange squares indicate flares detected by TESS~ \citep{2021arXiv211013155H}. In calculating the bolometric energy, we adopt $T_{\rm flare} =12000\mathrm{K}$ as a reference value, and $9,000$ and $15,000\,\mathrm{K}$ for computing the lower and upper limits, respectively. The upper and right panels show the histograms with respect to rise time and bolometric energy, respectively, where a factor 700 is multiplied to the number of the Tomo-e Gozen flares given the difference of the effective total observation time for active M dwarfs. The red line indicates a relation $E_\mathrm{bol} \propto t_\mathrm{rise}^3$, which is expected from a simple flare model (see section \ref{sec:discussion}). We show the case with a magnetic field strength of $B = 1,000\,\mathrm{G}$.}
\label{fig:trise_vs_Ebol}
\end{figure*}

Table \ref{table:flare_energy} summarizes the peak luminosity, length scale of the emission region, and total emitted energy. We adopt $T_{\rm flare}=12,000 \mathrm{K}$ as a reference value, and $T_{\rm flare}=9000 \mathrm{K}$ and $15,000 \mathrm{K}$ for computing lower and upper limits; in this case, the uncertainty of $E_\mathrm{bol}$ is about 20-30\%. We also plot them with respect to the flare timescales in figure \ref{fig:trise_vs_Ebol}. The peak luminosities and bolometric energies range from $10^{29}\,\mathrm{erg\,sec^{-1}} \lesssim L_\mathrm{peak} \lesssim 10^{31}\,\mathrm{erg\,sec^{-1}}$ and $10^{31}\,\mathrm{erg} \lesssim E_{\rm bol} \lesssim 10^{34}\,\mathrm{erg}$. 
In figure \ref{fig:trise_vs_Ebol}, we also plot the TESS flares reported by \cite{2021arXiv211013155H} where $T_\mathrm{flare} =  9000\,\mathrm{K}$ is assumed. We also show the histograms with respect to the rise time and the total emitted energy in the right and upper panels, respectively, where we multiply 700 for our flares given the difference of the effective flare search duration~\footnote{We omit TIC 21286088 from the histogram because it is not in our original target list~(see appendix \ref{sec:all_lcs})}. We see again that our survey and the TESS survey are complementary; our survey is more sensitive to flares with $t_\mathrm{rise} < 20-100\,\mathrm{sec}$ while the TESS survey is much more complete for flares with $t_\mathrm{rise} \gtrsim 20\,\mathrm{sec}$. The absence of TESS flares with $E_\mathrm{bol} < 10^{32}\,\mathrm{erg}$ is merely due to the selection criterion. From the histograms, the two populations are implied to be connected both in $t_\mathrm{rise}$ and $E_\mathrm{bol}$ although the uncertainties are still large mainly due to the limited sample size of our flares.  The red solid line in figure \ref{fig:trise_vs_Ebol} indicates the relation $E_\mathrm{bol} \propto t_\mathrm{rise}^3$, which is expected based on a dimensional argument for flares produced by reconnection of a magnetic loop~(see section \ref{sec:discussion}). The detected flares are broadly compatible with this relation.

\subsection{Complex flare} \label{sec:complex_lc}
Our model fitting suggests that some of the flares show deviations from classical flare models. Particularly, the flare light curves of TIC 358561826 and TIC 15904458 clearly show multiple peaks (figure \ref{fig:tic_complex}). We here comment on the properties of these complex flares. We also derive flare parameters, energy, and luminosities of the main peaks, and values are listed in table \ref{table:flare_paras}--\ref{table:flare_energy}. 

\subsubsection{TIC 15904458}
The light curve of TIC 15904458 shows two clear peaks with a time interval of $\sim 1$ min. We simultaneously fit the two peaks separately with the light curve model for the classical flare (equation \ref{eq:lc_fit_model}). The obtained flare parameters are not distinct from those of the classical flares. The residual of the light curve after subtracting the two main peaks is shown in the bottom panel of figure \ref{fig:tic_complex}. There can be the third and fourth peaks at around 150 sec and 330 sec after the first peak. These multiple peaks might correspond to quasi-periodic pulsations (QPP)~\citep{2009SSRv..149..119N,2016SoPh..291.3143V}. The periods of QPPs observed by TESS and Kepler range from a few minutes to a few tens of minutes \citep{2016MNRAS.459.3659P, 2021arXiv211013155H}. Since the time interval for the flare is about 1 minute, this event is one of the fastest QPP events observed in the optical bands. Similar flares with short intervals are also seen in the one-second cadence flare light curves of YZ CMi \citep{2016ApJ...820...95K}. 

\subsubsection{TIC 358561826}
\begin{figure*}
\begin{center}
\includegraphics[width=0.45 \linewidth]{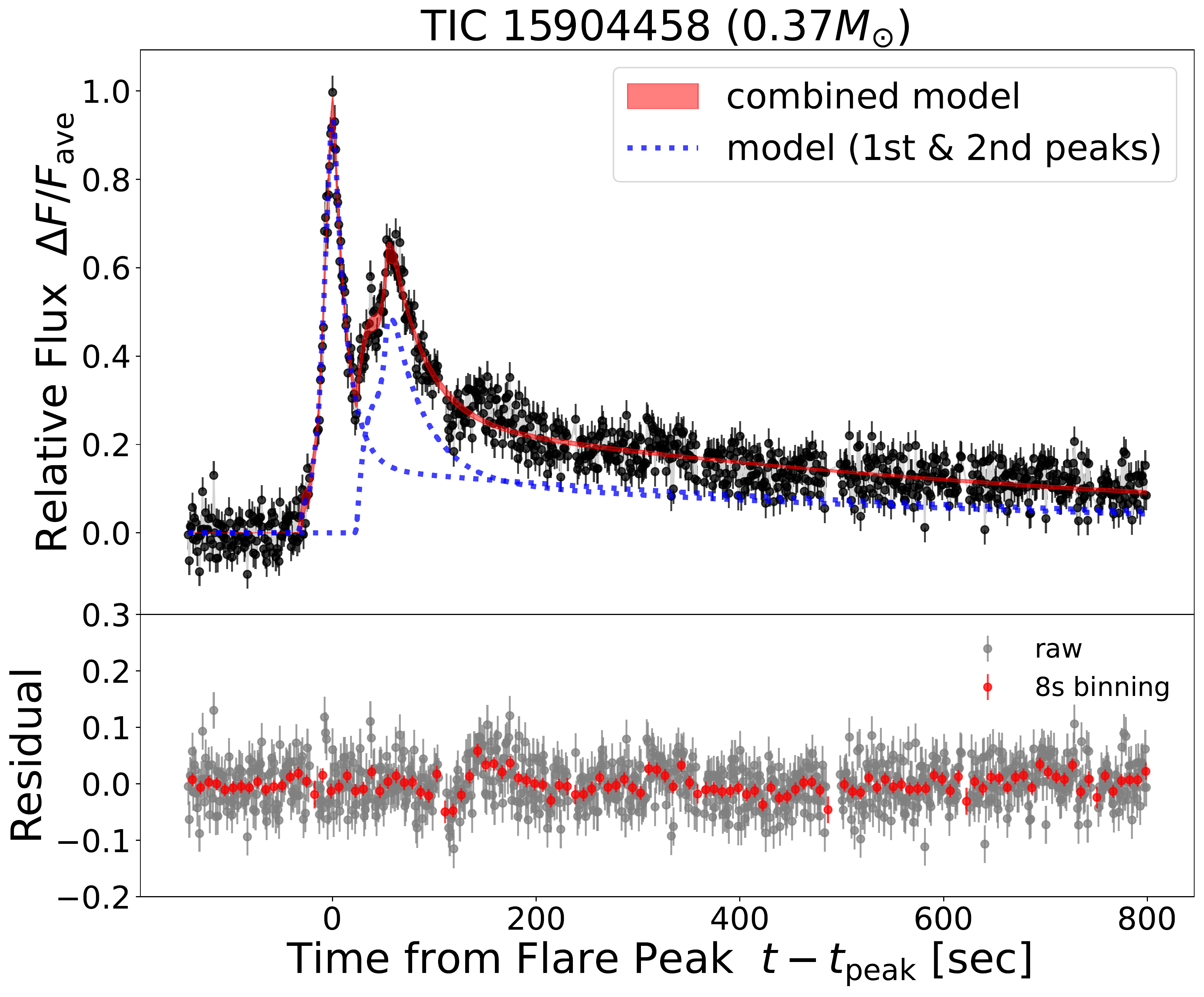}
\includegraphics[width=0.43 \linewidth]{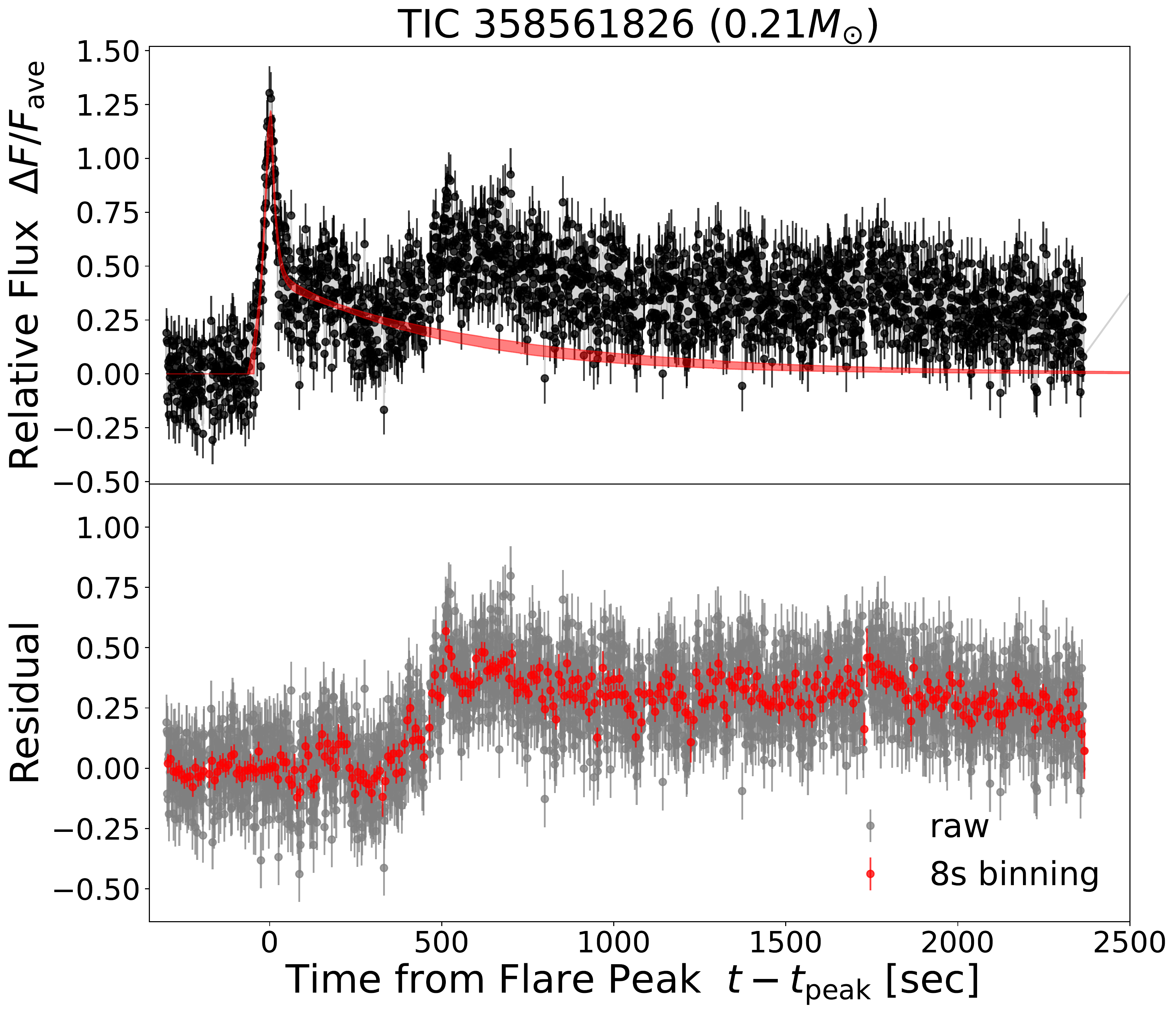}
\end{center}
\caption{(left) Flare light curve of TIC 15904458 (black points). The blue dotted lines show the fitted models for the first and second peaks. The red shaded region corresponds to the 90 \% credible interval of the light curve model (equation \ref{eq:lc_fit_model}) defined by the posterior distributions of the model parameters. The bottom panel shows the raw residuals after subtracting the fitted models (gray points) along with a binned light curve over $8$ sec (red points). (right) Flare light curve of TIC 358561826. Only the first peak can be fitted by the light curve model.  }
\label{fig:tic_complex}
\end{figure*}

The flare light curve of TIC 358561826 is composed of a sharp peak and a hump. Similar structures have been reported for optical flares detected by TESS~\citep{2020AJ....159...60G, 2021MNRAS.504.3246J, 2021arXiv211013155H}.  We fit the main peak with the light curve model for the classical flare (equation \ref{eq:lc_fit_model}), which is shown as the red line in figure \ref{fig:tic_complex}. Note that we only calculate energy and other parameters for the main peak because the flare likely continues after the end of our observation. Looking at the residual light curve, the hump can be divided into two additional peaks at around 500 sec and 1800 sec after the main peak. It is difficult to disentangle the two components due to their significant overlap, thus we do not conduct further analysis for the hump. One may notice a possible periodic feature with a period of $\sim 120$ sec in the residual light curve. However, this most likely corresponds to the period inherent to the observation mode; in the current data set, there are intervals of 7-8 sec after every 120 sec of observation.   

\subsubsection{Fraction of complex flares}
We identify 2 clear complex flares out of 22 events. Both of the complex flares have large amplitudes $\sim 100\%$ and long total duration of a few tens of minutes, suggesting that the long and energetic flares tend to have complex structure. The flares with relatively short duration do not show any multiple clear peaks even if they have large amplitudes (e.g. TIC 380197137). This is consistent with the previous finding  \citep{2014ApJ...797..122D,2021arXiv211013155H}. \cite{2014ApJ...797..122D} found that the fraction of complex flares increases with flare duration for GJ1243, and the rate is nearly zero for flares with a duration of 10 minutes. \cite{2021arXiv211013155H} also found that the fraction of complex flare is 42.3\% for the total flare samples in the TESS 20-second-cadence data, and the fraction increases to be 70\% for flares with $E_{\rm bol}>10^{35}\,\mathrm{erg}$. 

Though we do not identify clear multiple peaks for the other light curves, it can be due to the difficulty of identifying sub-structures in relatively low-precision data. Indeed, some of the flares have small or moderate deviations from classical shapes, exhibiting oscillation features or multiple peaks; e.g., the light curve of TIC 185243119 may include the second peak at $\sim$ 100 sec after the first peak. It will be important to confirm the true complex fraction by deeper observations and analyses.

\subsection{Stellar activities of the flare stars}
Flares occur more frequently at magnetically active and/or mid-late-type M dwarfs \citep{2017ApJ...849...36Y,2020ApJ...905..107M}. As shown in table~\ref{table:star_para} and figure~\ref{fig:fit_hr}, the flare stars are mid-late-type M dwarfs (M3-M6) with red colors, $G_{\rm bp} -G_{\rm rp}>2.5$. To study the stellar properties in detail, we investigate the rotation periods, the Rossby numbers, and the H$\alpha$ emission line intensities of the flare stars, all of which are known to be strongly correlated to the magnetic activity of M dwarfs.

\subsubsection{Rotational periods from the TESS data} \label{sec:rot_tess}
To determine the rotation periods, we search the available TESS data for both light curves and Full-Frame images (FFIs) of the 22 M dwarfs with flare detection. All of the flare stars have FFI data for at least one Sector. For TIC 17198188, TIC 18376490, TIC 20536892, TIC 21286574, TIC 55288759, TIC 119031076, and TIC 25172568, we find 2-minute cadence light curves and target pixel files presented by the Science Processing Operations Center (SPOC). After removing outliers in the light curve data using a sigma clipping of $5\sigma$, we remove the low-frequency trend using the Savitzky-Golay filter with a window width of $\sim 8$ day. Then, we calculate the periodograms to search for periodicities. We find that TIC 17198188, TIC 18376490, TIC 20536892, TIC 21286574, and TIC 119031076 show strong peaks at 0.39, 0.38, 2.1, 0.26, 1.57 days in their frequency power spectra, respectively. The periodogram of TIC 55288759 also shows a marginal peak at 5.5 days while TIC 251725681 shows no clear periodic signal. We apply differential imaging to the pixel data by comparing the images around the flux maxima and minima\footnote{https://docs.lightkurve.org/tutorials/3-science-examples/periodograms-verifying-the-location-of-a-signal.html} to confirm whether the centroids of the periodic signals are consistent with the target coordinates. We also calculate the periodograms of the light curves of nearby pixels to the centroids to see if the periodicity can be caused by systematic noise. Based on these analyses, we confirm that all of periodic signals originate from the target stars.

For the other M dwarfs without TESS light curve or target pixel files, we analyze FFI data. We make the light curves from the simple aperture photometry, in which we manually choose the aperture region. Then, we remove the data that show significant trends, in particular, those at the beginning and the end of the light curve. We found clear peaks for 7 stars: TIC 15904458, TIC 46305452, TIC 121638493, TIC 243017627, TIC 315617164, TIC 374272810, TIC 435904068. Though the light curves of TIC 46305452, TIC 243017627, and TIC 315617164 are affected by nearby stars, we confirm that the signals are from the targets again by differential imaging, centroid analyses, and periodograms of nearby pixels to the centroids. 

The obtained rotation periods are summarized in table \ref{table:star_para}. We estimate the convective turnover time by using the empirical relations $\tau_{\rm conv}$ and mass $\log_{10} [\tau_{\rm conv}/{\rm day}]=2.33^{+0.06}_{-0.05}+1.50^{+0.21}_{-0.20}(M_{\star} /M_{\odot} ) + 0.31^{+0.16}_{-0.17} (M_{\star} /M_{\odot} ) ^{2} $ \citep{2018MNRAS.479.2351W}, to calculate the Rossby number $R_{o} = P_{\rm rot}/\tau$. We find that the Rossby number is much smaller than 0.14 in all the 12 cases with clear signals of rotation periods, indicating that their stellar activities are likely at maximum levels~\citep{2018MNRAS.479.2351W}.

\subsubsection{$\mathrm{H}{\alpha}$ emission line intensity from the LAMOST data \label{sec:halpha_lamost}} 
We first cross-match the list of our flare stars with the Large Sky Area Multi-Object Fiber Spectroscopic Telescope (LAMOST) DR7 v2 catalog~\citep{2012RAA....12.1197C,2015RAA....15.1095L} and obtain spectroscopic data for 11 stars (see table \ref{table:star_para}). We then compute the ratio of the H$\alpha$ emission line luminosity to the bolometric luminosity $L_{\mathrm{H}{\alpha}}/L_{\rm bol}$ using the equivalent width of the  emission line. We find that 10 out of the 11 stars show strong H$\alpha$ emission with $L_{\mathrm{H}{\alpha}}/L_{\rm bol} > 1.5 \times 10^{-4}$. Only TIC 358573853 shows a relatively weak $\mathrm{H}{\alpha}$ emission with $L_{\mathrm{H}{\alpha}}/L_{\rm bol} \sim 1.5 \times 10^{-5}$. The results are summarized in table \ref{table:star_para}. According to the \cite{2017ApJ...834...85N}, the activity level of M dwarfs is saturated at $L_{\mathrm{H}{\alpha}}/L_{\rm bol} \gtrsim 1 \times 10^{-4}$ for fast rotators, or those with small Rossby numbers $R_{o} \lesssim 0.1$. In addition to the LAMOST data, we also search the previous references, and find that TIC 20536892 is an active star with $L_{\mathrm{H}{\alpha}}/L_\mathrm{bol} = 1.18 \times 10^{-4}$ and $L_\mathrm{NUV}/L_\mathrm{bol} = 2.5\times 10^{-3}$~\citep{2016ApJ...817....1J}. All of the 11 M dwarfs with strong H$\alpha$ emission satisfy this criterion. 

\subsubsection{Event rate of fast optical flares} \label{sec:event_rate}
In summary, we find that 12/22 of the flare stars are magnetically active based on their rotation periods measured from the TESS data and 10/11 of the flare stars based on the H$\alpha$ emission measured from the LAMOST data. Since detection of periodic signature associated with rotation is not guaranteed depending on the visibility of the stellar spots, the ratio obtained from the LAMOST data is more reliable. Accordingly, we conclude that $\sim$ 90\% of M dwarfs producing fast optical flares are magnetically active. 

We now calculate the occurrence rate of fast optical flares per active M dwarf. To this end, we first need to estimate the total observation time of active M dwarfs by our survey. To estimate the magnetically-active fraction in the observed M dwarfs, we search the 5,692 observed M dwarfs in the LAMOST low-resolution catalog\footnote{https://dr7.lamost.org/catalogue} and find 1392 objects. Among them, we find 276 active M dwarfs with equivalent widths of $\mathrm{H}{\alpha}$ emission line larger than $2\AA$ at the 3$\sigma$ level. This corresponds to $L_{\mathrm{H}{\alpha}}/L_{\rm bol} \simeq 1.5 \times 10^{-4}$ for a star with $T=3400\,\mathrm{K}$. Then, the magnetically-active fraction can be estimated as $276/1,392 \sim 0.2$, assuming that the 1,392 objects are selected unbiasedly. Given that the effective total observation time of our survey is $\sim 150$ M-dwarf days (see subsection \ref{sec:target_selection}), the effective total observation time for active M dwarfs is $\sim 150\,\mathrm{day}\times 0.2 \sim 30$ day. 

\begin{figure*}
\begin{center}
\includegraphics[width=0.45 \linewidth]{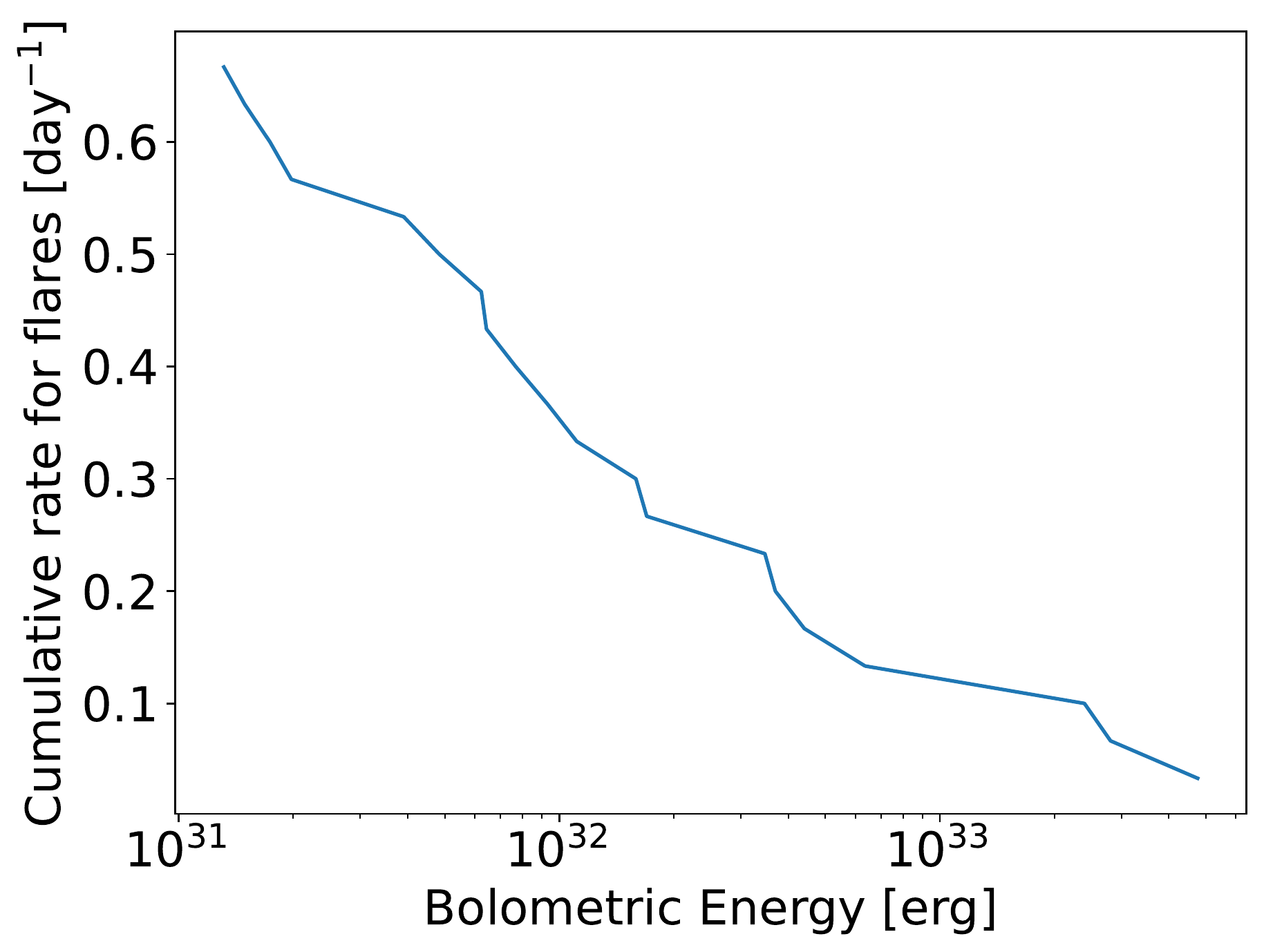}
\end{center}
\caption{Cumulative occurrence rate of fast optical flares per magnetically active M dwarfs.}
\label{fig:ccd_flare}
\end{figure*}

Figure \ref{fig:ccd_flare} shows the cumulative occurrence rate of fast optical flares based on our observation. In the calculation, we exclude TIC 21286088, TIC 21286574, and TIC 358573853; the first is apparently faint and not in our original target catalog, the data for the second system do not include the rising phase, and the last system is magnetically inactive. We find that the rates are 0.67 day$^{-1}$, 0.35 day$^{-1}$, and 0.12 day$^{-1}$ for flares with $E_{\rm bol}>10^{31}\,\mathrm{erg}, 10^{32}\,\mathrm{erg}$, and $10^{33} \mathrm{erg}$, respectively. These rates are broadly consistent with the flare frequency distribution obtained by previous optical flare surveys~\citep[e.g.][]{2020AJ....159...60G}, although the direct comparison is difficult since the cadence and sensitivity of the observations are different. 

\section{Summary and discussion \label{sec:discussion}} 

\subsection{Summary of our observations and comparison with previous studies}

In this study, we use Tomo-e Gozen to search for flares from M dwarfs at one-second cadence. By analyzing the obtained light curves of $\sim$ 300 M-dwarf days in total, we identify 22 fast optical flares; 20 classical flares with apparently one single peak and 2 complex flares consisting of multiple peaks. The flare light curves sharply rise up with a timescale of $5\, \mathrm{sec} \lesssim t_\mathrm{rise} \lesssim 100\,\mathrm{sec}$, tend to have a flat top structure with a duration of a few sec. The peak luminosity ranges from $10^{29}\,\mathrm{erg\,sec^{-1}} \lesssim L_\mathrm{peak} \lesssim 10^{31}\,\mathrm{erg\,sec^{-1}}$ assuming that the temperature of the blackbody component is $T_\mathrm{flare} \sim 9,000-15,000\,\mathrm{K}$. The decay phase of the flare can be fitted by two exponential functions and the decay time is tightly correlated with and systematically longer than the rise time. The total emitted energies are estimated to be $10^{31}\,\mathrm{erg} \lesssim E_{\rm bol} \lesssim 10^{34}\,\mathrm{erg}$. 

We compare the Tomo-e Gozen flares with those detected by the previous high cadence surveys with TESS and ULTRACAM, and find that the Tomo-e Gozen flares constitute the bright end of fast optical flares. We also compare the average profile of the detected flares with the template constructed from the Kepler one-minute cadence data by \cite{2014ApJ...797..122D}. The Tomo-e Gozen flares have steeper rise and shallower decay phases. The difference is genuinely associated with the flare duration, i.e., faster flares show steeper rise and shallower decay, originated from the difference of time resolution between the observations, i.e., the rise part and the flare-top feature seen in the Tomo-e Gozen flares is simply not resolved by Kepler, or the way of the data analysis. Indeed, \cite{2022arXiv220505706T} reanalyzed the same 1-minute cadence data as \cite{2014ApJ...797..122D}, and found that the model with the steeper rise is favored. In order to clarify this point, it will be useful (i) to compare templates of the flare light curves obtained from the same targets but with different data sampling rates, (ii) to construct several templates by dividing flares by durations and/or amplitudes, and (iii) to simultaneously observe flares with different cadences, e.g., with Tomo-e Gozen and TESS. 


We have confirmed that the current flares are from magnetically active stars based on the analyses of rotational period and $\mathrm{H\alpha}$ emission obtained from the TESS and LAMOST data. 
The estimated event rates of fast optical flares from active M dwarfs are $\sim 0.7\,{\rm day}^{-1}$ with $E_{\rm bol}>10^{31}\,\mathrm{erg}$, suggesting that these flares rather frequently occur. They could be more similar to the one with timescale of a few seconds observed in the UV and sub-millimeter bands~\citep{2021ApJ...911L..25M}. Similar flares would also be detected in NGTS survey \citep{2021MNRAS.504.3246J}, which searches for flares based on the 13 seconds cadence observations. GALEX also detected a number of very fast and energetic flares from solar-type stars~\citep{2016ApJ...817....1J}. These results would indicate that multi-wavelength fast flares are common in various types of main-sequence stars.

\subsection{Discussion on the flare mechanism\label{sec:time}}
The current observation successfully resolves the rise part, the flare top, and the decay part of the light curves of fast optical flare. Here we discuss the implications on the flare mechanism. 

In general, a flare is triggered by magnetic reconnection in the corona. The optical flares are considered to be powered by a downward particle beam injected towards the chromosphere and below. In this case, the energy injection occurs in an Alfv\'en transit time of the reconnected magnetic field (equation \ref{eq:time_A}). The rise time of the flare light curve should be comparable to or slightly longer than the Alfv\'en transit time, i.e., 
\begin{equation}\label{eq:trise_relation}
    t_\mathrm{rise} \gtrsim t_\mathrm{A}.
\end{equation}
and the total emitted energy is also bounded by the total magnetic energy of the reconnected fields, i.e., 
\begin{equation}\label{eq:Ebol_relation1}
    E_{\rm bol} \approx f \left(\frac{B^2}{8\pi}\right) l_\mathrm{loop}^3 \sim  4 \times 10^{30} \mathrm{erg} \left(\frac{f}{0.1}\right)\left(\frac{B}{1,000\,\mathrm{G}}\right)^{2} \left(\frac{l_{\rm loop}}{10^{9}\,\mathrm{cm}}\right)^{3}, \label{eq:tot_energy}
\end{equation}
where $f < 1$ is the conversion efficiency. From Eqs (\ref{eq:time_A}), (\ref{eq:trise_relation}), and (\ref{eq:Ebol_relation1}), the rise time and the total emitted energy need to satisfy
\begin{equation}
    t_{\rm rise}  \gtrsim 10.8\,\mathrm{sec}\,\left(\frac{f}{0.1}\right)^{-1/3} \left(\frac{M_\mathrm{A}}{0.1}\right)^{-1} \left(\frac{n}{10^{10}\,\mathrm{cm^{-3}}}\right) ^{1/2} \left(\frac{B}{1,000\, \mathrm{G}}\right)^{-5/3}  \left(\frac{  E_{\rm bol}}{10^{31}\,\mathrm{erg}}\right)^{1/3}.\label{eq:time_A_2}
\end{equation}
The red line in figure \ref{fig:trise_vs_Ebol} indicates this relation with $f = 0.1$, $M_\mathrm{A} = 0.1$, $n = 10^{10}\,\mathrm{cm^{-3}}$, and $B=1,000\,\mathrm{G}$. Note that these values are considered to be typical for magnetically active M dwarfs~\citep[e.g.,][]{2019A&A...626A..86S}. The detected flares are broadly consistent with this relation, and we conclude the rise time can be explained as the Alfv\'en transit time of a magnetic loop with a length scale of $l_\mathrm{loop} \sim 10^4\,\mathrm{km}$ and a field strength of $B \sim 1,000\,\mathrm{G}$. 

When a strong particle beam is injected towards the chromosphere and below, the chromosphere is compressed down by a shock to form a hot gas layer. \cite{2016ApJ...820...95K} showed that the WL flares can be mainly powered by the hydrogen recombination emission from this hot compressed region. In order to accurately calculate the resultant flare light curve, one needs to perform a radiative hydrodynamic simulation including non-local-thermodynamic-equilibrium~(LTE) effects. Instead, here we consider the flare energetics in a simplified manner, with which we can still estimate the key physical parameters, i.e., density and temperature of the hot compressed region, from the photometric light curves.

\begin{figure*}[h]
\begin{center}
\includegraphics[width=0.43\linewidth]{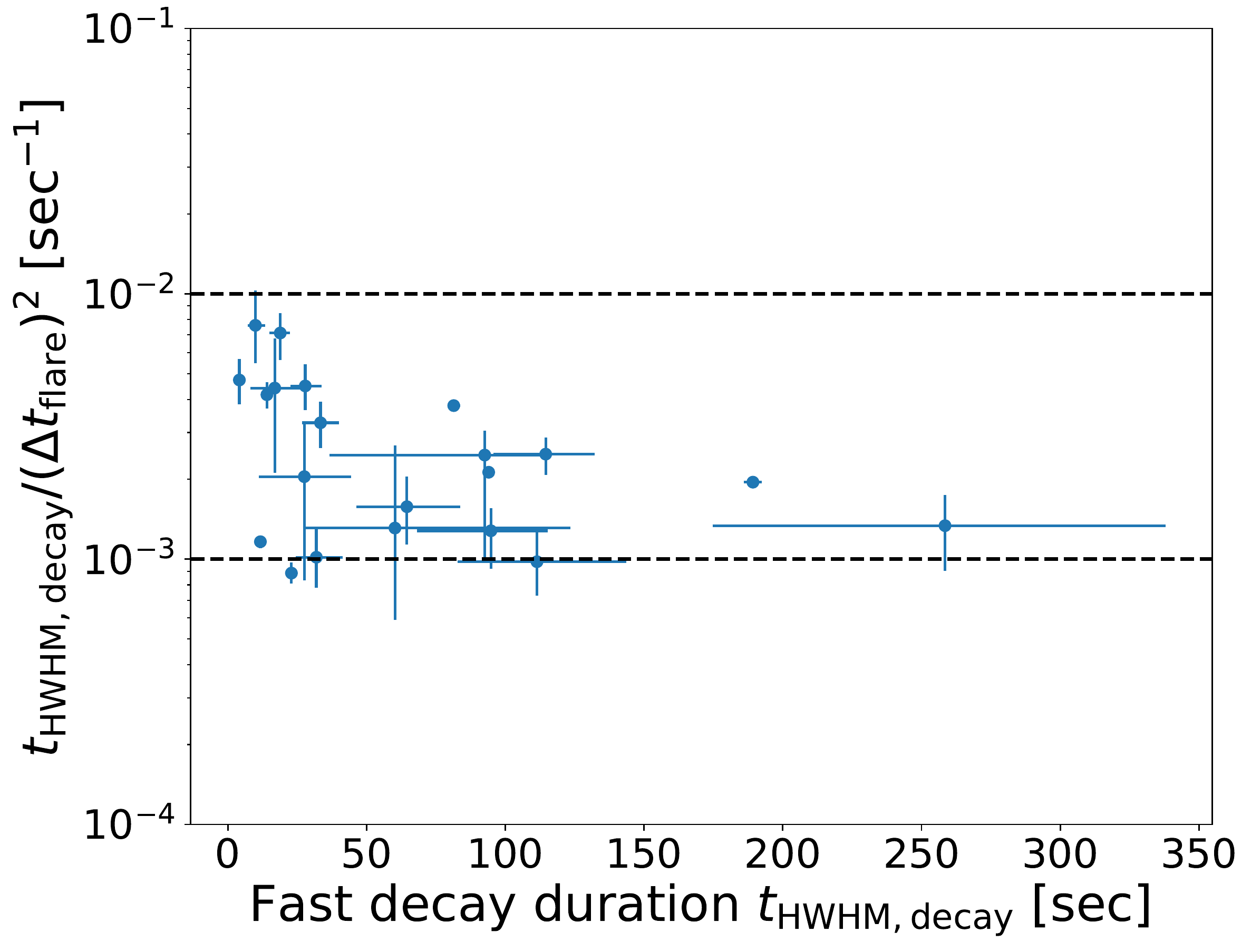}
\includegraphics[width=0.45\linewidth]{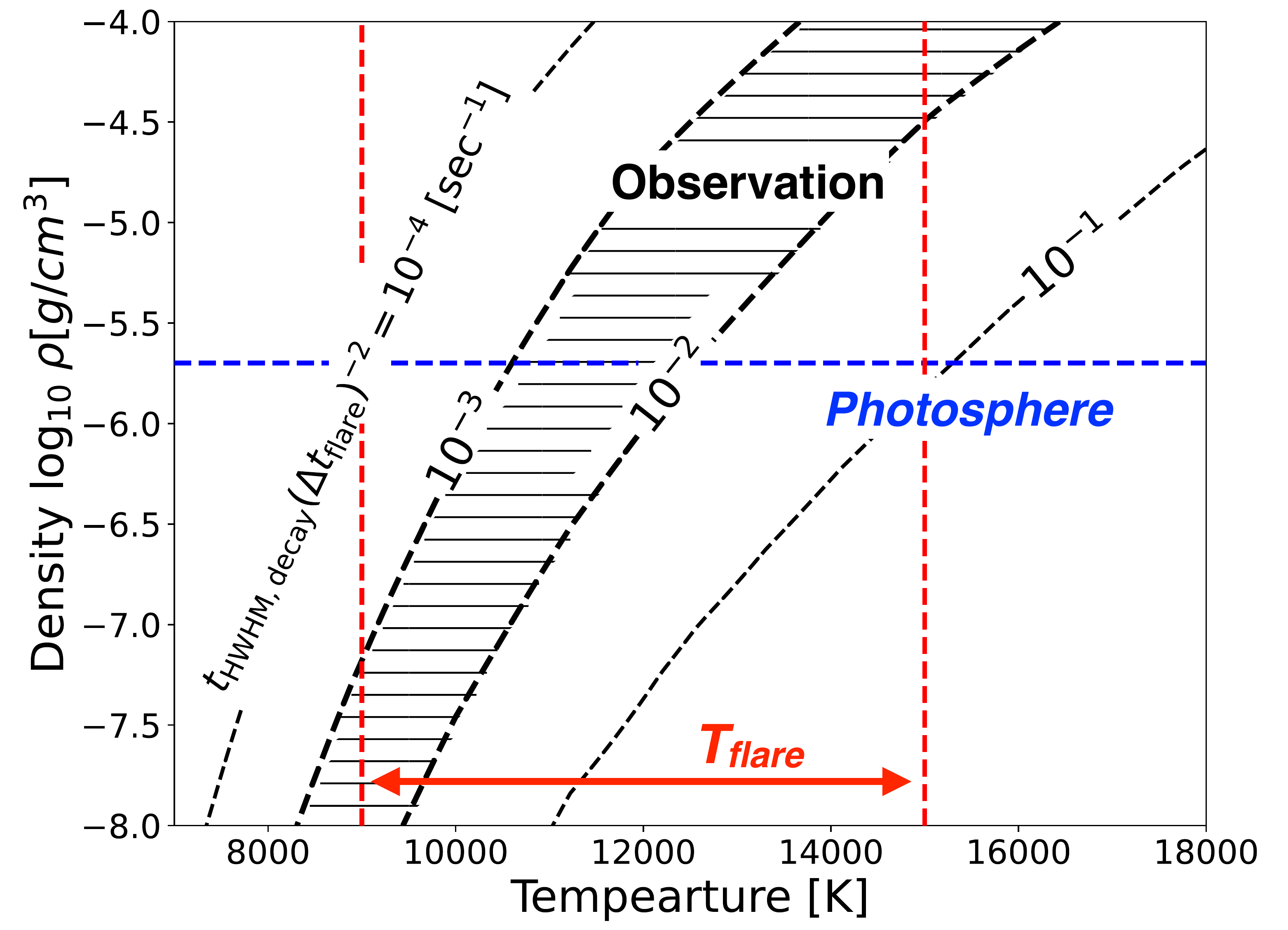}
\end{center}
\caption{(left) Plot of $t_{\rm HWHM, decay}$ and $t_{\rm HWHM, decay}/(\Delta t_{\rm flare})^{2}$. (right) Contours of $t_{\rm  HWHM, decay}/(\Delta t_{\rm flare})^{2}$ on the $(\rho, T_{\rm flare})$ plane.  }
\label{fig:fwhm_tdiff2}
\end{figure*}

Based on the chorompspheric compression model, we assume that the flare emission comes from a layer of hot gas with a width of $\Delta R$, a surface area of $A$, a column density of $\Sigma$, a temperature of $T_\mathrm{flare}$. In this case, the total emitted energy should be comparable to the internal energy of the compressed region;  
\begin{equation}\label{eq:E_bol}
    E_\mathrm{bol} \approx  \frac{3}{2}\left(\frac{ \Sigma A}{m_{\rm p}}\right)k_{\rm B}T_\mathrm{flare}, \label{eq:kin_sigma}
\end{equation}
where $k_{\rm B}$ is the Boltzman constant, $m_{\rm p}$ is the proton mass. The detection of the flare peak provides a closure relation between $A$ and $T_\mathrm{flare}$ as 
\begin{equation}\label{eq:Lpeak}
    L_\mathrm{peak} \approx \sigma_\mathrm{SB} A T_\mathrm{flare}^4.
\end{equation}
From equations (\ref{eq:E_bol}) and (\ref{eq:Lpeak}), the column density can be described as 
\begin{equation}\label{eq:sig}
\Sigma \approx \frac{2}{3}\frac{m_\mathrm{p}\sigma_{\rm SB}}{k_\mathrm{B}} T_\mathrm{flare}{}^3 \Delta t_\mathrm{flare} \sim  80 \; \mathrm{g\,cm^{-2}} \left(\frac{T_{\rm flare}}{12,000 \mathrm{K}} \right)^{3} \left(\frac{ \Delta t_{\rm flare}}{100\,\mathrm{sec}}\right),
\end{equation}
where $\Delta t_{\rm flare}$ is defined and calculated from the observed light curve as follows: 
\begin{equation}\label{eq:Delta_t}
    \Delta t_{\rm flare} \equiv \frac{E_\mathrm{bol}}{L_\mathrm{peak}} = \int \frac{f (t)}{f_{\rm peak}} dt. 
\end{equation}
The observed values range from $30\,\mathrm{sec} \lesssim \Delta t_{\rm flare} \lesssim 300\,\mathrm{sec}$ (table \ref{table:flare_energy}). 
The decay time of the flare corresponds to the radiative cooling time of the hot compressed region, i.e., 
\begin{eqnarray}
      t_{\rm HWHM, decay} &\approx& \frac{\tau \Delta R}{c} = \frac{\kappa \Sigma^{2}}{c\rho} \approx \frac{\kappa}{c\rho} \left(\frac{2}{3}\frac{m_\mathrm{p}\sigma_{\rm SB}}{k_\mathrm{B}} T_\mathrm{flare}{}^3 \Delta t_\mathrm{flare}\right)^2  \\
      &\sim& 55 \,\mathrm{sec} \left(\frac{\kappa}{530\,\mathrm{cm}^{2}\,\mathrm{g^{-1}}}\right) \left(\frac{T_{\rm flare}}{12,000 \, \mathrm{K}}\right)^{6}  \left(\frac{\rho}{2\times 10^{-6}\, \mathrm{g\,cm^{-3}}}\right) ^{-1}\left(\frac{ \Delta t_{\rm flare}}{100\,\mathrm{sec}}\right)^{2}. \label{eq:t_diff} 
\end{eqnarray}
where $\tau = \kappa\rho \Delta R$ is the optical depth with $\kappa$ being the opacity, which also depends on the temperature and density. The reference value for $\rho \simeq 2\times 10^{-6} \mathrm{g\,cm^{-3}}$ is taken to be the density at the photosphere for a M3.5 star \citep{2021ApJ...919...29S}. 
From equation (\ref{eq:t_diff}), we can obtain an observational constraint; 
\begin{equation}
      \frac{ t_{\rm HWHM, decay}}{(\Delta t_{\rm flare})^{2}} \simeq 5.5 \times 10^{-3} \,\mathrm{sec}^{-1} \left(\frac{\kappa (\rho, T_\mathrm{flare})}{530\,\mathrm{cm}^{2}\,\mathrm{g^{-1}}}\right) \left(\frac{T_{\rm flare}}{12,000 \, \mathrm{K}}\right)^{6}  \left(\frac{\rho}{2\times 10^{-6}\, \mathrm{g\,cm^{-3}} }\right) ^{-1}. \label{eq:t_diff_2} 
\end{equation}
The left hand side of equation (\ref{eq:t_diff_2}) consists of the observed quantities while the right hand side can be calculated for a given density and temperature of the emitting region. 
The left panel in figure \ref{fig:fwhm_tdiff2} shows the observed values of $t_\mathrm{HWHM, decay}/(\Delta t_{\rm flare})^{2}$ for the Tomo-e Gozen flares, which range from $10^{-3}\,\mathrm{sec}^{-1}$ to $10^{-2}\,\mathrm{sec}^{-1}$. The right panel in figure \ref{fig:fwhm_tdiff2} shows contours of $t_\mathrm{HWHM, decay}/(\Delta t_{\rm flare})^{2}$ on the $(\rho, T_\mathrm{flare})$ plane. As for $\kappa$, we calculate the Rosseland mean opacity by utilizing the OPAL opacity table\footnote{http://cdsweb.u-strasbg.fr/topbase/OpacityTables.html} with assuming the chemical composition of the solar photosphere~\citep{2005ASPC..336...25A}. 
We note that the H$^{-}$ opacity is important in the parameter range of interest. The observed range is indicated by the striped region. 
Assuming a range of temperature $T_\mathrm{flare} = 9,000-15,000\,\mathrm{K}$ inferred from the previous observations of fast optical flares~\citep[e.g.][]{1992ApJS...78..565H,2003ApJ...597..535H, 2013ApJS..207...15K}, the density of the hot compressed region is estimated to be $\rho = 10^{-8} \mathrm{g\,cm^{-3}}-2\times 10^{-5} \mathrm{g\,cm^{-3}}$, which is consistent with the density for the compressed chromosphere or the photosphere. We conclude that the decay time of the detected flares can be determined by the radiative cooling of the compressed chromosphere down to near the photosphere with a temperature of $\gtrsim10,000\,\mathrm{K}$. 

As above, the rise and decay times are explained by independent physical processes. On the other hand, figure \ref{fig:trise_vs_t_12} indicates a correlation between them. Assuming that the density structure and magnetic field strength are similar among the flare stars, the rise time mainly depends on the loop size as $t_\mathrm{rise} \propto l_\mathrm{loop}$ (see equations \ref{eq:time_A} and \ref{eq:trise_relation}). On the other hand, from equations (\ref{eq:Ebol_relation1}), (\ref{eq:Lpeak}), (\ref{eq:Delta_t}), and (\ref{eq:t_diff}), the decay time is $t_\mathrm{HWHM, decay} \propto \Delta t_\mathrm{flare}{}^2 \propto l_\mathrm{loop}{}^2$ where we assume $A \propto l_\mathrm{loop}{}^2$. Resultantly we expect $t_\mathrm{HWHM, decay} \propto t_\mathrm{rise}{}^2$, which is broadly consistent with figure \ref{fig:trise_vs_t_12}. This may also support the chromospheric compression model for the detected fast optical flares. 

The limitation of the current model is that we neglect energy loss of the compressed atmosphere during the energy injection. Indeed, the energy deposition from the electron beams can be regulated by the radiative losses, leading to the decrease in the internal energy of the compressed stellar atmospheres \citep{2005ApJ...630..573A,2015SoPh..290.3487K}. This would result in the the effective decrease in the total accumulated energy $E_{\rm bol}$ in equation (\ref{eq:kin_sigma}). Hence, decaying timescales in equation (\ref{eq:t_diff}) can be shorter than the current estimate. For the more detailed discussion, the radiative hydrodynamic flare modeling is required. 

We finally note that the slow decay phase, which is characterized by $\tau_{\rm slow}$ ranging from several tens to a few thousand seconds, can be explained by radiative cooling of a lower density region than the compressed chromosphere, e.g. upper chromosphere or flare loops~\citep[e.g.][]{2011LRSP....8....6S, 2018ApJ...859..143H}.


\subsection{Future work and implication}

The current study is the first attempt to search for stellar flares using the Tomo-e Gozen camera, which identifies a new population of M-dwarf flares owing to its wide FOV and one-second cadence. We can expand the research in several ways. 

\begin{itemize}
    \item We searched for optical flares only photometrically. Simultaneous multi-wavelength observations e.g. in radio and X-ray bands, and/or spectroscopic observations capable to detect e.g., $\mathrm{H}{\alpha}$, $\mathrm{H}{\beta}$ and Balmer lines, will be useful to clarify the mechanisms and consequences of the flares~\citep{2013ApJS..207...15K,2020PASJ...72...68N,2021ApJ...911L..25M,2021NatAs.tmp..246N,2022arXiv220403481A}. 

    \item We have shown that the light curve shapes of the fast flares are different from that of relatively slow flares detected by lower-cadence observations. In order to confirm whether the difference is genuinely associated with the flare duration or originated from the difference of the time resolution, it would be useful to simultaneously detect flares by e.g., TESS and Tomo-e Gozen, which can be done by utilizing the information of Sectors and target lists of ongoing and upcoming TESS observations. 
   
    \item While we here focus on flares from M dwarfs, we can also search for flares from different types of stars. In fact, GALEX had identified energetic flares with short durations from solar-type stars~\citep{2019ApJ...883...88B}. Tomo-e Gozen will be useful for searching the optical counterparts. 
    
    \item In this paper, we blindly search flares from field stars, but it might be also interesting to monitor for variabilities from specific areas of the sky, e.g. open clusters or star-forming regions \citep{2003PASJ...55..653I,2020ApJ...893...67M,2021MNRAS.507.2103G}. Thanks to the wide FOV of Tomo-e Gozen, multiple young stellar objects can be simultaneously captured, which enable us to resolve flares more precisely.

    \item The theoretical model we presented in subsection \ref{sec:time} is fairly simple. Further studies based on numerical simulations of the atmospheric response after an injection of particle beam~\citep[e.g.,][]{2005ApJ...630..573A, 2015SoPh..290.3487K,2020PASJ...72...68N} will be necessary to confirm the origins of the timescales of the fast optical flares. 
    
    \item The very fast flares from M dwarfs can be a contamination source for a blind search of fast optical transient of unknown type, e.g., optical counterpart of fast radio bursts \citep{2020ApJ...896..142B,2022arXiv220412334N,2022ApJ...929..139L}. Indeed, the brightest flare in our sample is originally detected as a contamination to a target star and is eventually confirmed to be from a nearby faint M dwarf with $G=18.63$.

\end{itemize}

\begin{ack}
The authors thank Munetake Momose, Toshiyuki Mizuki, Shinsuke Takasao, and Takeru Suzuki for the discussion. This work has been also supported by the Japan Society for the Promotion of Science (JSPS) KAKENHI grants, JP16H06341, JP17H06363, JP18H05223, JP18H01261, JP20K04010, JP20K14512, JP20H01904, and JP21H04491. This work is supported in part by the Optical and Near-Infrared Astronomy Inter-University Cooperation Program. Guoshoujing Telescope (the Large Sky Area Multi-Object Fiber Spectroscopic Telescope LAMOST) is a National Major Scientific Project built by the Chinese Academy of Sciences. Funding for the project has been provided by the National Development and Reform Commission. LAMOST is operated and managed by the National Astronomical Observatories, Chinese Academy of Sciences. 
\end{ack}

\appendix
\section{Systematic noise reduction from raw light curves} \label{sec:cot}
We apply a cotrending method to the raw light curves in order to remove common systematic variability mostly due to atmospheric noises. First, for each observation date, we divide detected objects (target stars and reference stars) into groups according to the CMOS sensor IDs they are on; member stars in each group are located close in the sky and have the same time series of the light-curve data. 
Second, for each group, we analyze the raw light curves of the member stars to construct a set of basis vectors representing common systematic trends. The cotrending basis vectors for each group are constructed as follows~\citep{2012PASP..124.1000S}.
\begin{enumerate}
    \item First we normalize each light curve by the mean value of the flux and calculate the standard deviation $\sigma_{y}$. We then select objects with $\sigma_{y}$ smaller than the median of the group in order to eliminate stars with large intrinsic variability, significant contamination from nearby stars, and/or bad pixels in the aperture.
    \item For the light curves of the selected objects, we calculate the Pearson correlation coefficients $\sigma_{\mathrm{Pearson}, i,j}$, which represent the strength of the correlation between $i$ and $j$-th stars. We measure the strength of correlation of i-th star's light curve with others by taking the median of $\sigma_{\mathrm{Pearson}, i,j}$ over the index $j$, and select top 70\% of the light curves in terms of the correlation strength. In the above procedures, roughly 35 \% of the detected objects, typically 70-100 stars per group, are selected. 
    \item From a set of the light curves of the remaining stars, we construct a $M\times N$ matrix, where $M$ is the number of the stars and $N$ is the number of the time series, and apply singular value decomposition (SVD) to it. We then obtain a set of basis vectors, singular values, and coefficient matrices. From the coefficient matrices, we compute $p$ values for whether each basis vector is dominated by contribution from a specific star. We select at most 7 basis vectors in order of singular value. We stop the selection process when the $p$ value of a selected basis vector is smaller than 0.1.  

\end{enumerate}

After determining the cotrending basis vectors, we subtract their contributions from the light curves by computing inner products between the basis components and the raw light curves. Note that, in principle, we can adopt more sophisticated procedures to optimize the weights for subtracting the basis vector components based on the Bayesian formalism~\citep{2012PASP..124.1000S}. However, we adopt the current method since the typical number of objects in each group is not large enough to construct a reliable prior. 

\section{Classical flare light curves} \label{sec:all_lcs}
\input{fig_summary.tex}
Figure \ref{fig:lc_flares} shows the 18 light curves of classical flares detected by our survey. 
As for TIC 20536892, TIC 21286088, and TIC 243017627, the detrended light curves are still partially contaminated by noise and we analyze them independently as follows.

\begin{figure*}[h]
\begin{center}
\includegraphics[width=0.550 \linewidth]{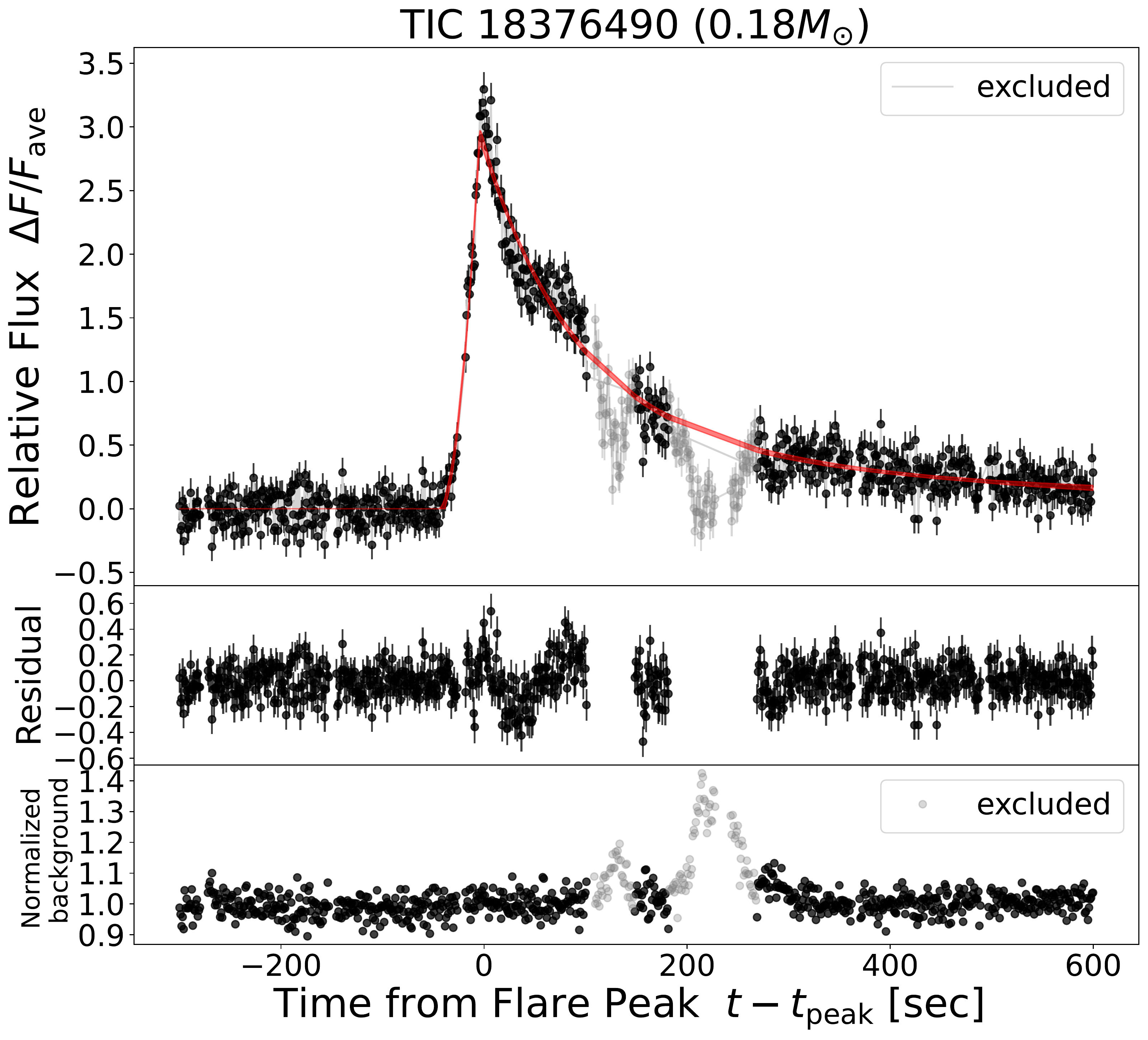}
\end{center}

\caption{Flare light curve of TIC 20536892. The gray points indicate the data with contamination, and the red shaded region corresponds to the 90 \% credible interval of the fitting model (equation \ref{eq:lc_fit_model}). The middle and bottom panels show the residual after subtracting the best-fit model and the normalized background level, respectively.}
\label{fig:tic_20536892}
\end{figure*}

The flaring event from TIC 20536892 is significantly affected by atmospheric noise during its slow decay phase. We investigate the background level during the flaring event, and find that the apparent dips in the decay phase are caused by the change of the background level. Figure \ref{fig:tic_20536892} shows the light curve, the fitting model, and the normalized background level. The background level is high for the blue points, which we do not use for the fitting. Although there may be an additional structure at $50\,\mathrm{sec}\lesssim t-t_\mathrm{peak} \lesssim 100\,\mathrm{sec}$, we stop the analysis at this point given the contamination due to the bad weather.

\begin{figure*}[h]
\begin{center}
\includegraphics[width=0.550\linewidth]{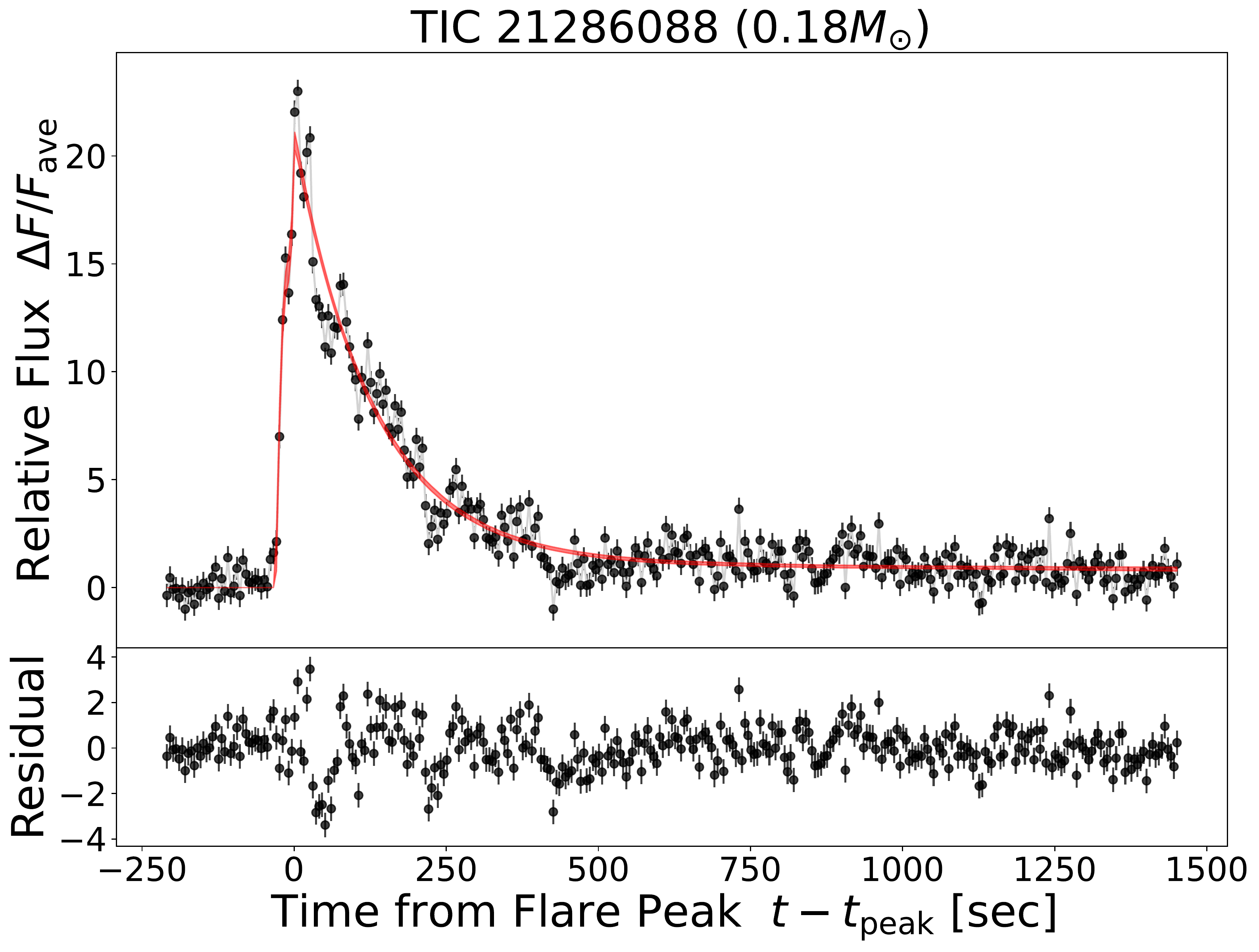}
\end{center}
\caption{Flare light curve of TIC 21286088. The red shaded region corresponds to the 90 \% credible interval of the fitting model (equation \ref{eq:lc_fit_model}).}
\label{fig:tic_21286088}
\end{figure*}

TIC 21286088 ($G=18.63$) is about 10 times fainter than a nearby target star TIC 21286087 ($G=16.26$) at a separation of $\sim 11\,\arcsec$. We use the point-spread-function (PSF) photometry with a Moffat profile~ \citep{1969A&A.....3..455M} to calculate the light curves of TIC 21286087 and TIC 21286088 with an appropriate background correction. For the PSF fitting, we use 5-second integrated images. Figure \ref{fig:tic_21286088} shows the light curve of TIC 21286088. The flare amplitude is estimated to be 2000 \%, which is the largest flare among the current sample. 

The light curve of TIC 243017627 ($G=15.83)$ is contaminated by a nearby object TIC 243017628 ($G=16.26)$ at a separation of $\sim 6 \arcsec$. We model the PSFs of the both object, and compute the flux leakage from TIC 243017628 to TIC 243017627 in the 5-pixel aperture. 
We find that $\sim 20\%$ of the detected flux from TIC 243017627 during the non-flaring phase comes from TIC 243017628. We subtract this contamination from the light curve to correctly estimate the flare amplitude;  
\begin{equation}
\frac{F_{\rm flare} (t)}{F_{1}}= \left(\frac{( F_{1} + F_{2} + F_{\rm flare} (t))- (F_{1} + F_{2}) }{F_{1} + F_{ 2}}\right) \frac{F_{1} + F_{2}}{F_{1}}, 
\end{equation}
where $F_{1}$ is the stellar flux from the target star, $F_{2}$ is the flux from the contaminating star, $F_{\rm flare}(t)$ is the flare light curve. Note that $F_{1} + F_{2} + F_{\rm flare} (t)$ is the total flux, and $F_{1} + F_{2} $ is the flux before the flare. 

\newpage
\bibliography{ref}

\end{document}

%% file: table_stellar.tex
\addtolength{\tabcolsep}{-1pt} \begin{table} 
\caption{Stellar parameters of 22 M dwarfs with flares.  } 
  \label{table:star_para}  
\begin{tabular}{ccccccccccc} 
\hline 
TIC ID& $T_{\rm mag}$ \footnotemark[$1$]& $G_{\rm mag}$& SpT & $R_{\rm \star} (R_{\rm \odot})$& $M_{\rm \star} (M_{\rm \odot})$& $T_{\rm eff}$ (K)& $P_{\rm rot}$ (day) \footnotemark[$2$]& $L_{\rm \alpha}/L_{\rm \star}$ \footnotemark[$3$]& $\tau_{\rm conv}$ (day) \footnotemark[$4$]& flux type \footnotemark[$5$] \\ 
\hline 
$15904458$& $13.50$& $14.90$&M4V& $0.38$& $0.37$&$3125$&$1.03$&$3.1 \times 10^{-4}$& $66.27$&\texttt{FLUX\_r7} \\ 
$17198188$& $15.25$& $16.86$&M6V& $0.14$& $0.11$&$2857$&$0.39$& $\cdot \cdot \cdot$& $145.17$&\texttt{FLUX\_r5} \\ 
$18376490$& $14.15$& $15.57$&M5V& $0.21$& $0.18$&$3085$&$0.38$&$3.0 \times 10^{-4}$& $117.12$&\texttt{FLUX\_r7} \\ 
$20536892$& $15.47$& $16.84$&M4V& $0.39$& $0.38$&$3170$&$2.10$&$1.2 \times 10^{-4}$& $63.06$&\texttt{FLUX\_r7} \\ 
$21286088$& $17.13$& $18.63$&M4V& $0.21$& $0.18$&$3133$& $\cdot \cdot \cdot$& $\cdot \cdot \cdot$& $118.06$&$\cdot \cdot \cdot$ \\ 
$21286574$& $11.95$& $13.19$&M3V& $0.38$& $0.37$&$3369$&$0.26$&$3.8 \times 10^{-4}$& $66.07$&\texttt{FLUX\_r14} \\ 
$46305452$& $14.81$& $16.11$&M3V& $0.35$& $0.33$&$3317$&$0.44$&$1.8 \times 10^{-4}$& $73.47$&\texttt{FLUX\_r7} \\ 
$55288759$& $11.66$& $12.91$&M3V& $0.34$& $0.33$&$3350$& $\cdot \cdot \cdot$&$1.9 \times 10^{-4}$& $75.00$&\texttt{FLUX\_r14} \\ 
$98751507$& $13.18$& $14.74$&M4V& $0.27$& $0.25$&$3166$& $\cdot \cdot \cdot$& $\cdot \cdot \cdot$& $94.75$&\texttt{FLUX\_r14} \\ 
$119031076$& $12.29$& $13.62$&M4V& $0.27$& $0.25$&$3215$&$1.57$&$1.7 \times 10^{-4}$& $94.84$&\texttt{FLUX\_r7} \\ 
$121638493$& $15.17$& $16.44$&M3V& $0.30$& $0.27$&$3300$&$1.21$& $\cdot \cdot \cdot$& $88.04$&\texttt{FLUX\_r5} \\ 
$185243119$& $15.30$& $16.62$&M4V& $0.34$& $0.32$&$3209$& $\cdot \cdot \cdot$& $\cdot \cdot \cdot$& $76.06$&\texttt{FLUX\_r7} \\ 
$203228080$& $15.37$& $16.78$&M4V& $0.39$& $0.38$&$3131$& $\cdot \cdot \cdot$&$3.1 \times 10^{-4}$& $64.21$&\texttt{FLUX\_r5} \\ 
$243017627$& $14.50$& $15.83$&M4V& $0.34$& $0.33$&$3246$&$1.67$&$3.4 \times 10^{-4}$& $74.83$&\texttt{FLUX\_r5} \\ 
$251725681$& $13.62$& $14.90$&M4V& $0.39$& $0.38$&$3278$& $\cdot \cdot \cdot$& $\cdot \cdot \cdot$& $64.34$&\texttt{FLUX\_r5} \\ 
$315617164$& $15.50$& $16.82$&M4V& $0.31$& $0.29$&$3224$&$0.59$& $\cdot \cdot \cdot$& $84.57$&\texttt{FLUX\_r7} \\ 
$358561826$& $14.68$& $16.15$&M5V& $0.26$& $0.24$&$3032$& $\cdot \cdot \cdot$&$2.8 \times 10^{-4}$& $98.40$&\texttt{FLUX\_r7} \\ 
$358573853$& $13.86$& $15.20$&M4V& $0.24$& $0.21$&$3225$& $\cdot \cdot \cdot$&$<1.5 \times 10^{-5}$& $107.53$&\texttt{FLUX\_r7} \\ 
$366485128$& $15.59$& $16.82$&M3V& $0.36$& $0.35$&$3354$& $\cdot \cdot \cdot$& $\cdot \cdot \cdot$& $70.14$&\texttt{FLUX\_r5} \\ 
$374272810$& $13.95$& $15.31$&M4V& $0.20$& $0.17$&$3136$&$0.70$& $\cdot \cdot \cdot$& $122.39$&\texttt{FLUX\_r7} \\ 
$380197137$& $15.00$& $16.42$&M5V& $0.20$& $0.17$&$3094$& $\cdot \cdot \cdot$& $\cdot \cdot \cdot$& $122.24$&\texttt{FLUX\_r7} \\ 
$435904068$& $14.28$& $15.62$&M4V& $0.40$& $0.39$&$3230$&$1.36$&$1.7 \times 10^{-4}$& $62.15$&\texttt{FLUX\_r7} \\ 
 \hline  \end{tabular}  
\begin{tabnote} 
\footnotemark[$1$] TESS magnitude.  
\footnotemark[$2$] Rotational periods are estimated from TESS data in sub-subsection \ref{sec:rot_tess} 
\footnotemark[$3$] H${\rm \alpha}$ luminosity is estimated from LAMOST data in sub-subsection \ref{sec:halpha_lamost} except for TIC 20536892. We adopt the value from \cite{2016ApJ...817....1J} for TIC 20536892.\footnotemark[$4$] Convection timescale deterimined from empirical relation as discussed in sub-subsection \ref{sec:rot_tess}. 
\footnotemark[$5$] Flux types used for making lightcurves in this study. For TIC 21286088, we adopt a PSF photometry as discussed in appendix \ref{sec:all_lcs}.\end{tabnote} 
\end{table}  \addtolength{\tabcolsep}{1pt} 

%% file: flare_mcmc.tex
\begin{table} 
\caption{Fitted parameters for 22 flares. }   
 \begin{tabular}{ccccccc} 
\hline 
TIC ID& $f_{\rm peak}$& $t_{\rm rise}$ [sec]& $\Delta t_{\rm peak}$ [sec]& $\tau_{\rm fast}$ [sec]& $\tau_{\rm slow}$ [sec]&  $\cal C$ \\ 
\hline 
15904458 \footnotemark[$1$]& $0.93 \pm ^{0.03}_{-0.02}$& $31.21 \pm ^{1.64}_{-1.62}$& $2.96 \pm ^{1.14}_{-1.83}$& $13.27 \pm ^{1.35}_{-1.77}$& $713.49 \pm ^{36.09}_{-32.04}$& $0.85 \pm ^{0.03}_{-0.06}$ \\ 
15904458 \footnotemark[$2$]& $0.49 \pm ^{0.03}_{-0.04}$& $30.51 \pm ^{1.21}_{-1.06}$& $7.17 \pm ^{2.04}_{-2.78}$& $33.06 \pm ^{3.13}_{-2.80}$& $713.49 \pm ^{36.09}_{-32.04}$& $0.74 \pm ^{0.08}_{-0.05}$ \\ 
17198188 & $0.39 \pm ^{0.04}_{-0.04}$& $13.65 \pm ^{4.42}_{-5.15}$& $3.62 \pm ^{4.15}_{-2.58}$& $39.66 \pm ^{8.17}_{-8.54}$& $2453.19 \pm ^{1183.70}_{-1179.49}$& $0.87 \pm ^{0.02}_{-0.03}$ \\ 
18376490 & $0.27 \pm ^{0.02}_{-0.02}$& $35.71 \pm ^{4.16}_{-3.55}$& $4.30 \pm ^{4.58}_{-3.09}$& $24.16 \pm ^{7.89}_{-6.43}$& $839.07 \pm ^{353.42}_{-171.13}$& $0.66 \pm ^{0.04}_{-0.04}$ \\ 
20536892 \footnotemark[$1$]& $2.90 \pm ^{0.04}_{-0.04}$& $41.42 \pm ^{3.85}_{-3.45}$& $1.48 \pm ^{1.50}_{-1.03}$& $89.25 \pm ^{6.91}_{-7.23}$& $420.00 \pm ^{123.31}_{-70.85}$& $0.77 \pm ^{0.06}_{-0.06}$ \\ 
21286088 & $20.88 \pm ^{0.30}_{-0.23}$& $30.46 \pm ^{0.72}_{-0.95}$& $2.11 \pm ^{1.32}_{-1.43}$& $119.30 \pm ^{2.71}_{-2.32}$& $1224.66 \pm ^{90.60}_{-61.75}$& $0.90 \pm ^{0.01}_{-0.01}$ \\ 
21286574 & $0.32 \pm ^{0.00}_{-0.00}$& $\cdots$& $\cdots$& $31.11 \pm ^{1.51}_{-1.39}$& $300.36 \pm ^{14.57}_{-14.36}$& $0.72 \pm ^{0.01}_{-0.01}$ \\ 
46305452 & $0.14 \pm ^{0.01}_{-0.01}$& $79.23 \pm ^{22.28}_{-31.12}$& $53.13 \pm ^{32.17}_{-29.65}$& $80.13 \pm ^{29.02}_{-25.79}$& $3556.35 \pm ^{3499.99}_{-2542.64}$& $0.90 \pm ^{0.05}_{-0.06}$ \\ 
55288759 & $0.05 \pm ^{0.01}_{-0.00}$& $6.24 \pm ^{3.63}_{-1.51}$& $1.91 \pm ^{2.03}_{-1.37}$& $23.30 \pm ^{4.82}_{-5.32}$& $1162.01 \pm ^{839.70}_{-758.81}$& $0.90 \pm ^{0.02}_{-0.04}$ \\ 
98751507 & $0.15 \pm ^{0.02}_{-0.02}$& $8.42 \pm ^{3.97}_{-2.95}$& $8.50 \pm ^{4.72}_{-4.85}$& $8.52 \pm ^{6.58}_{-2.65}$& $50.75 \pm ^{39.22}_{-16.46}$& $0.69 \pm ^{0.18}_{-0.21}$ \\ 
119031076 & $0.11 \pm ^{0.00}_{-0.00}$& $13.47 \pm ^{3.62}_{-2.63}$& $6.32 \pm ^{1.67}_{-1.84}$& $13.79 \pm ^{2.18}_{-2.01}$& $154.76 \pm ^{23.86}_{-18.89}$& $0.75 \pm ^{0.04}_{-0.03}$ \\ 
121638493 & $0.22 \pm ^{0.10}_{-0.04}$& $26.58 \pm ^{26.63}_{-13.45}$& $23.93 \pm ^{53.38}_{-21.11}$& $32.12 \pm ^{26.48}_{-22.57}$& $892.48 \pm ^{1989.75}_{-770.64}$& $0.91 \pm ^{0.05}_{-0.21}$ \\ 
185243119 & $3.36 \pm ^{0.09}_{-0.07}$& $77.81 \pm ^{4.41}_{-4.01}$& $16.64 \pm ^{7.38}_{-9.27}$& $265.46 \pm ^{6.83}_{-9.23}$& $347.50 \pm ^{9600.55}_{-63.68}$& $0.91 \pm ^{0.08}_{-0.43}$ \\ 
203228080 & $0.32 \pm ^{0.04}_{-0.03}$& $33.71 \pm ^{13.84}_{-9.95}$& $23.50 \pm ^{22.14}_{-16.23}$& $141.14 \pm ^{32.13}_{-44.20}$& $4644.70 \pm ^{6392.41}_{-4338.90}$& $0.90 \pm ^{0.06}_{-0.32}$ \\ 
243017627 & $0.11 \pm ^{0.01}_{-0.01}$& $27.08 \pm ^{12.77}_{-8.24}$& $96.35 \pm ^{12.30}_{-17.45}$& $79.01 \pm ^{74.34}_{-67.32}$& $249.69 \pm ^{2244.42}_{-56.10}$& $0.47 \pm ^{0.49}_{-0.22}$ \\ 
251725681 & $0.04 \pm ^{0.01}_{-0.01}$& $21.64 \pm ^{13.76}_{-12.97}$& $16.60 \pm ^{11.20}_{-11.02}$& $23.90 \pm ^{49.54}_{-14.31}$& $925.25 \pm ^{2730.89}_{-404.35}$& $0.57 \pm ^{0.10}_{-0.12}$ \\ 
315617164 & $0.20 \pm ^{0.05}_{-0.04}$& $11.01 \pm ^{9.74}_{-4.80}$& $19.16 \pm ^{13.16}_{-15.47}$& $20.43 \pm ^{13.48}_{-12.53}$& $270.32 \pm ^{1351.24}_{-223.78}$& $0.92 \pm ^{0.05}_{-0.34}$ \\ 
358561826 & $1.14 \pm ^{0.06}_{-0.05}$& $54.60 \pm ^{7.94}_{-9.02}$& $8.90 \pm ^{4.63}_{-6.03}$& $13.25 \pm ^{1.95}_{-2.12}$& $549.68 \pm ^{79.63}_{-61.54}$& $0.61 \pm ^{0.03}_{-0.03}$ \\ 
358573853 & $0.10 \pm ^{0.02}_{-0.01}$& $78.10 \pm ^{16.67}_{-11.83}$& $8.59 \pm ^{55.43}_{-5.83}$& $102.19 \pm ^{46.74}_{-95.99}$& $782.60 \pm ^{8110.77}_{-575.61}$& $0.82 \pm ^{0.13}_{-0.44}$ \\ 
366485128 & $0.20 \pm ^{0.03}_{-0.02}$& $79.38 \pm ^{27.69}_{-29.02}$& $37.46 \pm ^{35.60}_{-26.28}$& $189.43 \pm ^{159.25}_{-145.29}$& $4978.39 \pm ^{14451.15}_{-4206.61}$& $0.67 \pm ^{0.15}_{-0.31}$ \\ 
374272810 & $0.17 \pm ^{0.01}_{-0.01}$& $33.72 \pm ^{6.48}_{-3.50}$& $51.79 \pm ^{9.74}_{-13.10}$& $49.70 \pm ^{61.38}_{-25.97}$& $908.32 \pm ^{4909.85}_{-226.32}$& $0.52 \pm ^{0.18}_{-0.08}$ \\ 
380197137 & $1.09 \pm ^{0.06}_{-0.06}$& $10.76 \pm ^{2.24}_{-2.69}$& $5.77 \pm ^{0.98}_{-1.22}$& $4.08 \pm ^{1.00}_{-0.94}$& $73.90 \pm ^{15.16}_{-11.91}$& $0.76 \pm ^{0.04}_{-0.04}$ \\ 
435904068 & $0.18 \pm ^{0.02}_{-0.02}$& $31.22 \pm ^{16.05}_{-8.05}$& $5.69 \pm ^{5.89}_{-3.90}$& $36.10 \pm ^{8.50}_{-8.53}$& $1407.44 \pm ^{1634.29}_{-1156.65}$& $0.93 \pm ^{0.03}_{-0.07}$ \\ 
 \hline  \end{tabular}   
\label{table:flare_paras}  
\begin{tabnote} 
\footnotemark[$1$] First peak of flare.  
\footnotemark[$2$] Second peak of flare. 
\end{tabnote} 
\end{table}   

%% file: flare_fwhm.tex
 
\begin{table} 
\caption{Full width at half maximum, and half width at half maximum for rising and decaying phases.  }   
 \begin{tabular}{cccc} 
\hline 
TIC ID& $t_{\rm FWHM}$ [sec]& $t_{\rm rise, HWHM}$ [sec]& $t_{\rm decay, HWHM}$ [sec] \\ 
\hline 
15904458 \footnotemark[$1$]& $22.95 \pm ^{0.84}_{-1.04}$& $8.40 \pm ^{0.73}_{-0.71}$& $11.73 \pm ^{0.65}_{-0.66}$ \\ 
15904458 \footnotemark[$2$]& $50.21 \pm ^{2.42}_{-2.71}$& $20.21 \pm ^{1.23}_{-1.30}$& $22.92 \pm ^{2.17}_{-1.94}$ \\ 
17198188 & $42.22 \pm ^{9.22}_{-7.66}$& $4.20 \pm ^{2.80}_{-1.74}$& $33.40 \pm ^{6.68}_{-6.60}$ \\ 
18376490 & $43.92 \pm ^{9.90}_{-8.07}$& $6.88 \pm ^{1.71}_{-1.16}$& $31.89 \pm ^{9.56}_{-7.36}$ \\ 
20536892 \footnotemark[$1$]& $95.38 \pm ^{2.80}_{-2.71}$& $12.21 \pm ^{0.89}_{-0.78}$& $81.41 \pm ^{2.35}_{-2.16}$ \\ 
21286088 & $117.74 \pm ^{1.96}_{-2.30}$& $21.44 \pm ^{0.83}_{-0.86}$& $94.00 \pm ^{1.64}_{-1.36}$ \\ 
21286574 & $\cdots$& $\cdots$& $32.99 \pm ^{0.94}_{-0.96}$ \\ 
46305452 & $148.47 \pm ^{19.39}_{-19.32}$& $25.01 \pm ^{16.30}_{-8.53}$& $64.53 \pm ^{19.17}_{-18.16}$ \\ 
55288759 & $23.42 \pm ^{4.23}_{-4.28}$& $2.34 \pm ^{1.14}_{-0.89}$& $18.89 \pm ^{3.53}_{-3.97}$ \\ 
98751507 & $24.07 \pm ^{3.42}_{-3.05}$& $5.19 \pm ^{3.42}_{-3.21}$& $9.93 \pm ^{3.53}_{-2.78}$ \\ 
119031076 & $23.53 \pm ^{1.64}_{-1.49}$& $3.07 \pm ^{0.91}_{-0.71}$& $14.07 \pm ^{1.59}_{-1.58}$ \\ 
121638493 & $80.97 \pm ^{25.00}_{-34.68}$& $8.79 \pm ^{11.29}_{-4.39}$& $27.55 \pm ^{16.91}_{-16.36}$ \\ 
185243119 & $234.18 \pm ^{8.35}_{-9.82}$& $28.12 \pm ^{3.77}_{-3.16}$& $189.24 \pm ^{3.27}_{-3.36}$ \\ 
203228080 & $157.57 \pm ^{23.61}_{-24.18}$& $14.78 \pm ^{9.43}_{-6.05}$& $114.55 \pm ^{17.61}_{-18.76}$ \\ 
243017627 & $204.46 \pm ^{17.45}_{-19.65}$& $15.47 \pm ^{9.04}_{-6.63}$& $94.81 \pm ^{20.41}_{-26.62}$ \\ 
251725681 & $84.23 \pm ^{66.85}_{-30.72}$& $6.79 \pm ^{8.96}_{-3.42}$& $60.21 \pm ^{63.34}_{-33.13}$ \\ 
315617164 & $42.97 \pm ^{10.67}_{-10.45}$& $5.30 \pm ^{5.09}_{-2.48}$& $16.89 \pm ^{9.14}_{-8.79}$ \\ 
358561826 & $46.45 \pm ^{3.84}_{-3.46}$& $16.79 \pm ^{4.38}_{-3.83}$& $20.91 \pm ^{3.07}_{-2.68}$ \\ 
358573853 & $126.14 \pm ^{25.72}_{-39.25}$& $15.60 \pm ^{5.83}_{-3.60}$& $92.57 \pm ^{22.02}_{-55.99}$ \\ 
366485128 & $332.53 \pm ^{88.98}_{-88.35}$& $32.08 \pm ^{18.48}_{-16.42}$& $258.48 \pm ^{79.54}_{-83.80}$ \\ 
374272810 & $169.45 \pm ^{26.65}_{-21.29}$& $8.50 \pm ^{2.48}_{-1.98}$& $111.41 \pm ^{32.26}_{-28.60}$ \\ 
380197137 & $12.60 \pm ^{0.96}_{-0.87}$& $2.62 \pm ^{1.16}_{-0.71}$& $4.14 \pm ^{0.84}_{-0.79}$ \\ 
435904068 & $45.87 \pm ^{10.20}_{-8.03}$& $11.05 \pm ^{6.31}_{-4.46}$& $27.89 \pm ^{5.80}_{-5.27}$ \\ 
 \hline  \end{tabular}   
\label{table:flare_fwhm}  
\begin{tabnote} 
\footnotemark[$1$] First peak of flare.  
\footnotemark[$2$] Second peak of flare. 
\end{tabnote} 
\end{table}   

%% file: flare_mcmc_energy.tex
\begin{table} 
\caption{Flare luminosity, size, and luminosity estimated in sub-subsection \ref{sec:flareenergy}, and $\Delta t_{\rm flare}$ in subsection \ref{sec:time}. Flare temperatures are assumed to be $9000, 12000$, and $15000$K.  }   
 \begin{tabular}{cccccc} 
\hline 
TIC ID& $L_{\rm max}$[erg/s]& $l_{\rm max}$[km]& $E_{\rm bol}$[erg]& $E_{\rm tomoe}$[erg]& $\Delta t_{\rm flare}$[sec] \\ 
\hline 
15904458 \footnotemark[$1$]& $4.39 \pm ^{1.07}_{-0.79} \times 10^{30}$& $1.93 \pm ^{0.59}_{-0.35} \times 10^{4}$& $4.41 \pm ^{1.08}_{-0.80} \times 10^{32}$& $8.9 \times 10^{31}$& 100.4 \\ 
15904458 \footnotemark[$2$]& $2.30 \pm ^{0.56}_{-0.42} \times 10^{30}$& $1.40 \pm ^{0.42}_{-0.25} \times 10^{4}$& $3.70 \pm ^{0.91}_{-0.67} \times 10^{32}$& $7.5 \times 10^{31}$& 160.8 \\ 
17198188 & $1.30 \pm ^{0.32}_{-0.23} \times 10^{29}$& $3.32 \pm ^{1.01}_{-0.60} \times 10^{3}$& $1.31 \pm ^{0.32}_{-0.24} \times 10^{31}$& $2.6 \times 10^{30}$& 101.1 \\ 
18376490 & $3.63 \pm ^{0.89}_{-0.66} \times 10^{29}$& $5.56 \pm ^{1.69}_{-1.00} \times 10^{3}$& $6.44 \pm ^{1.58}_{-1.16} \times 10^{31}$& $1.3 \times 10^{31}$& 177.1 \\ 
20536892 \footnotemark[$1$]& $1.64 \pm ^{0.40}_{-0.30} \times 10^{31}$& $3.73 \pm ^{1.13}_{-0.67} \times 10^{4}$& $2.40 \pm ^{0.59}_{-0.43} \times 10^{33}$& $4.8 \times 10^{32}$& 146.6 \\ 
21286088 & $3.07 \pm ^{0.75}_{-0.55} \times 10^{31}$& $5.11 \pm ^{1.55}_{-0.92} \times 10^{4}$& $6.45 \pm ^{1.58}_{-1.17} \times 10^{33}$& $1.3 \times 10^{33}$& 210.3 \\ 
21286574 & $2.55 \pm ^{0.63}_{-0.46} \times 10^{30}$& $1.47 \pm ^{0.45}_{-0.27} \times 10^{4}$& $\cdots$& $\cdots$& $\cdots$ \\ 
46305452 & $8.39 \pm ^{2.06}_{-1.52} \times 10^{29}$& $8.45 \pm ^{2.56}_{-1.52} \times 10^{3}$& $1.70 \pm ^{0.42}_{-0.31} \times 10^{32}$& $3.4 \times 10^{31}$& 202.3 \\ 
55288759 & $3.37 \pm ^{0.83}_{-0.61} \times 10^{29}$& $5.35 \pm ^{1.62}_{-0.96} \times 10^{3}$& $1.74 \pm ^{0.43}_{-0.31} \times 10^{31}$& $3.5 \times 10^{30}$& 51.5 \\ 
98751507 & $4.13 \pm ^{1.01}_{-0.75} \times 10^{29}$& $5.93 \pm ^{1.80}_{-1.07} \times 10^{3}$& $1.49 \pm ^{0.37}_{-0.27} \times 10^{31}$& $3.0 \times 10^{30}$& 36.1 \\ 
119031076 & $3.40 \pm ^{0.83}_{-0.61} \times 10^{29}$& $5.38 \pm ^{1.63}_{-0.97} \times 10^{3}$& $1.98 \pm ^{0.48}_{-0.36} \times 10^{31}$& $4.0 \times 10^{30}$& 58.1 \\ 
121638493 & $9.57 \pm ^{2.35}_{-1.73} \times 10^{29}$& $9.02 \pm ^{2.74}_{-1.63} \times 10^{3}$& $1.11 \pm ^{0.27}_{-0.20} \times 10^{32}$& $2.2 \times 10^{31}$& 116.1 \\ 
185243119 & $1.53 \pm ^{0.37}_{-0.28} \times 10^{31}$& $3.60 \pm ^{1.09}_{-0.65} \times 10^{4}$& $4.76 \pm ^{1.17}_{-0.86} \times 10^{33}$& $9.6 \times 10^{32}$& 311.5 \\ 
203228080 & $1.62 \pm ^{0.40}_{-0.29} \times 10^{30}$& $1.17 \pm ^{0.36}_{-0.21} \times 10^{4}$& $3.47 \pm ^{0.85}_{-0.63} \times 10^{32}$& $7.0 \times 10^{31}$& 214.6 \\ 
243017627 & $5.84 \pm ^{1.43}_{-1.06} \times 10^{29}$& $7.05 \pm ^{2.14}_{-1.27} \times 10^{3}$& $1.59 \pm ^{0.39}_{-0.29} \times 10^{32}$& $3.2 \times 10^{31}$& 272.3 \\ 
251725681 & $2.91 \pm ^{0.71}_{-0.53} \times 10^{29}$& $4.97 \pm ^{1.51}_{-0.90} \times 10^{3}$& $6.24 \pm ^{1.53}_{-1.13} \times 10^{31}$& $1.3 \times 10^{31}$& 214.4 \\ 
315617164 & $7.82 \pm ^{1.92}_{-1.41} \times 10^{29}$& $8.15 \pm ^{2.47}_{-1.47} \times 10^{3}$& $4.84 \pm ^{1.19}_{-0.87} \times 10^{31}$& $9.8 \times 10^{30}$& 61.9 \\ 
358561826 & $2.08 \pm ^{0.51}_{-0.38} \times 10^{30}$& $1.33 \pm ^{0.40}_{-0.24} \times 10^{4}$& $2.81 \pm ^{0.69}_{-0.51} \times 10^{33}$& $5.7 \times 10^{32}$& 1347.9 \\ 
358573853 & $2.26 \pm ^{0.55}_{-0.41} \times 10^{29}$& $4.38 \pm ^{1.33}_{-0.79} \times 10^{3}$& $4.38 \pm ^{1.07}_{-0.79} \times 10^{31}$& $8.8 \times 10^{30}$& 193.7 \\ 
366485128 & $1.45 \pm ^{0.35}_{-0.26} \times 10^{30}$& $1.11 \pm ^{0.34}_{-0.20} \times 10^{4}$& $6.36 \pm ^{1.56}_{-1.15} \times 10^{32}$& $1.3 \times 10^{32}$& 439.7 \\ 
374272810 & $2.28 \pm ^{0.56}_{-0.41} \times 10^{29}$& $4.40 \pm ^{1.33}_{-0.79} \times 10^{3}$& $7.68 \pm ^{1.88}_{-1.39} \times 10^{31}$& $1.5 \times 10^{31}$& 337.6 \\ 
380197137 & $1.32 \pm ^{0.32}_{-0.24} \times 10^{30}$& $1.06 \pm ^{0.32}_{-0.19} \times 10^{4}$& $3.90 \pm ^{0.96}_{-0.71} \times 10^{31}$& $7.9 \times 10^{30}$& 29.6 \\ 
435904068 & $1.18 \pm ^{0.29}_{-0.21} \times 10^{30}$& $1.00 \pm ^{0.30}_{-0.18} \times 10^{4}$& $9.30 \pm ^{2.28}_{-1.68} \times 10^{31}$& $1.9 \times 10^{31}$& 78.8 \\ 
 \hline  \end{tabular}   
\label{table:flare_energy}  
\begin{tabnote} 
\footnotemark[$1$] First peak of flare.  
\footnotemark[$2$] Second peak of flare. 
\end{tabnote} 
\end{table}   

%% file: fig_summary.tex
\begin{figure*}[htbp]\begin{center} 
 
 \includegraphics[width=0.49 \linewidth]{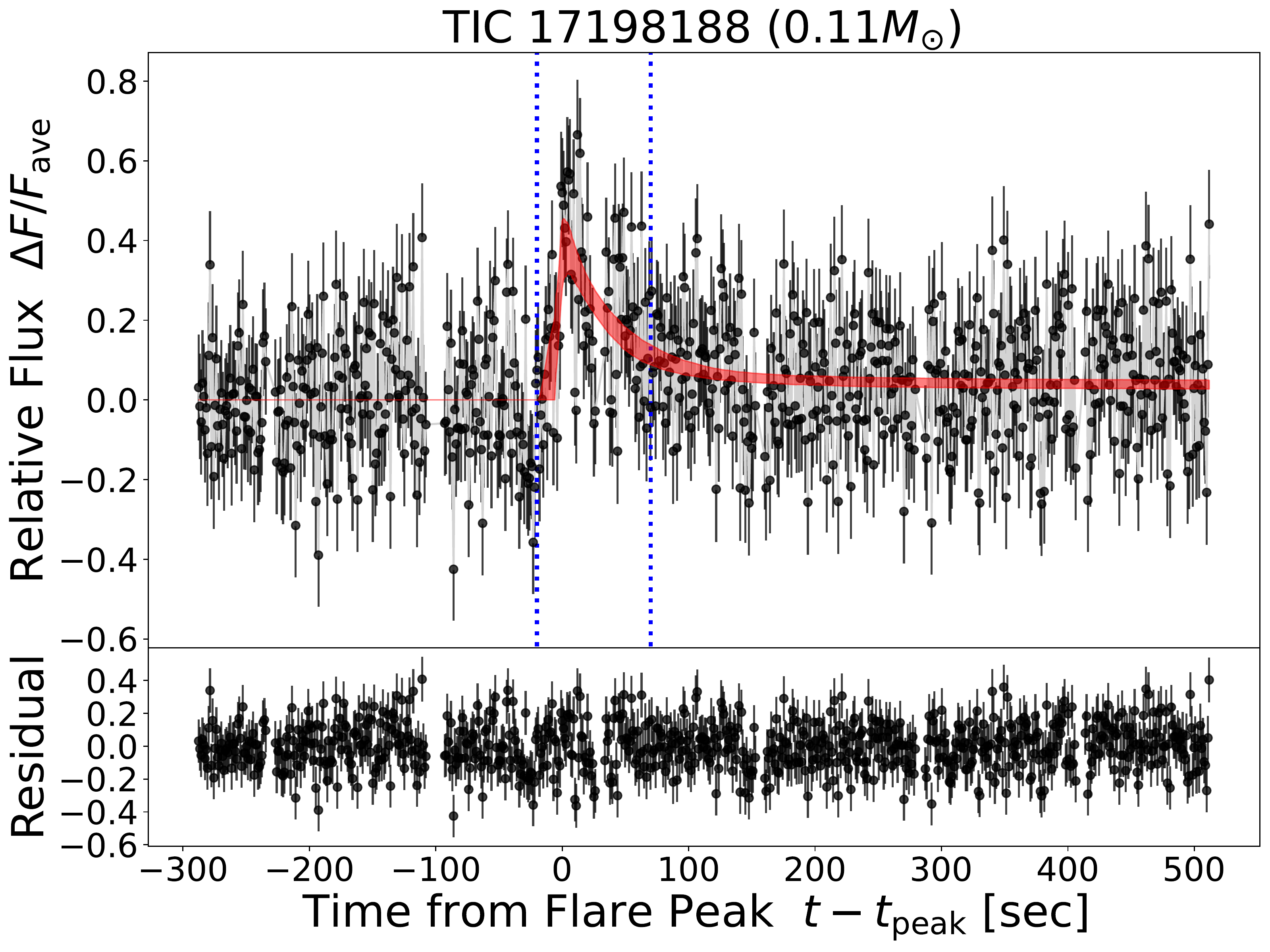}\hfill 
\includegraphics[width=0.49 \linewidth]{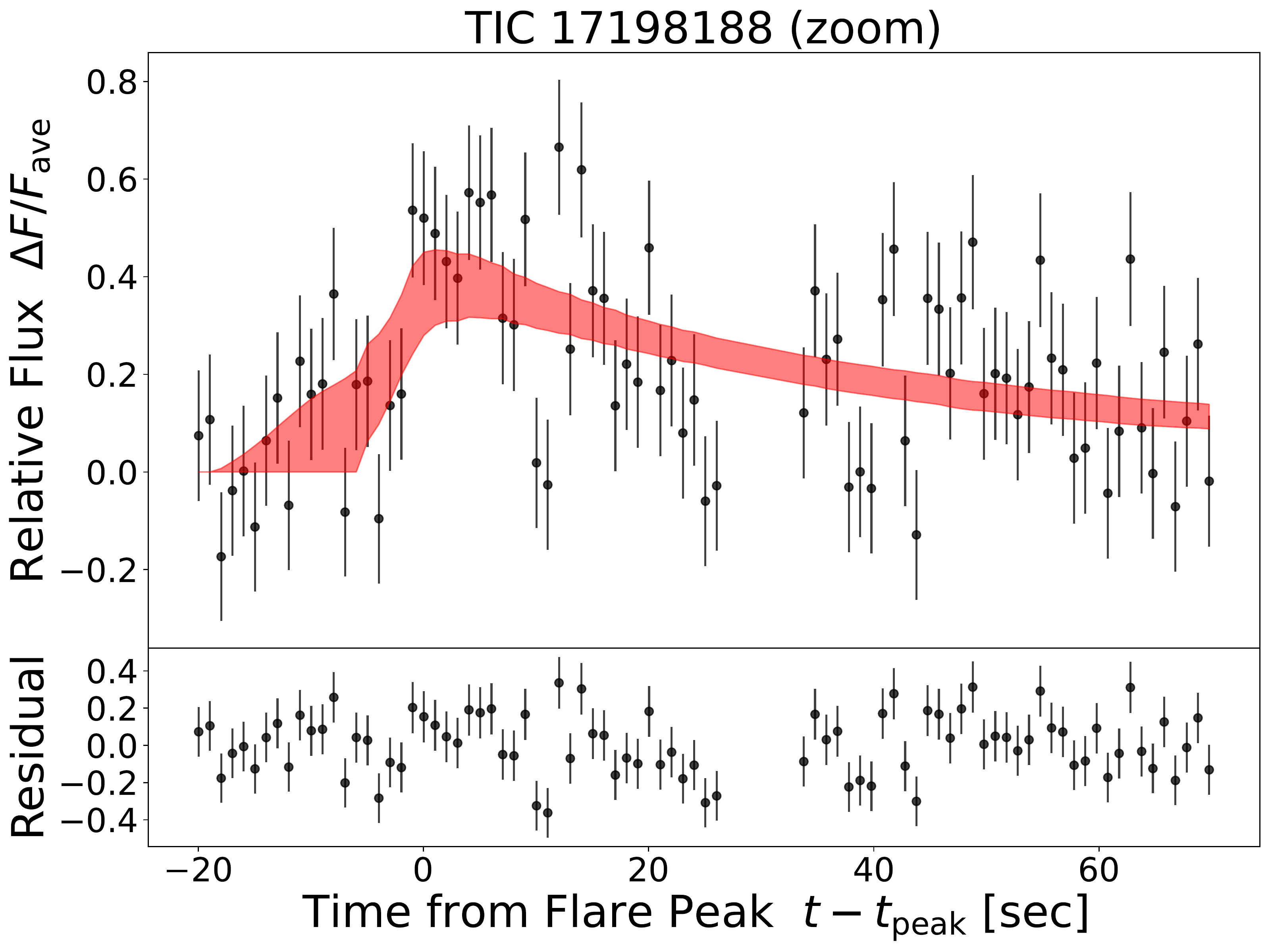}\hfill 
\includegraphics[width=0.49 \linewidth]{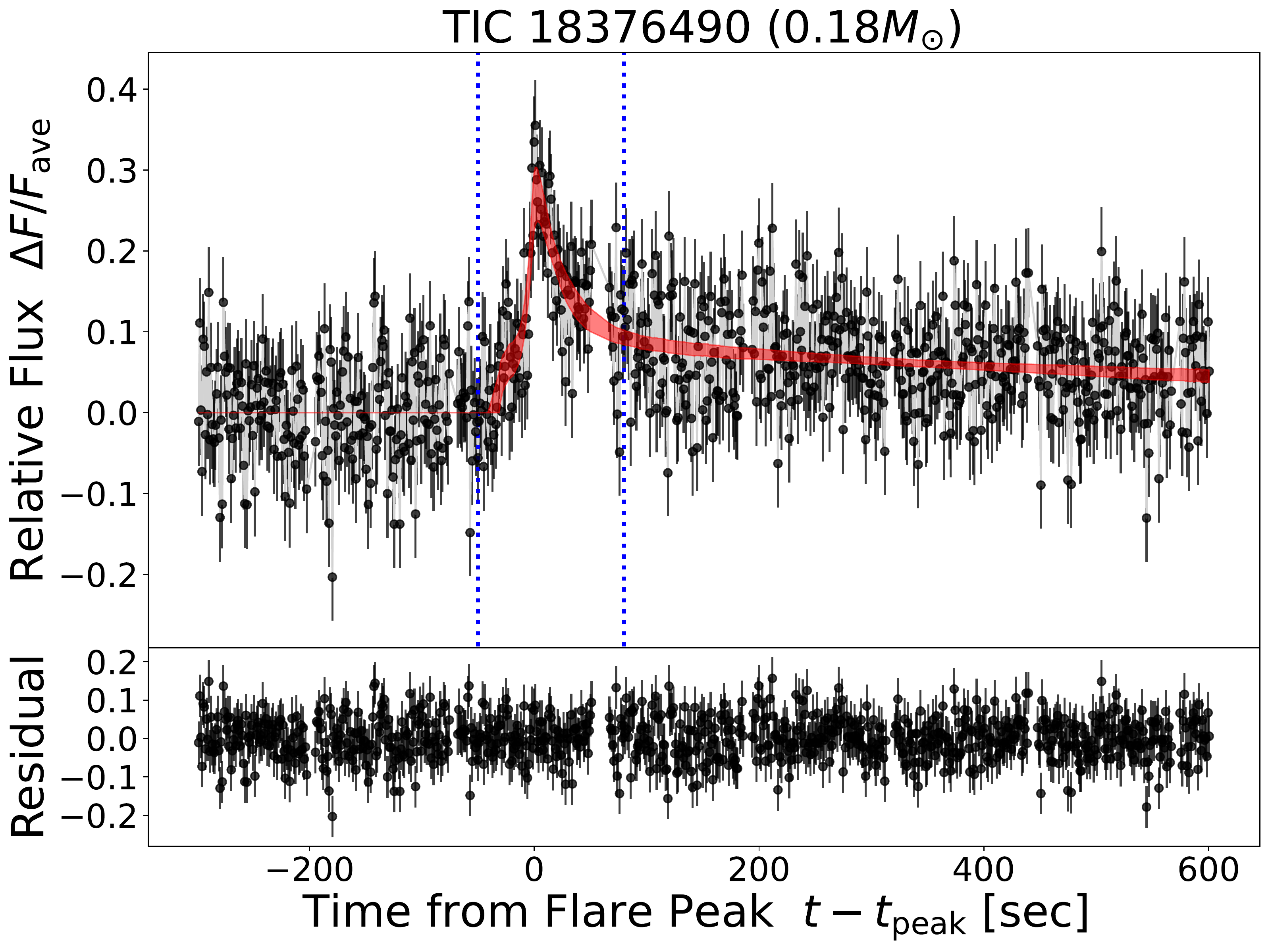}\hfill 
\includegraphics[width=0.49 \linewidth]{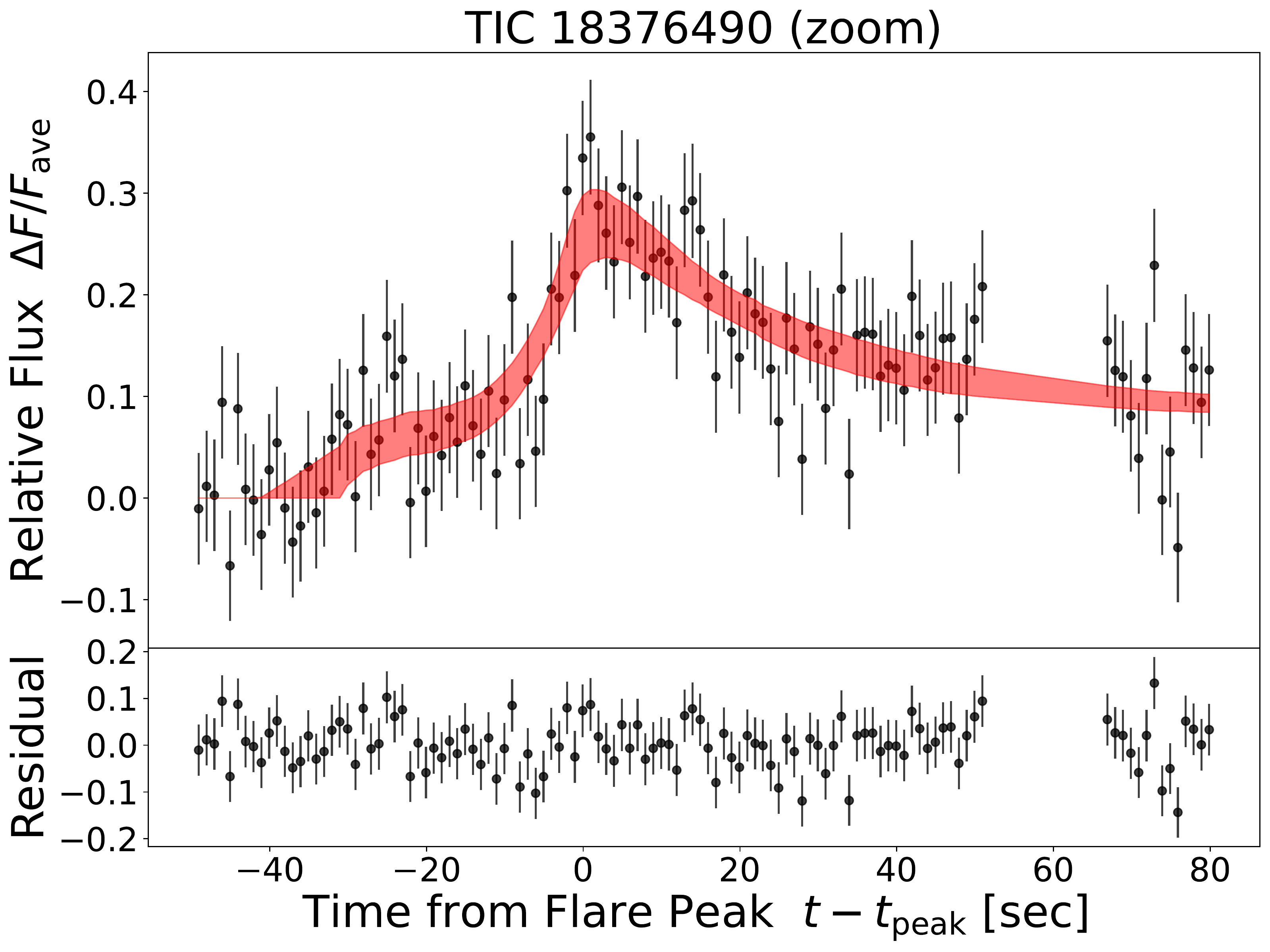}\hfill 
\includegraphics[width=0.49 \linewidth]{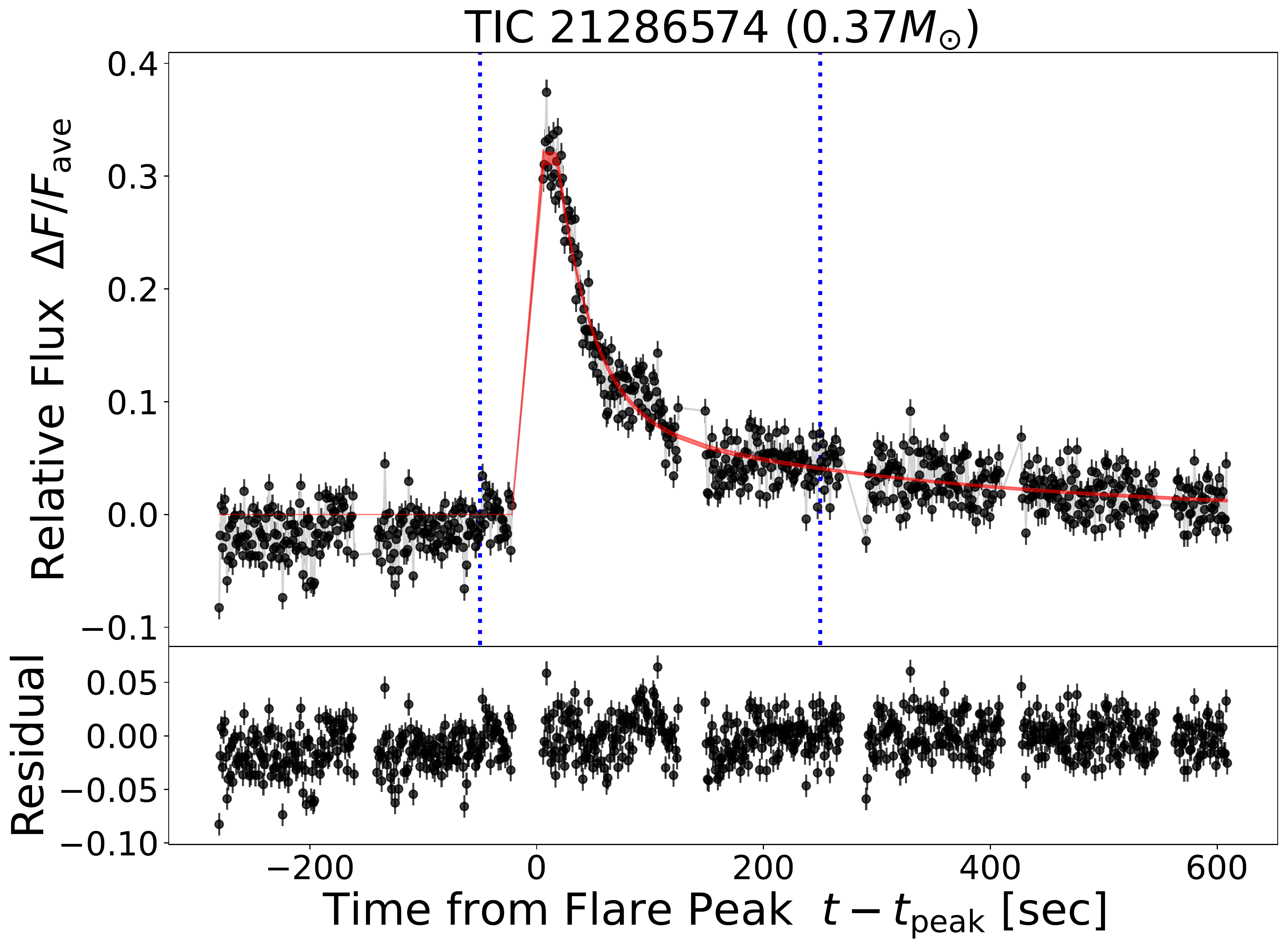}\hfill 
\includegraphics[width=0.49 \linewidth]{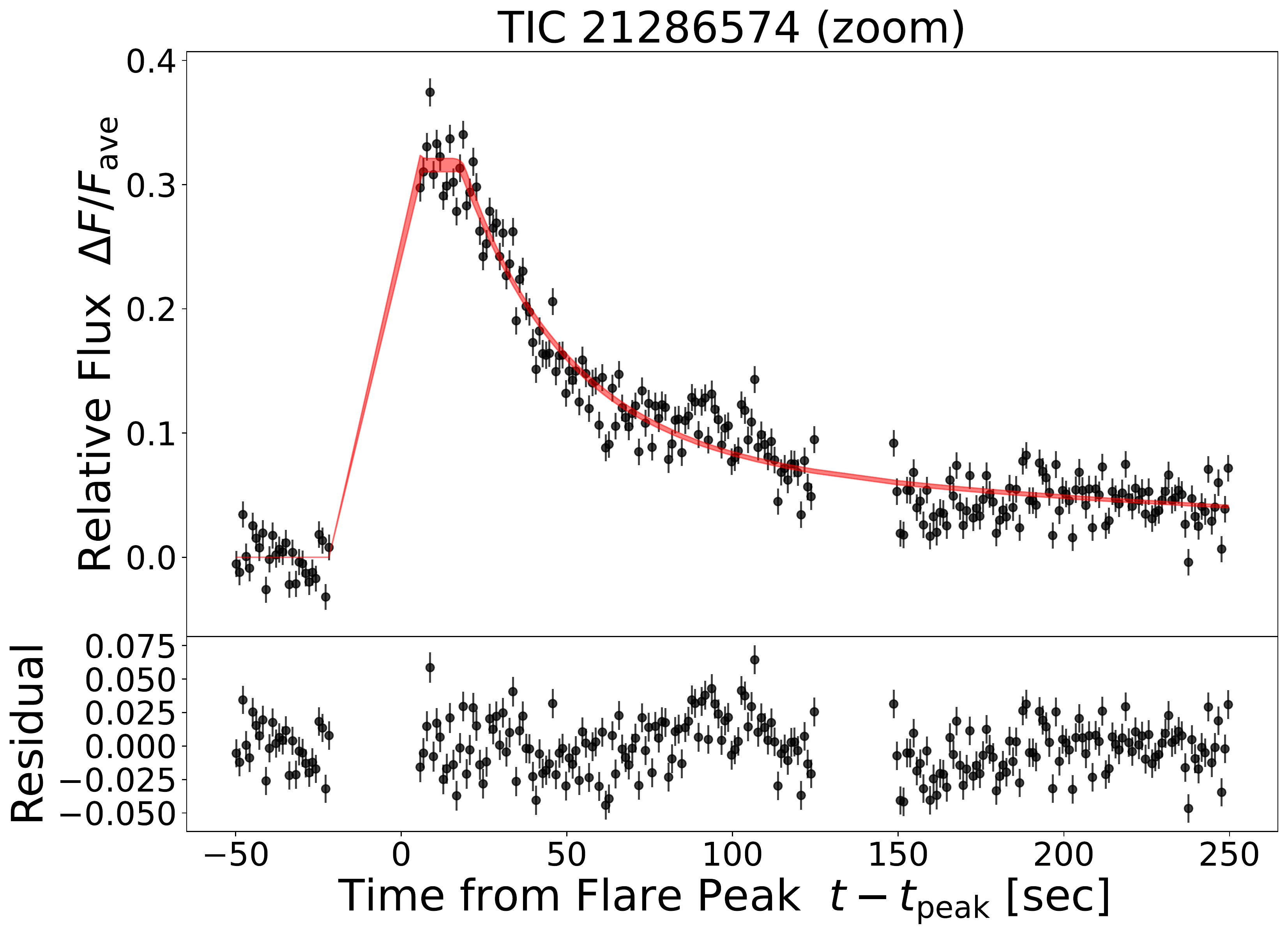}\hfill 
 \end{center}
\caption{ Classical flare light curves and the residuals after subtracting the best fit models. The left panels show the whole views, and the right panels show the zoom-up views of the region indicated by the blue dashed lines in the left panels. The red shaded region corresponds to the 90 \% credible interval of the fitting model (equation \ref{eq:lc_fit_model}).  \label{fig:lc_flares}} \end{figure*}  
 \addtocounter{figure}{-1}\begin{figure*}[htbp]\begin{center} 
 
 \includegraphics[width=0.49 \linewidth]{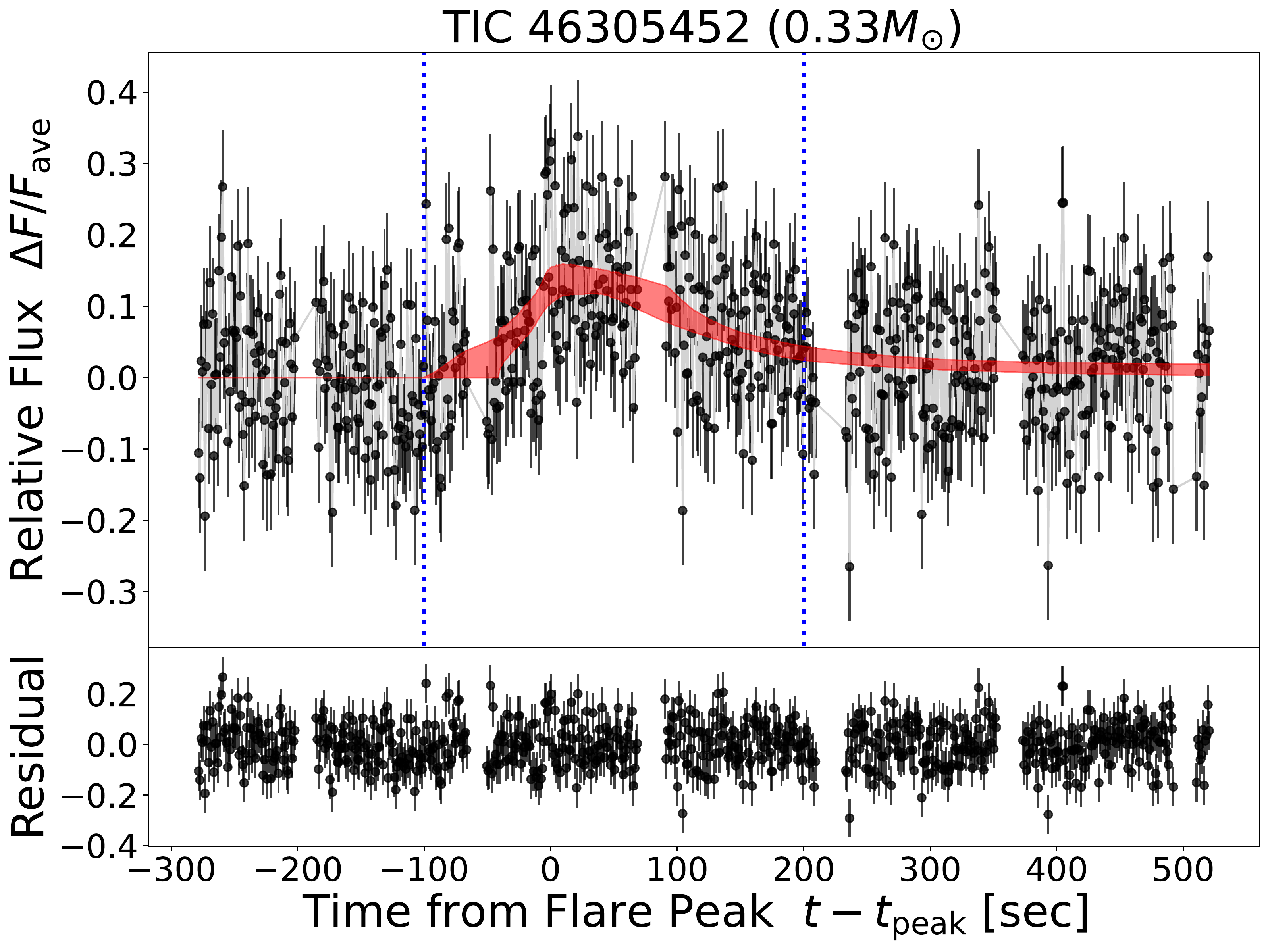}\hfill 
\includegraphics[width=0.49 \linewidth]{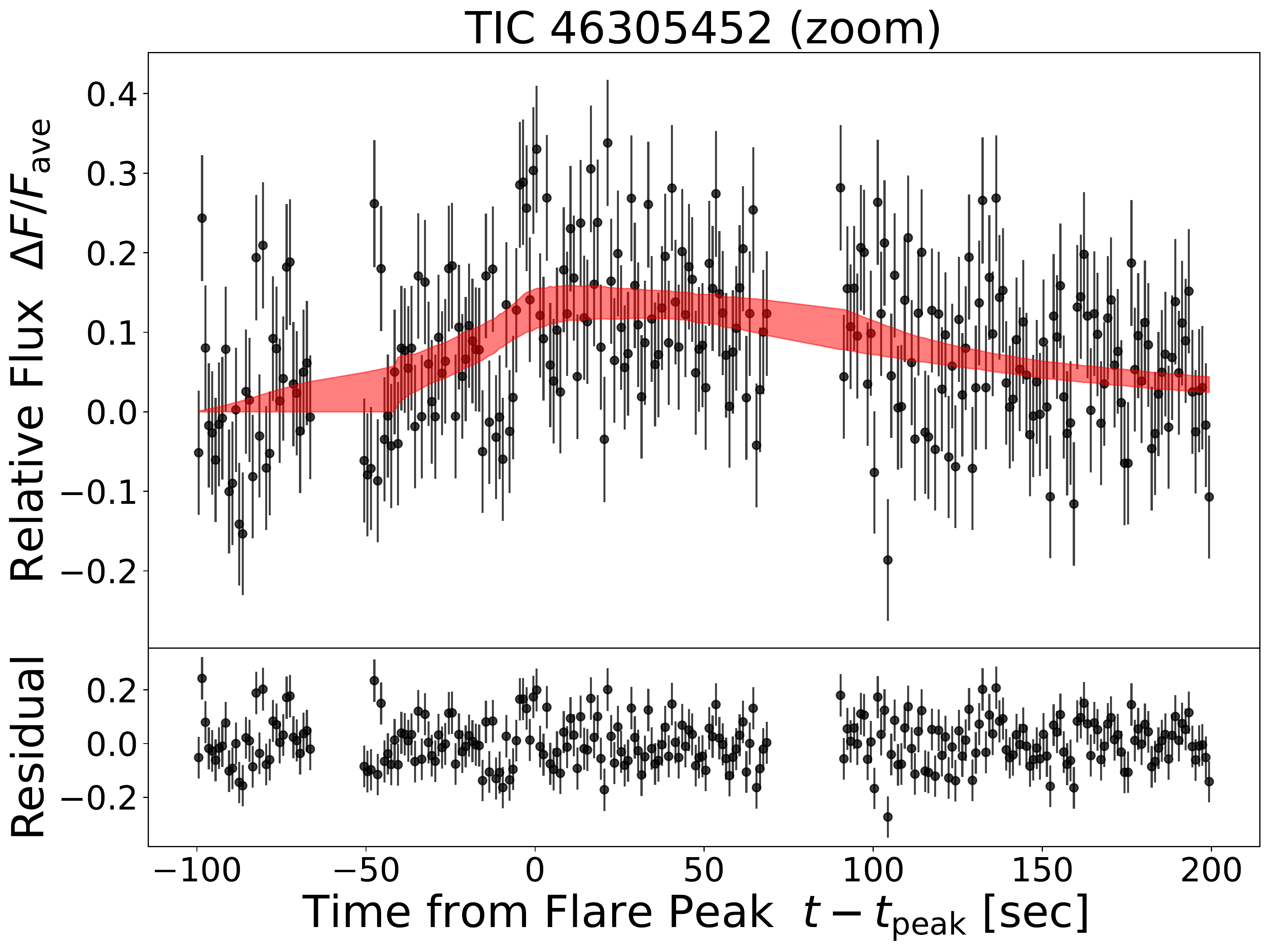}\hfill 
\includegraphics[width=0.49 \linewidth]{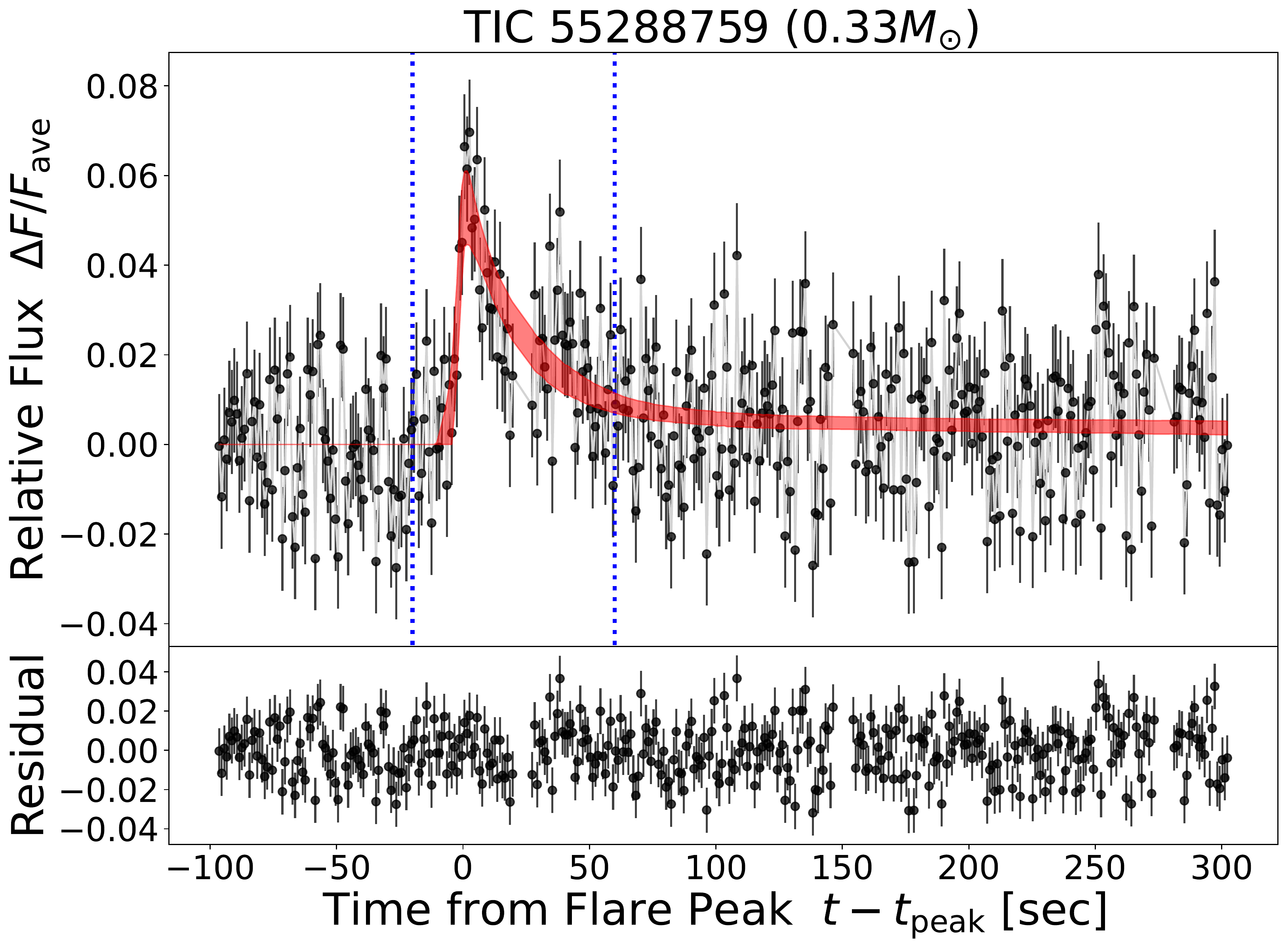}\hfill 
\includegraphics[width=0.49 \linewidth]{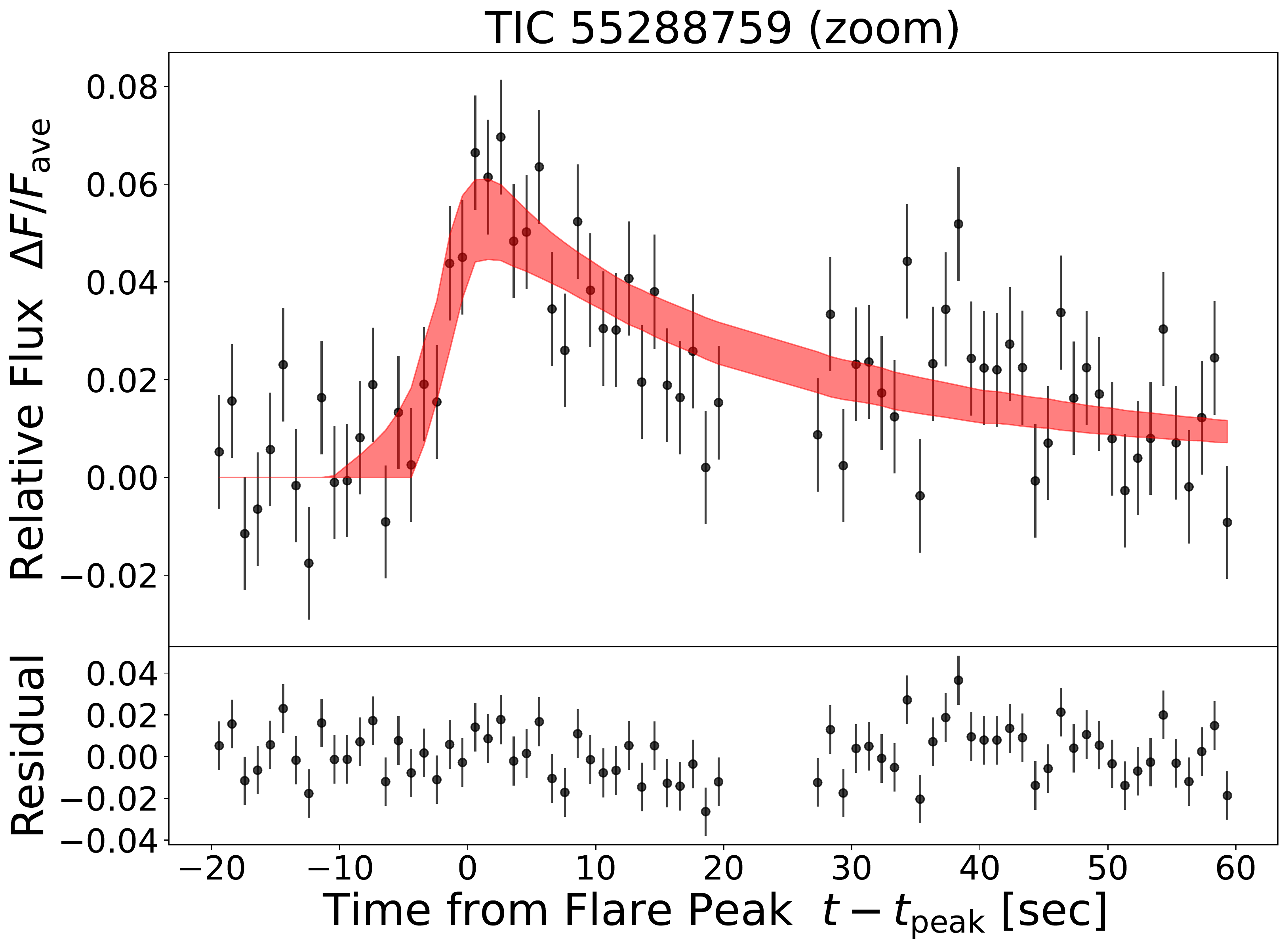}\hfill 
\includegraphics[width=0.49 \linewidth]{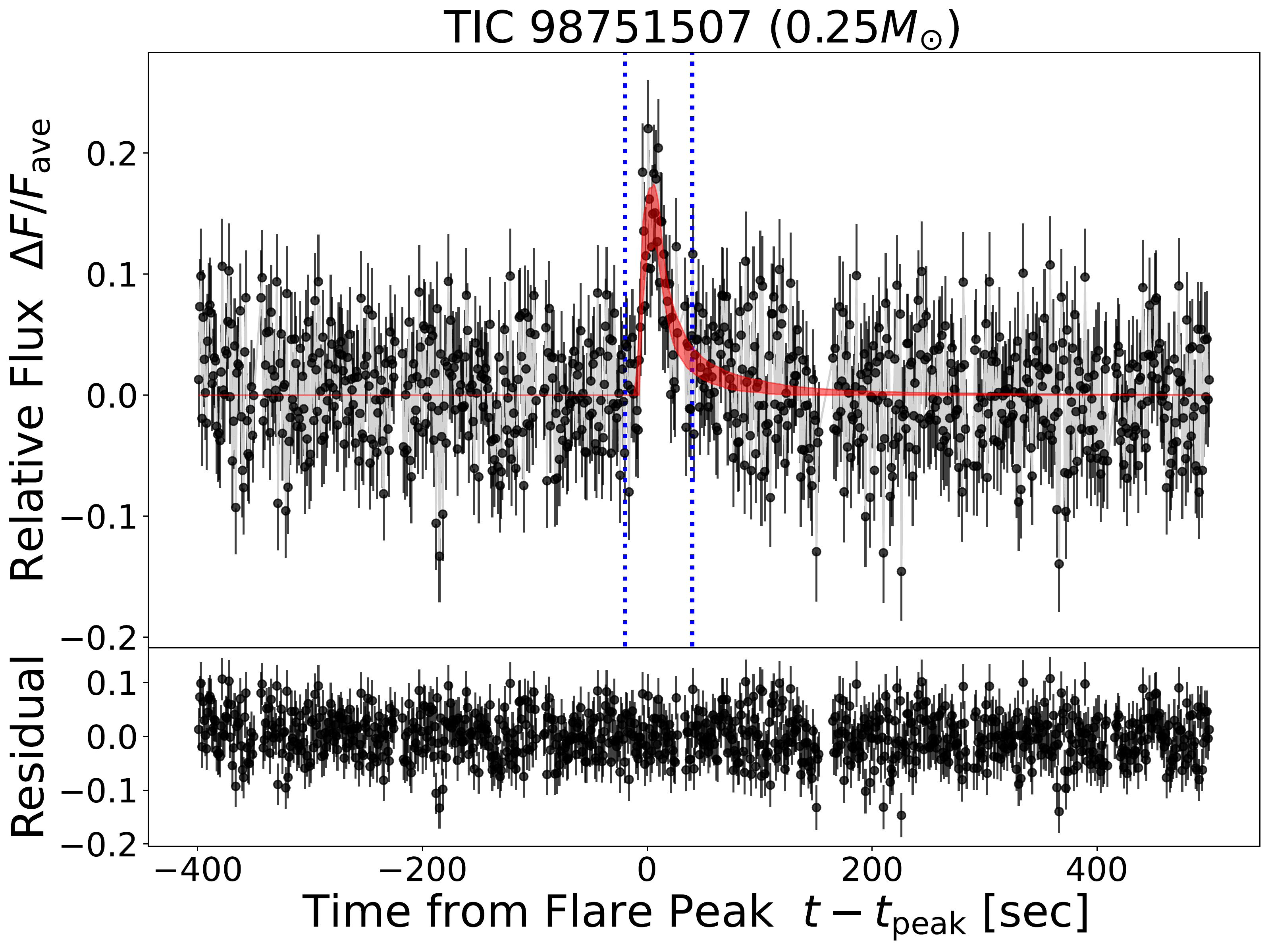}\hfill 
\includegraphics[width=0.49 \linewidth]{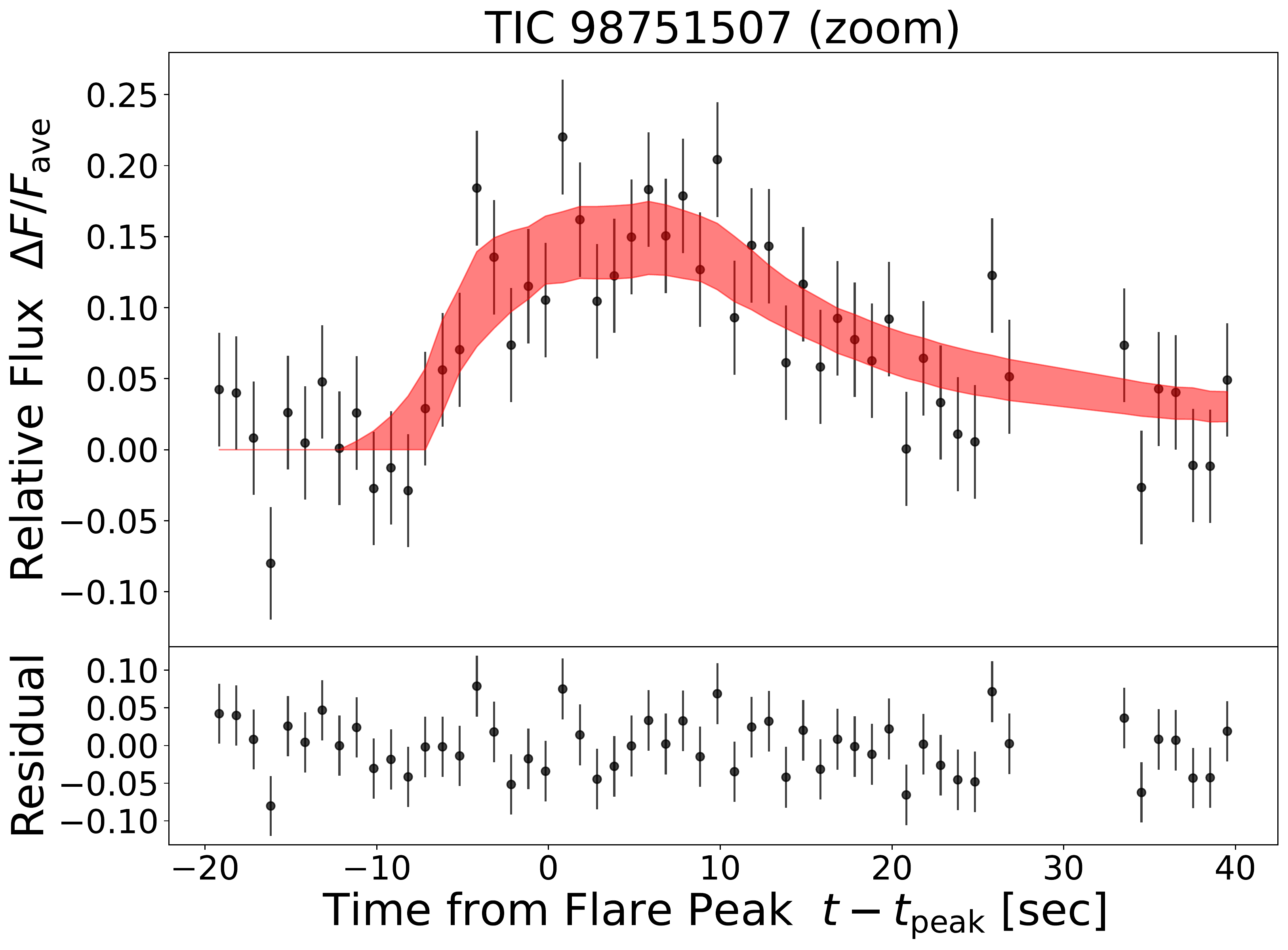}\hfill 
\caption{(Continue) }  \end{center}\end{figure*}  
 \addtocounter{figure}{-1}\begin{figure*}[htbp]\begin{center} 
 
 \includegraphics[width=0.49 \linewidth]{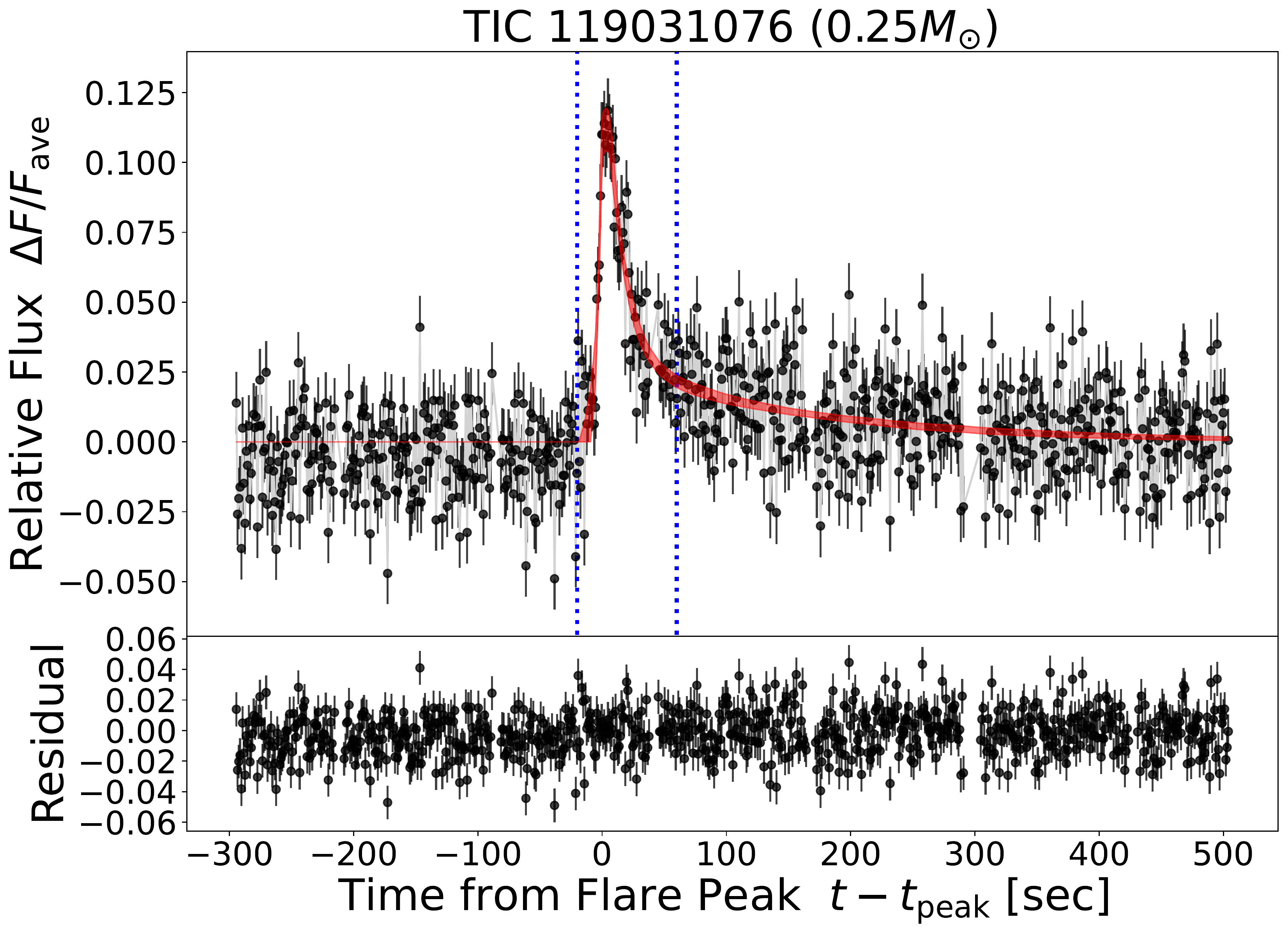}\hfill 
\includegraphics[width=0.49 \linewidth]{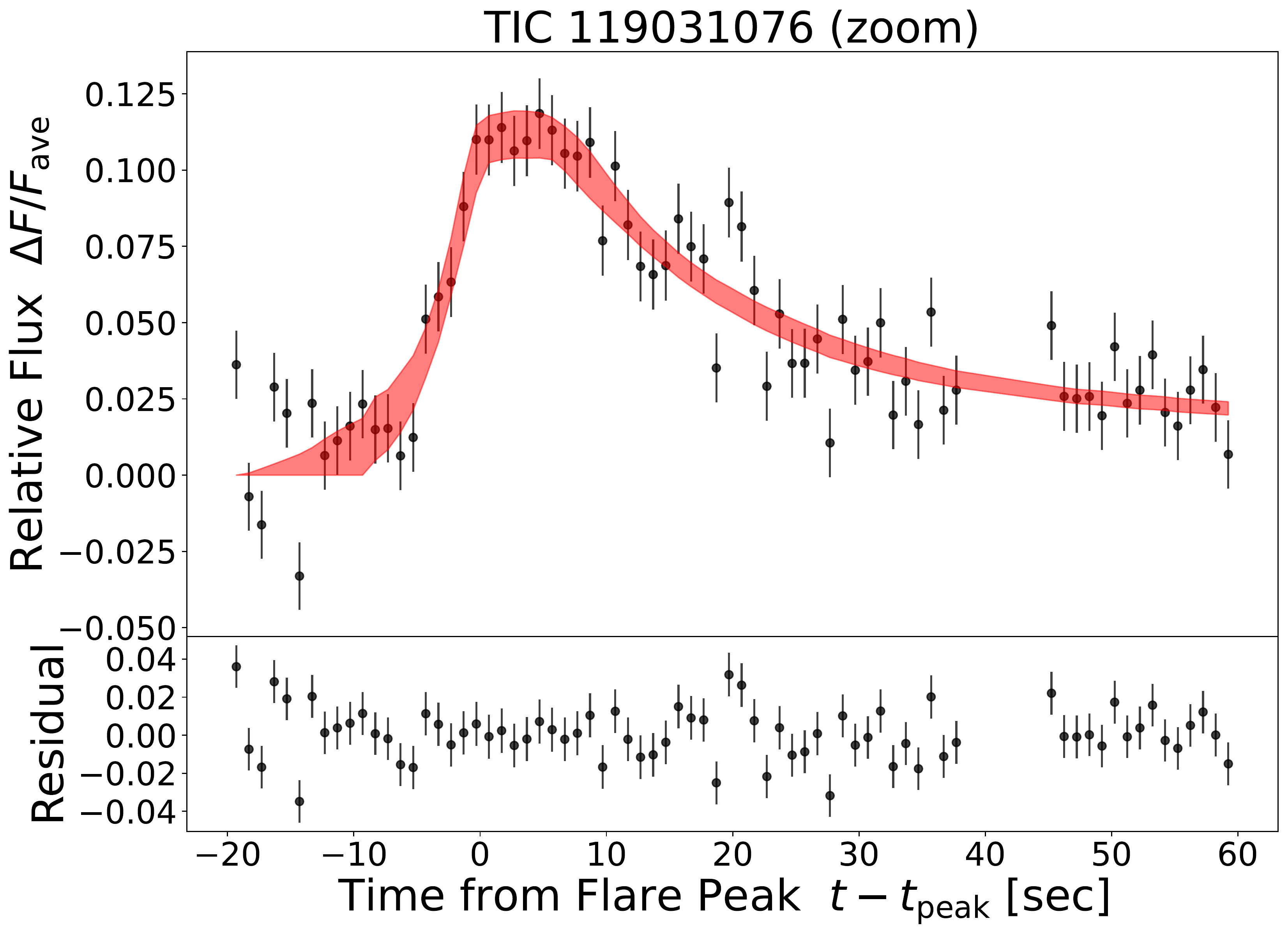}\hfill 
\includegraphics[width=0.49 \linewidth]{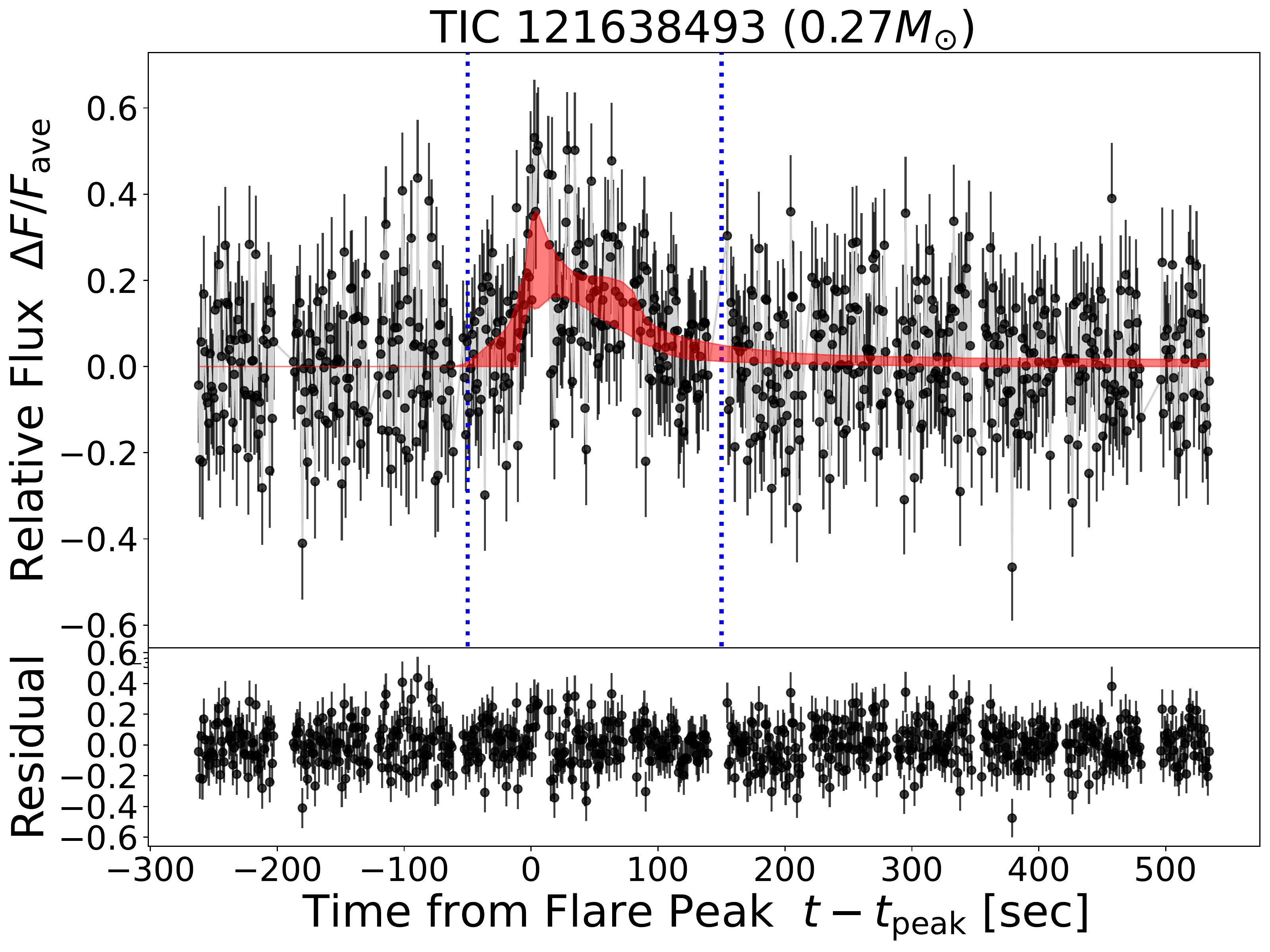}\hfill 
\includegraphics[width=0.49 \linewidth]{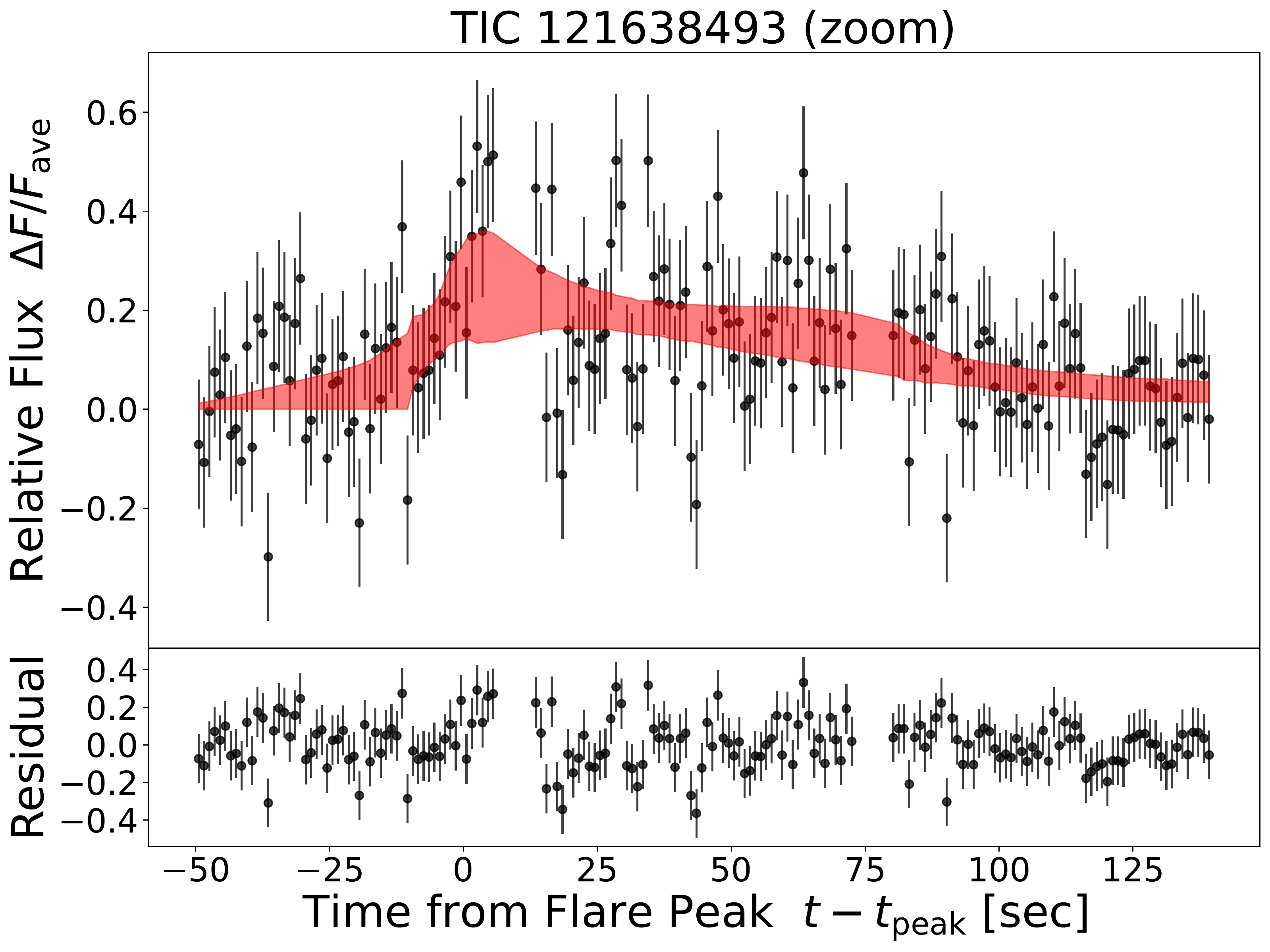}\hfill 
\includegraphics[width=0.49 \linewidth]{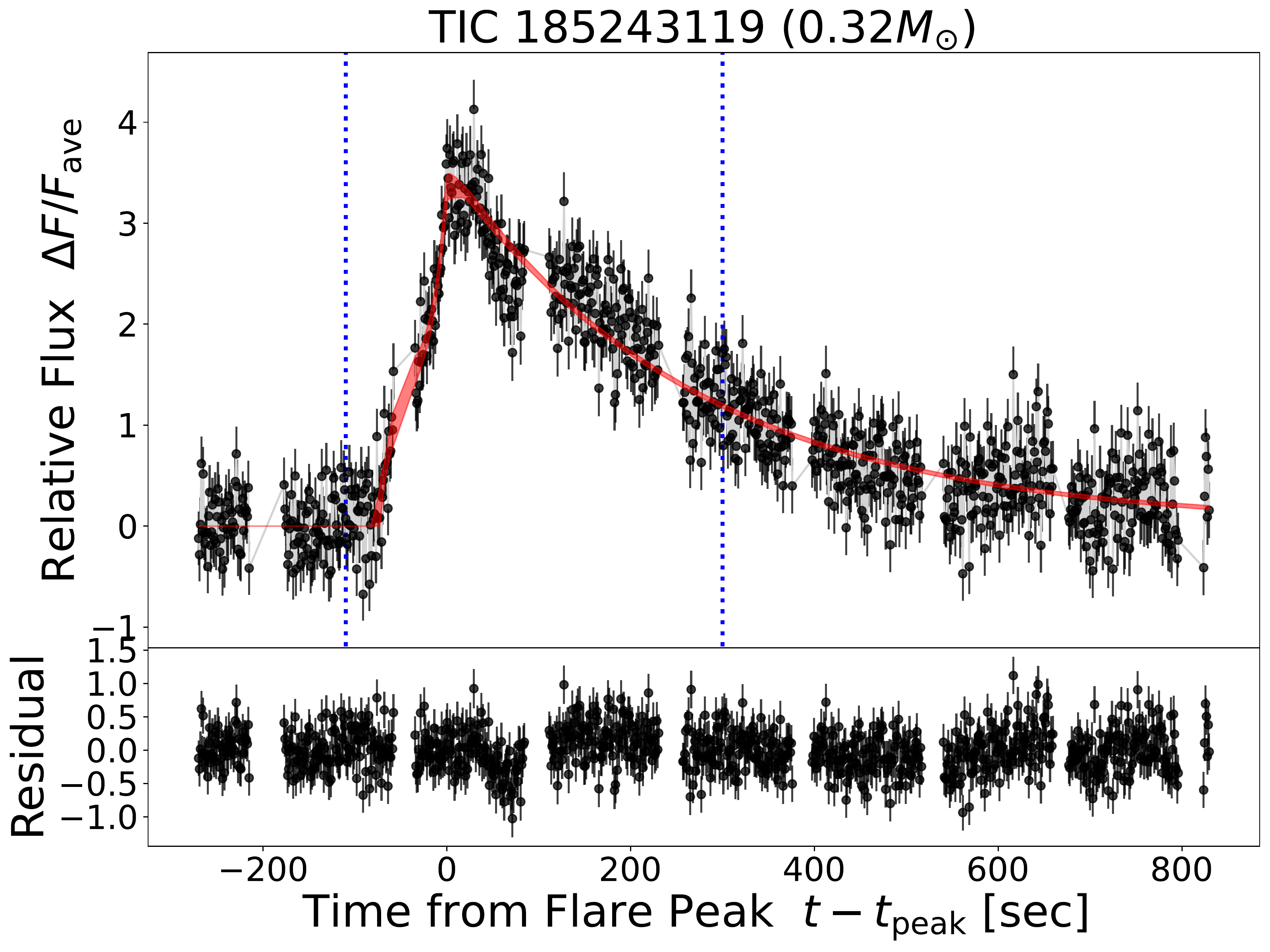}\hfill 
\includegraphics[width=0.49 \linewidth]{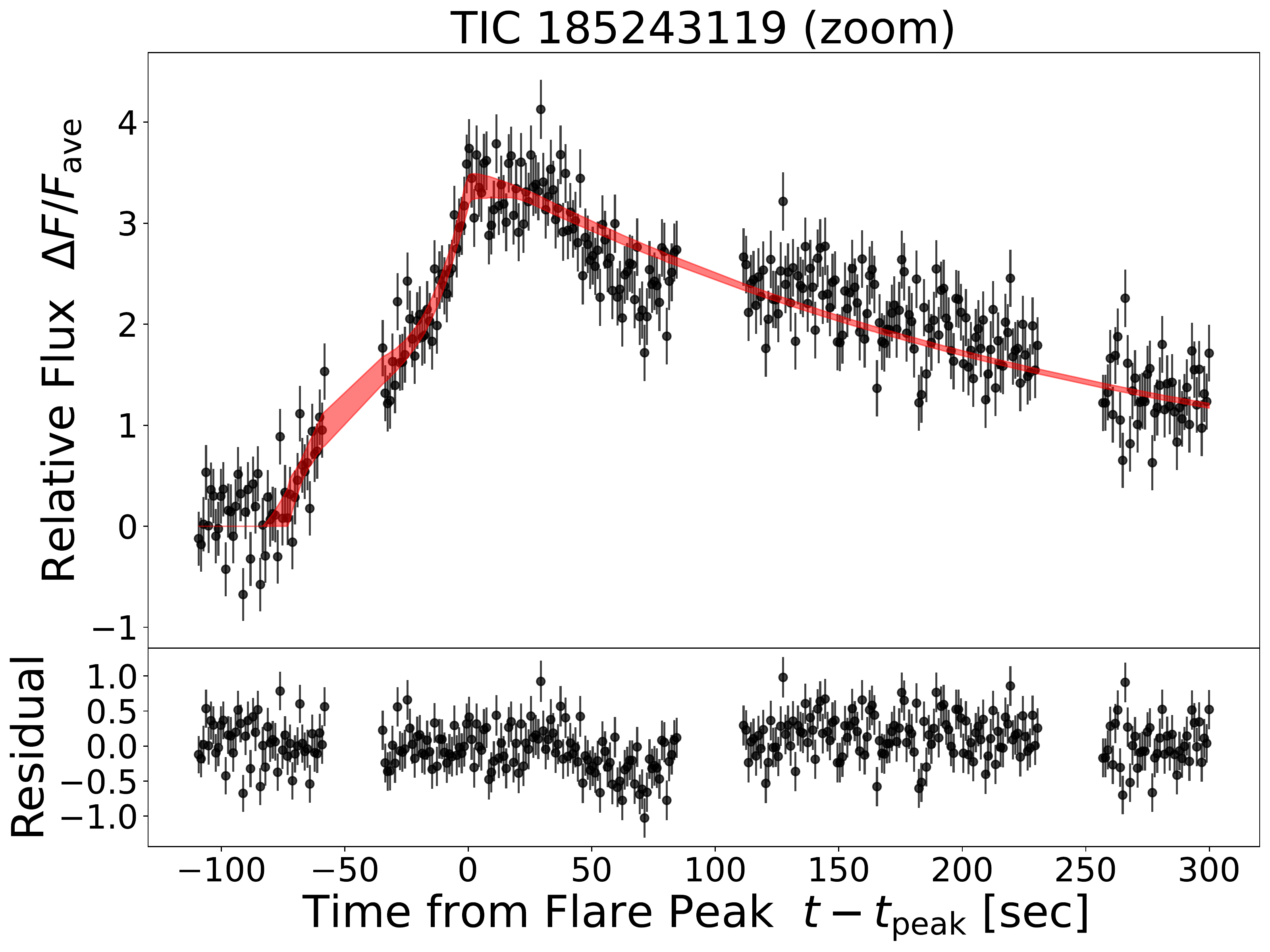}\hfill 
\caption{ (Continue) }  \end{center}\end{figure*}  
 \addtocounter{figure}{-1}\begin{figure*}[htbp]\begin{center} 
 
 \includegraphics[width=0.49 \linewidth]{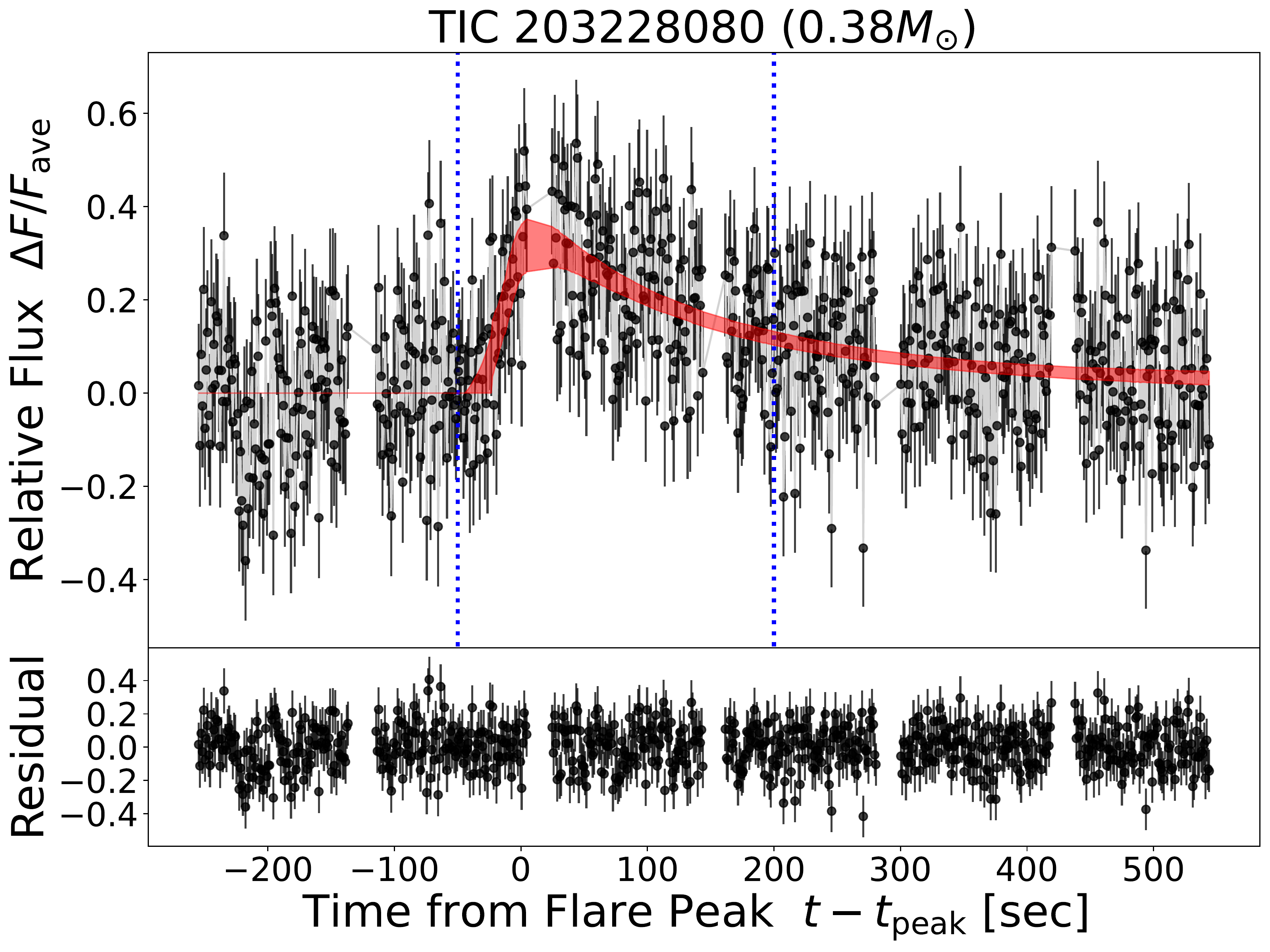}\hfill 
\includegraphics[width=0.49 \linewidth]{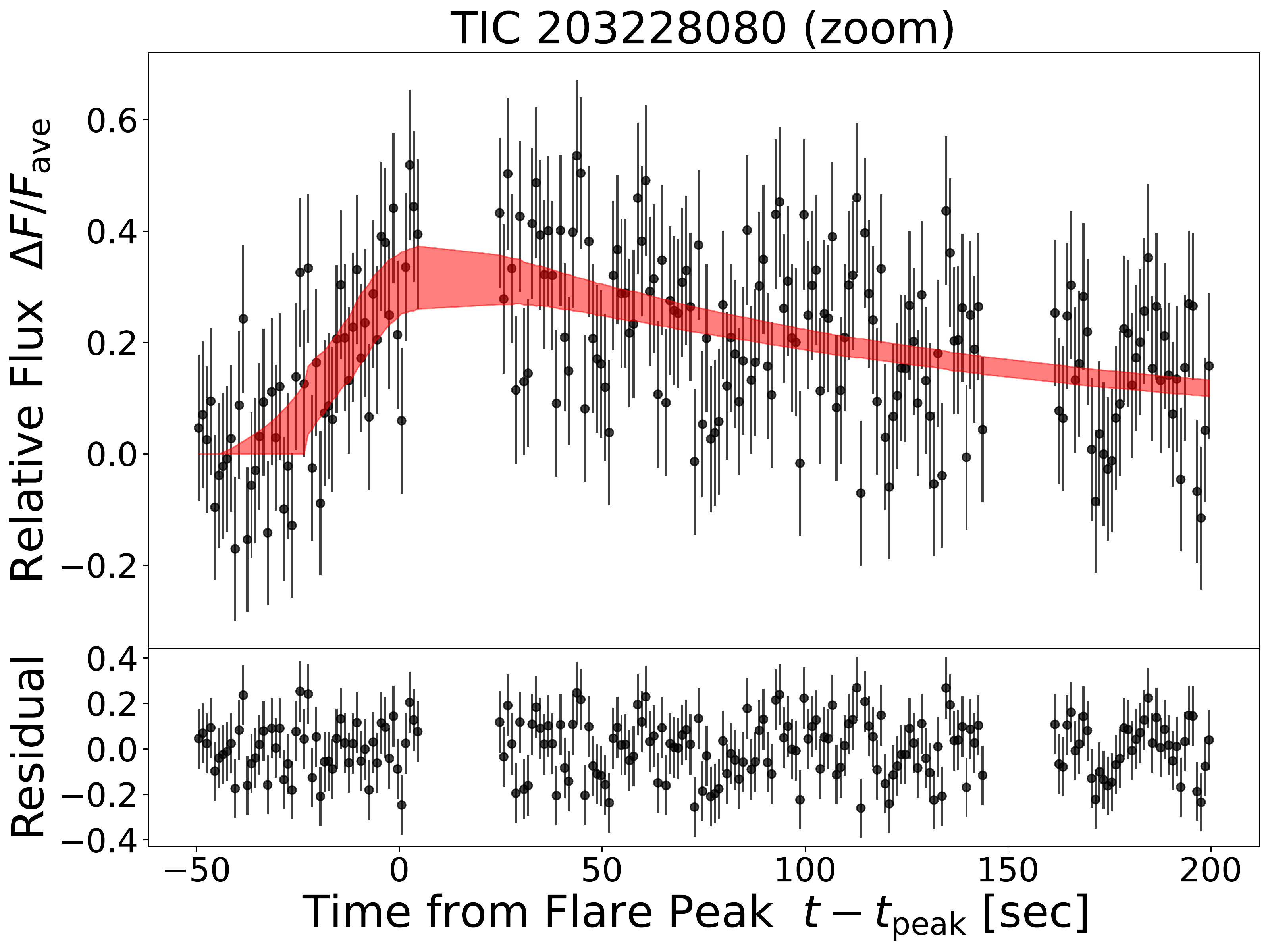}\hfill 
\includegraphics[width=0.49 \linewidth]{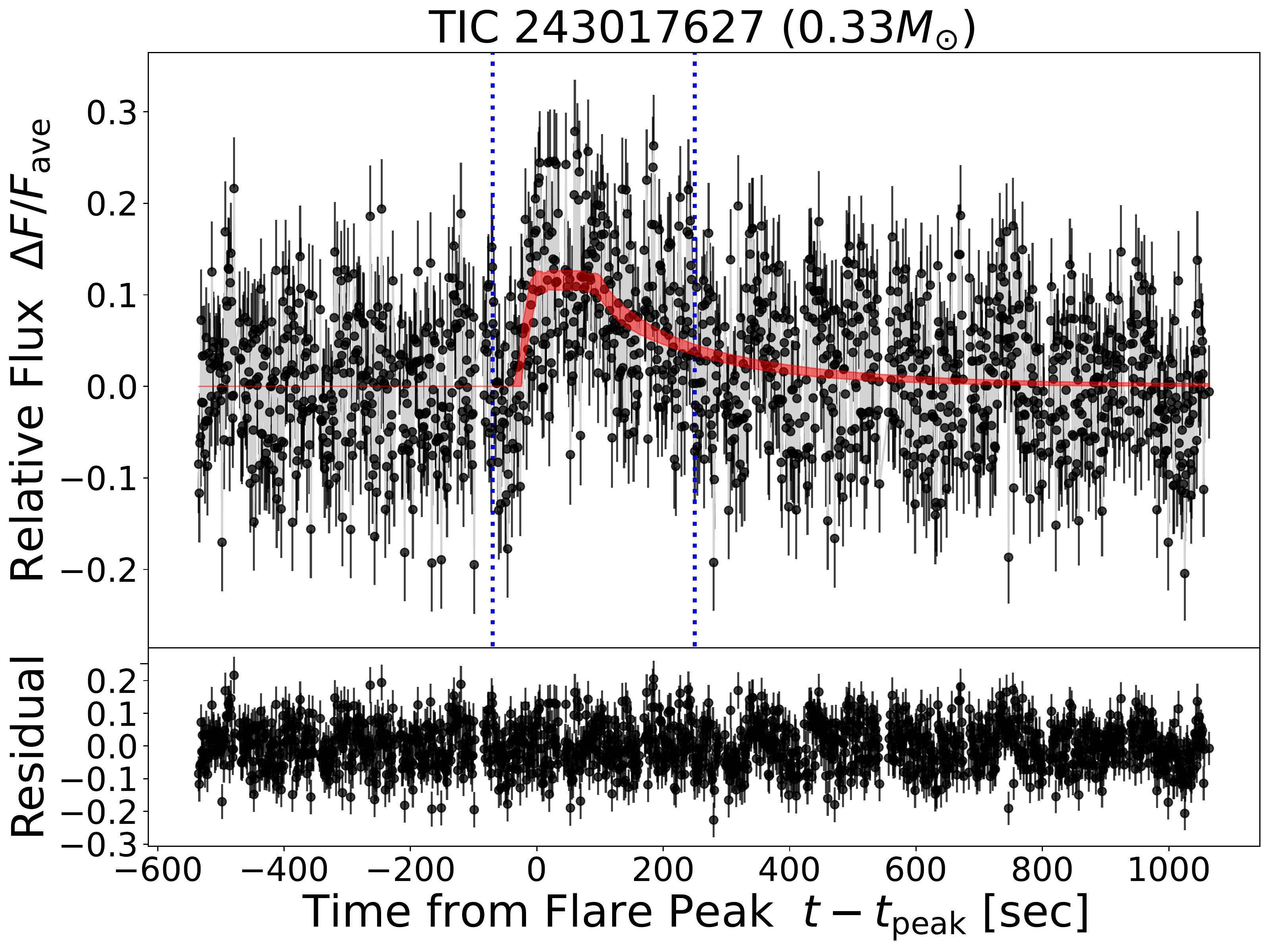}\hfill 
\includegraphics[width=0.49 \linewidth]{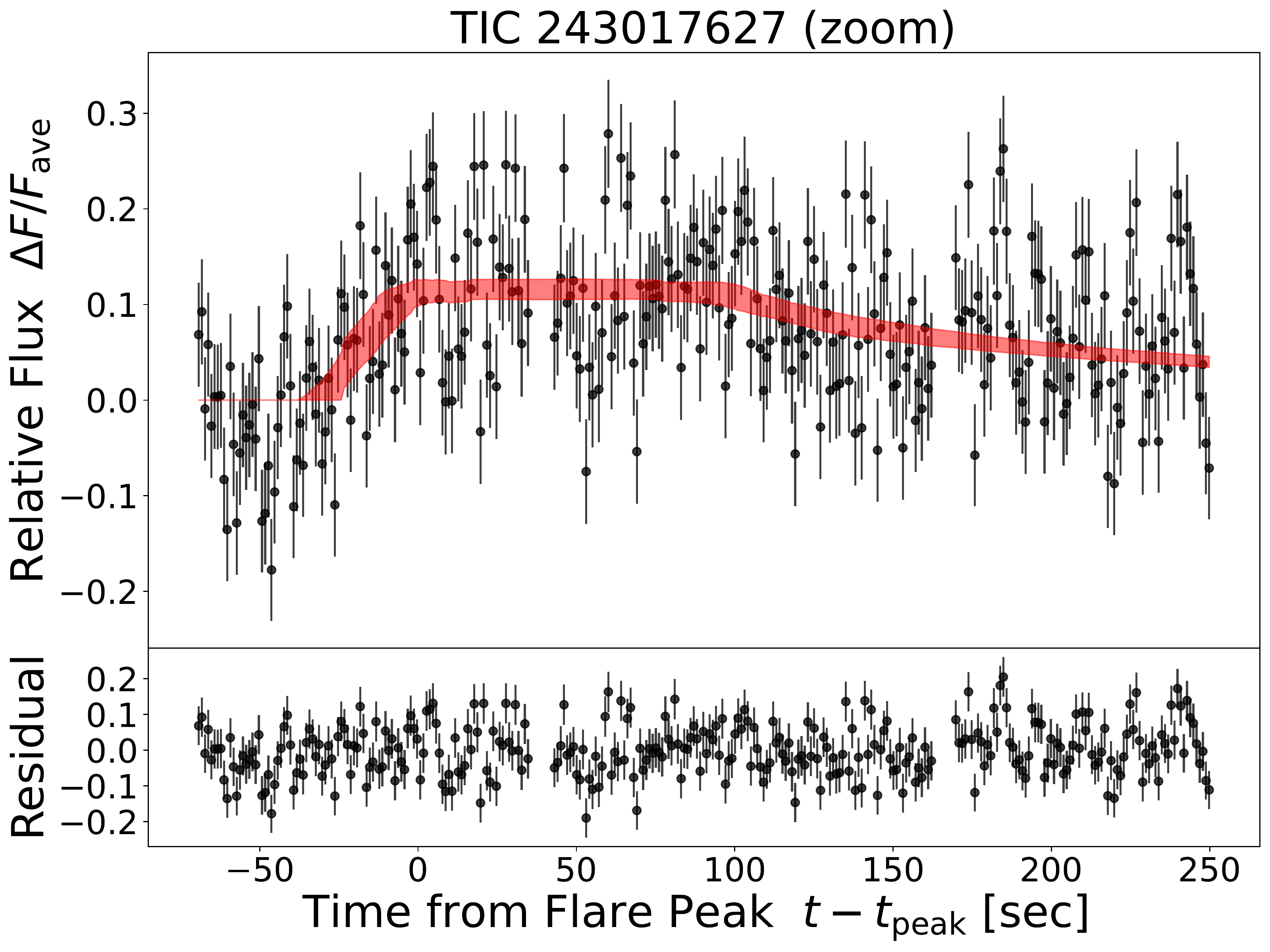}\hfill 
\includegraphics[width=0.49 \linewidth]{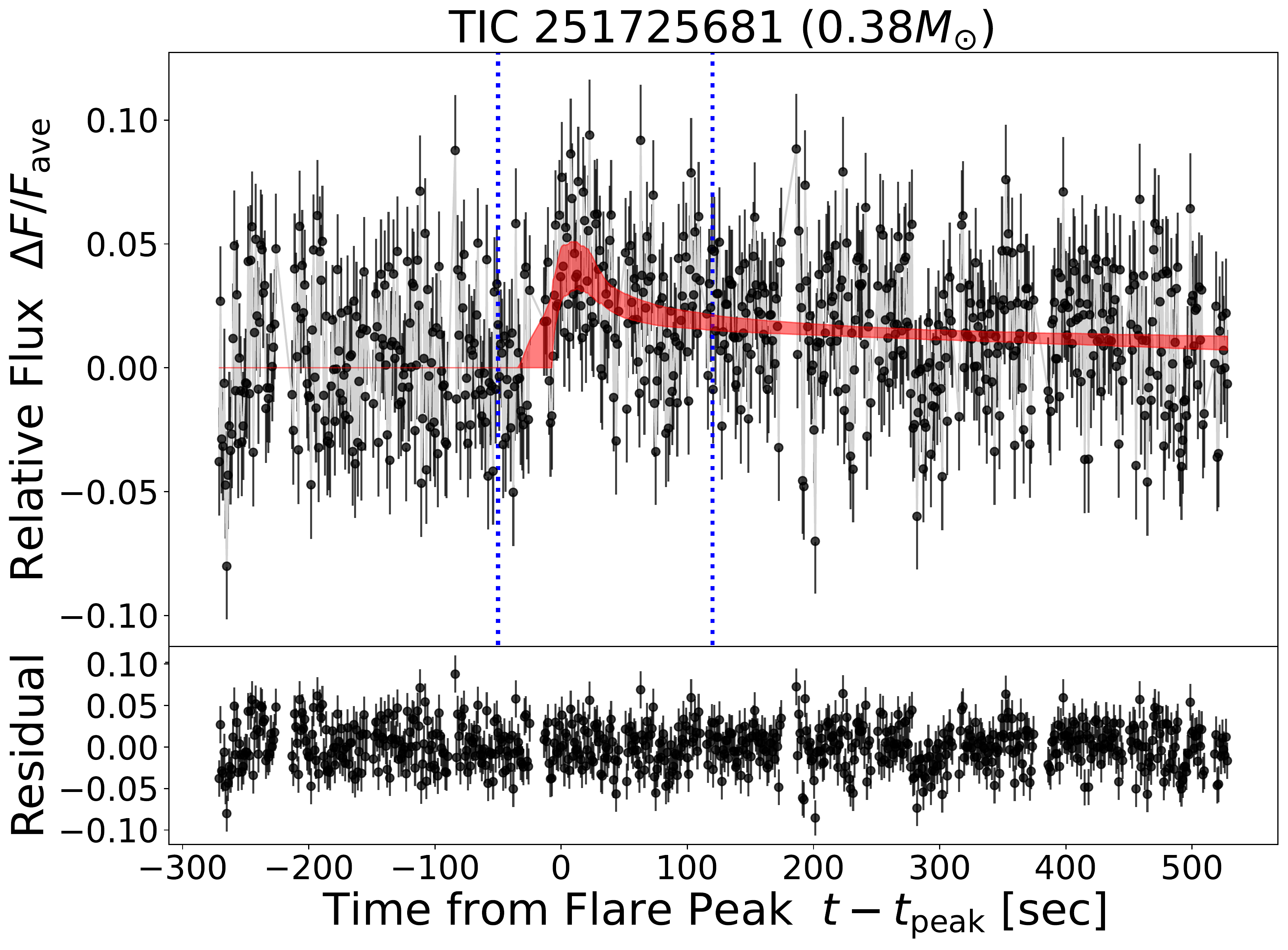}\hfill 
\includegraphics[width=0.49 \linewidth]{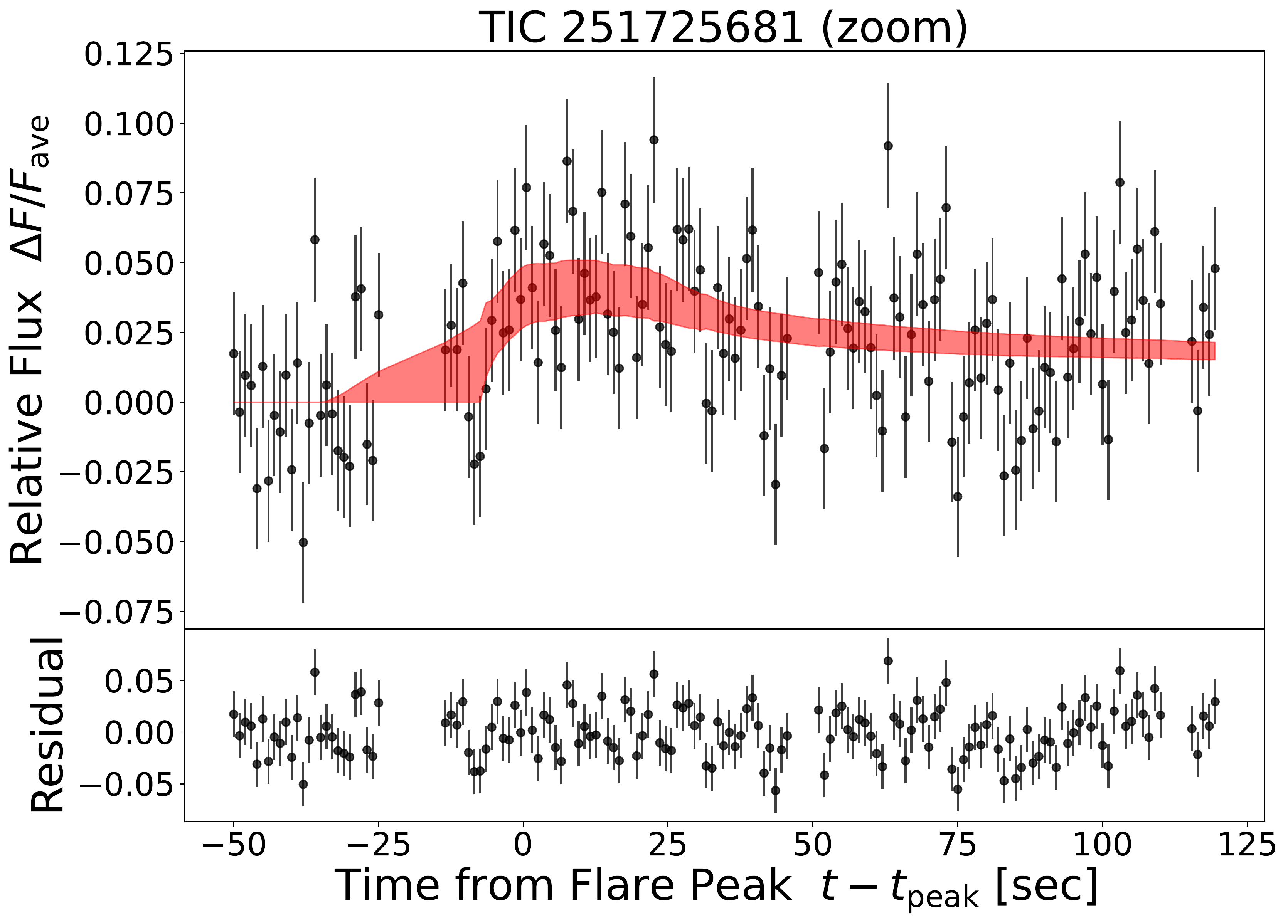}\hfill 
\caption{ (Continue) }  \end{center}\end{figure*}  
 \addtocounter{figure}{-1}\begin{figure*}[htbp]\begin{center} 
 
 \includegraphics[width=0.49 \linewidth]{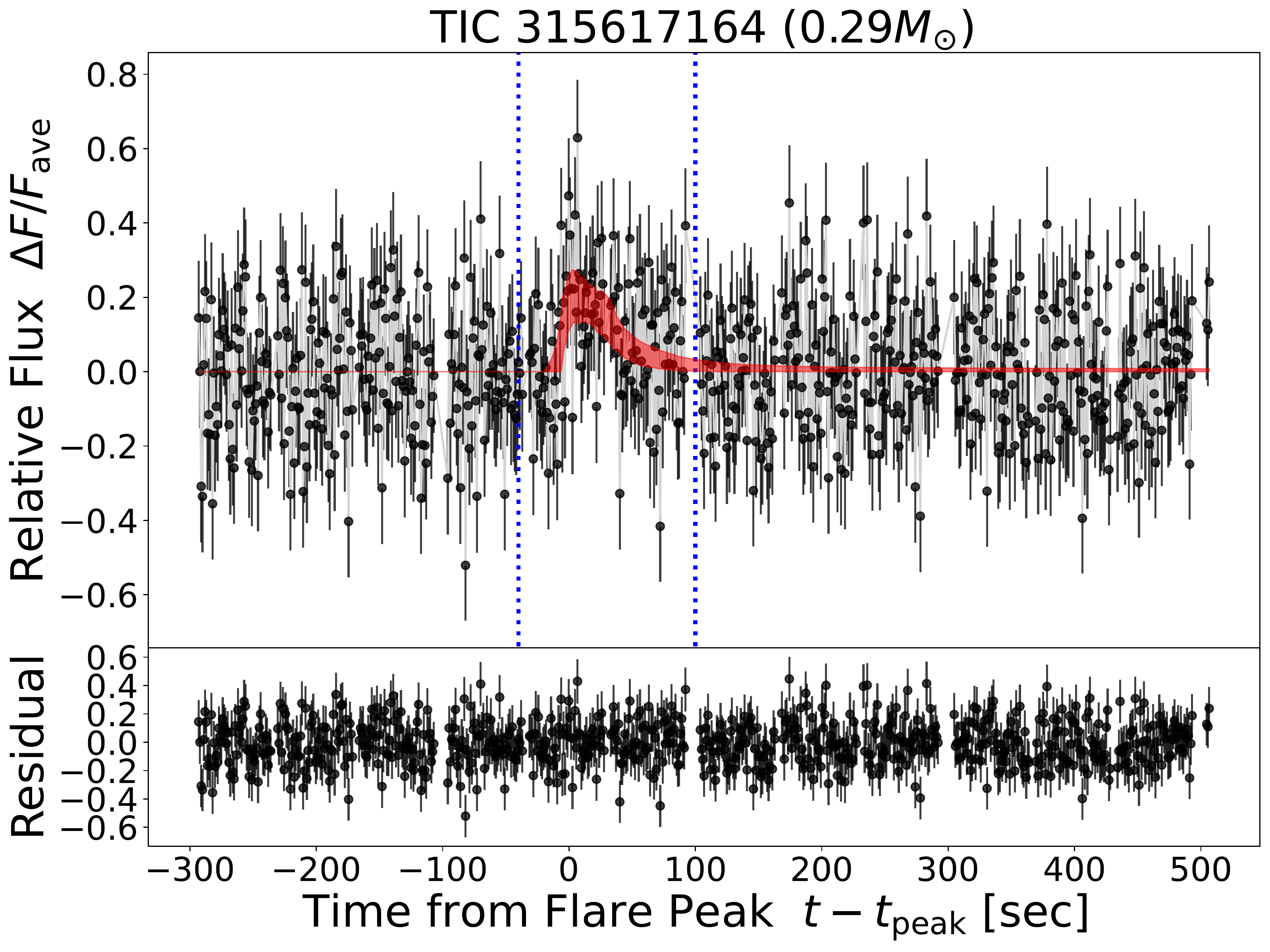}\hfill 
\includegraphics[width=0.49 \linewidth]{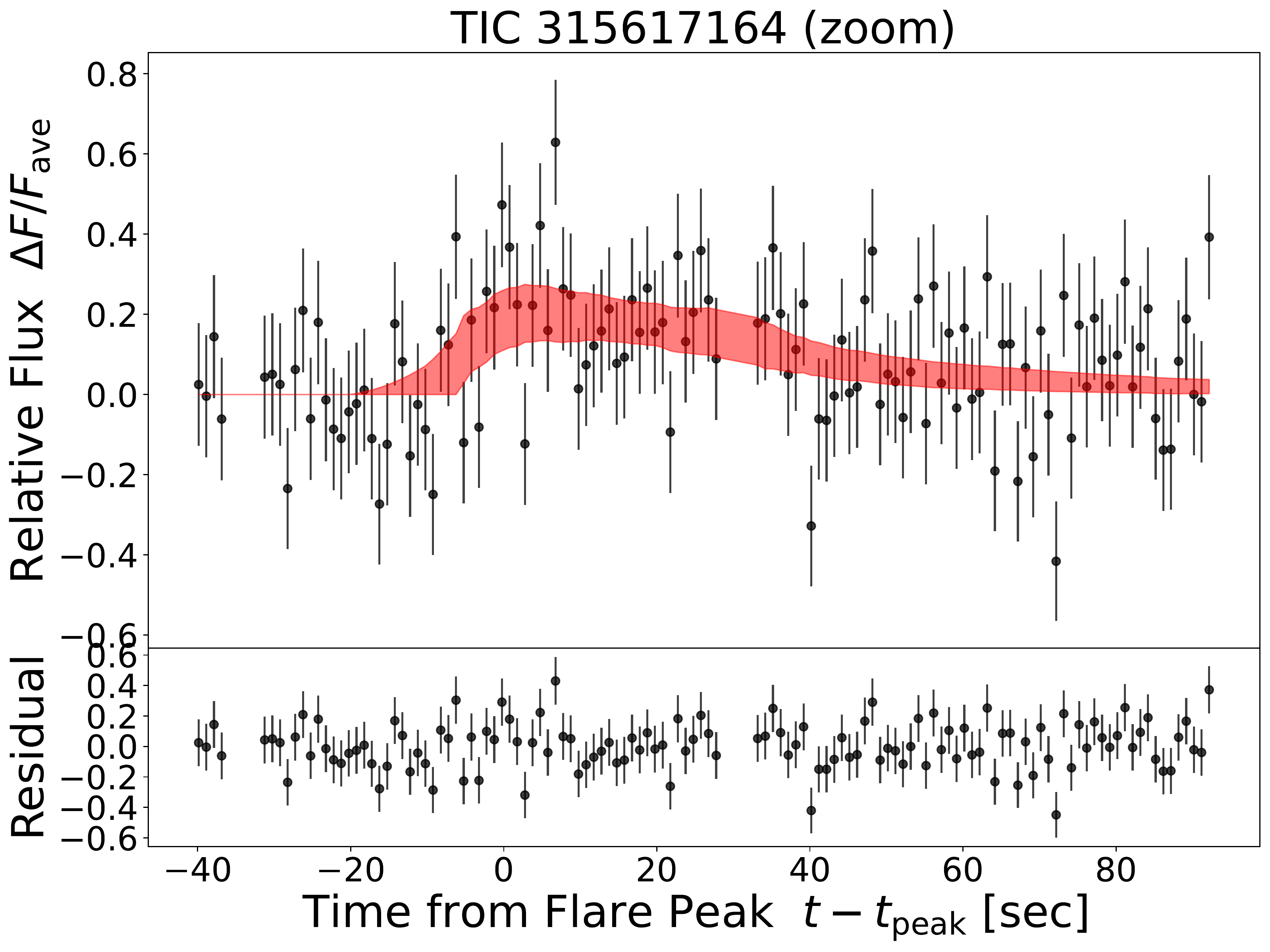}\hfill 
\includegraphics[width=0.49 \linewidth]{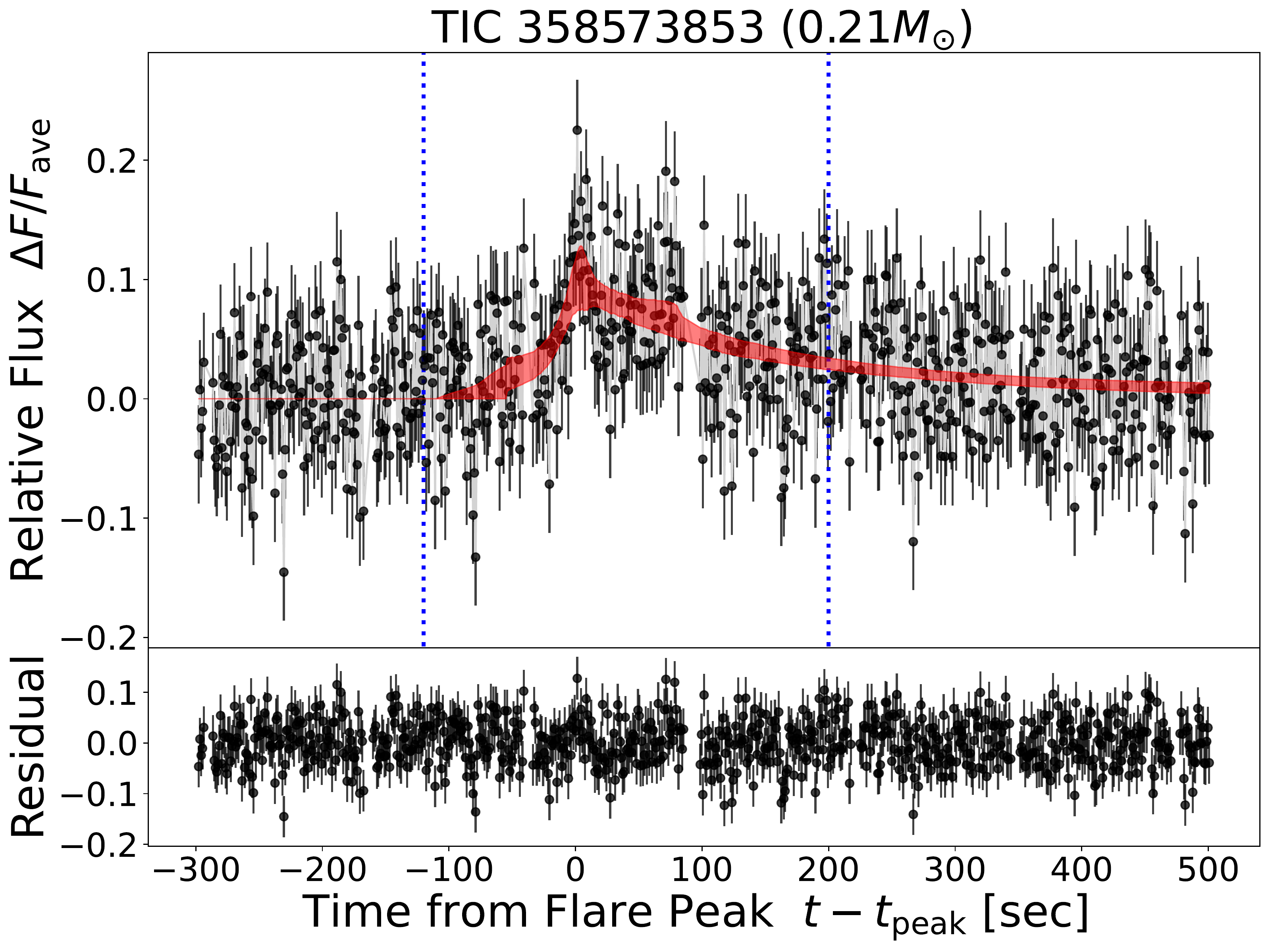}\hfill 
\includegraphics[width=0.49 \linewidth]{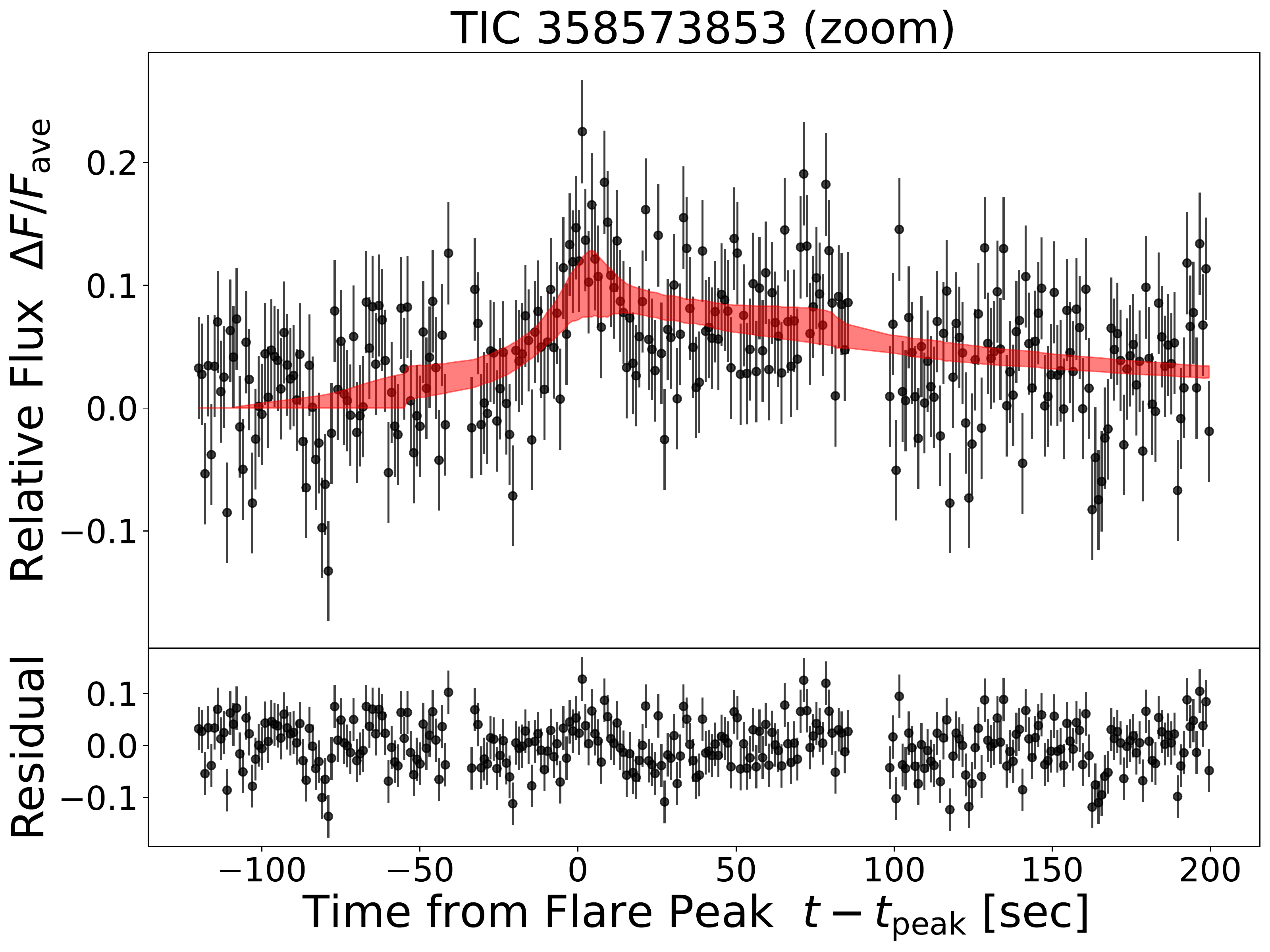}\hfill 
\includegraphics[width=0.49 \linewidth]{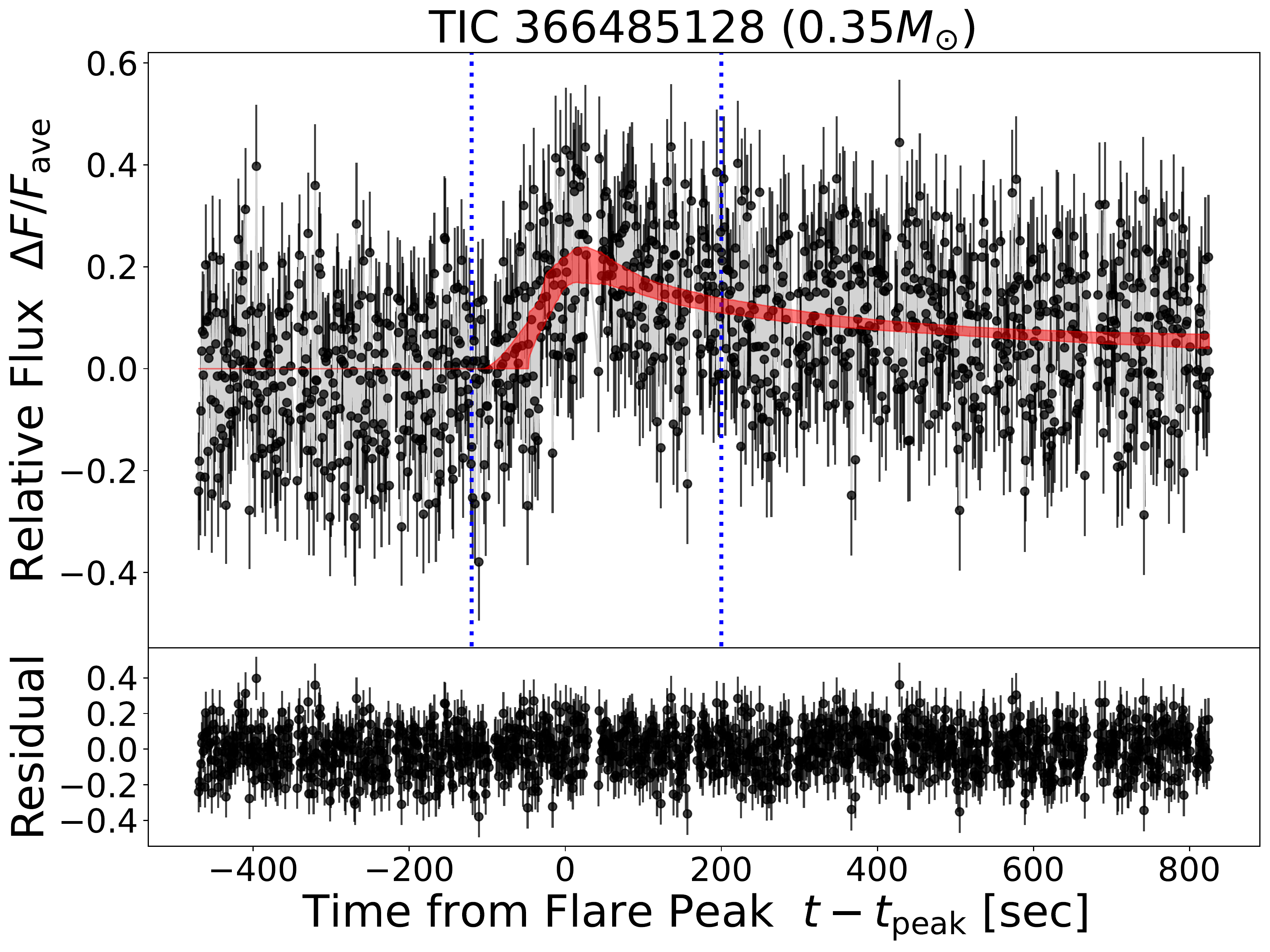}\hfill 
\includegraphics[width=0.49 \linewidth]{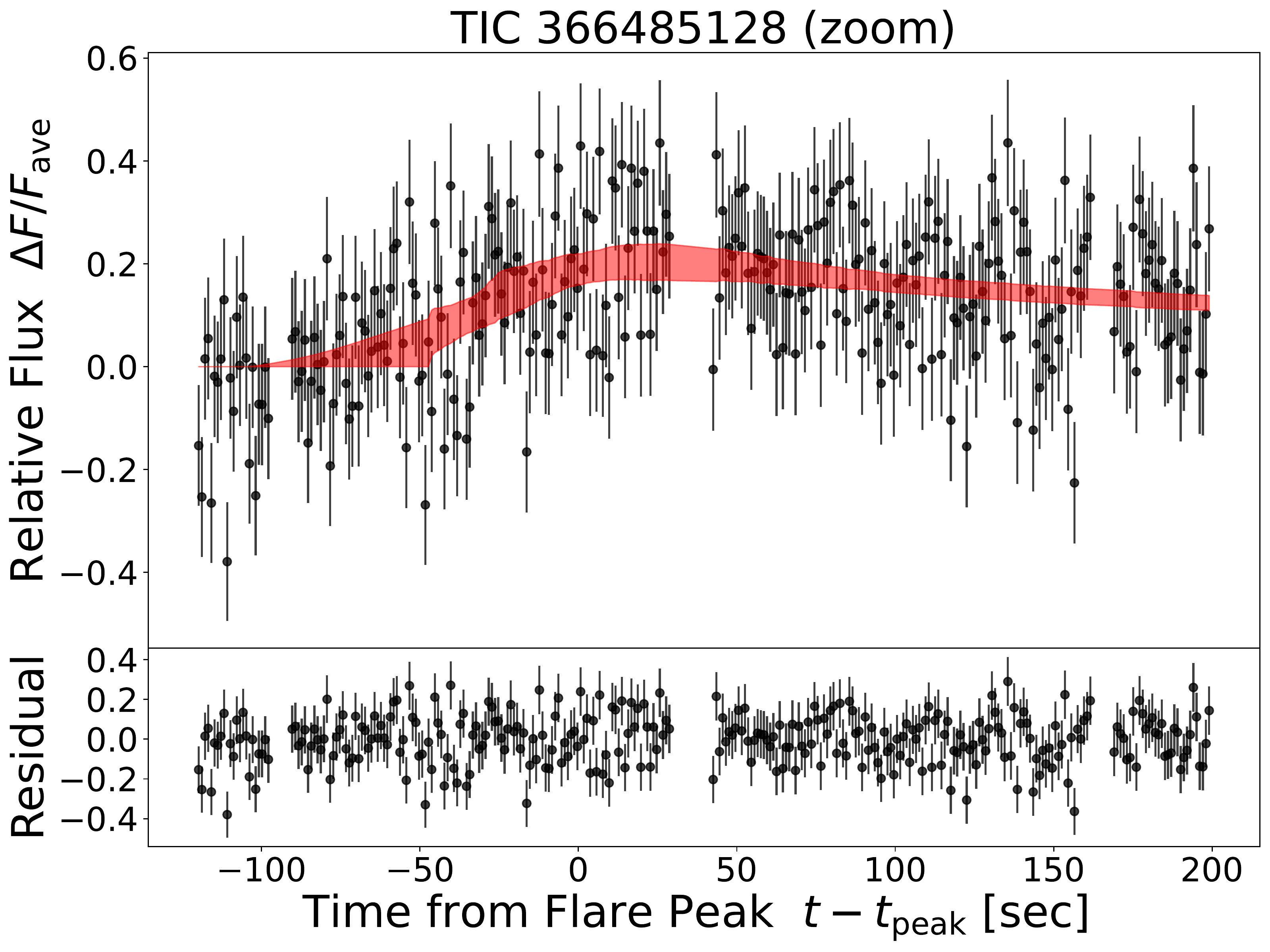}\hfill 
\caption{ (Continue) }  \end{center}\end{figure*}  
 \addtocounter{figure}{-1}\begin{figure*}[htbp]\begin{center} 
 
 \includegraphics[width=0.49 \linewidth]{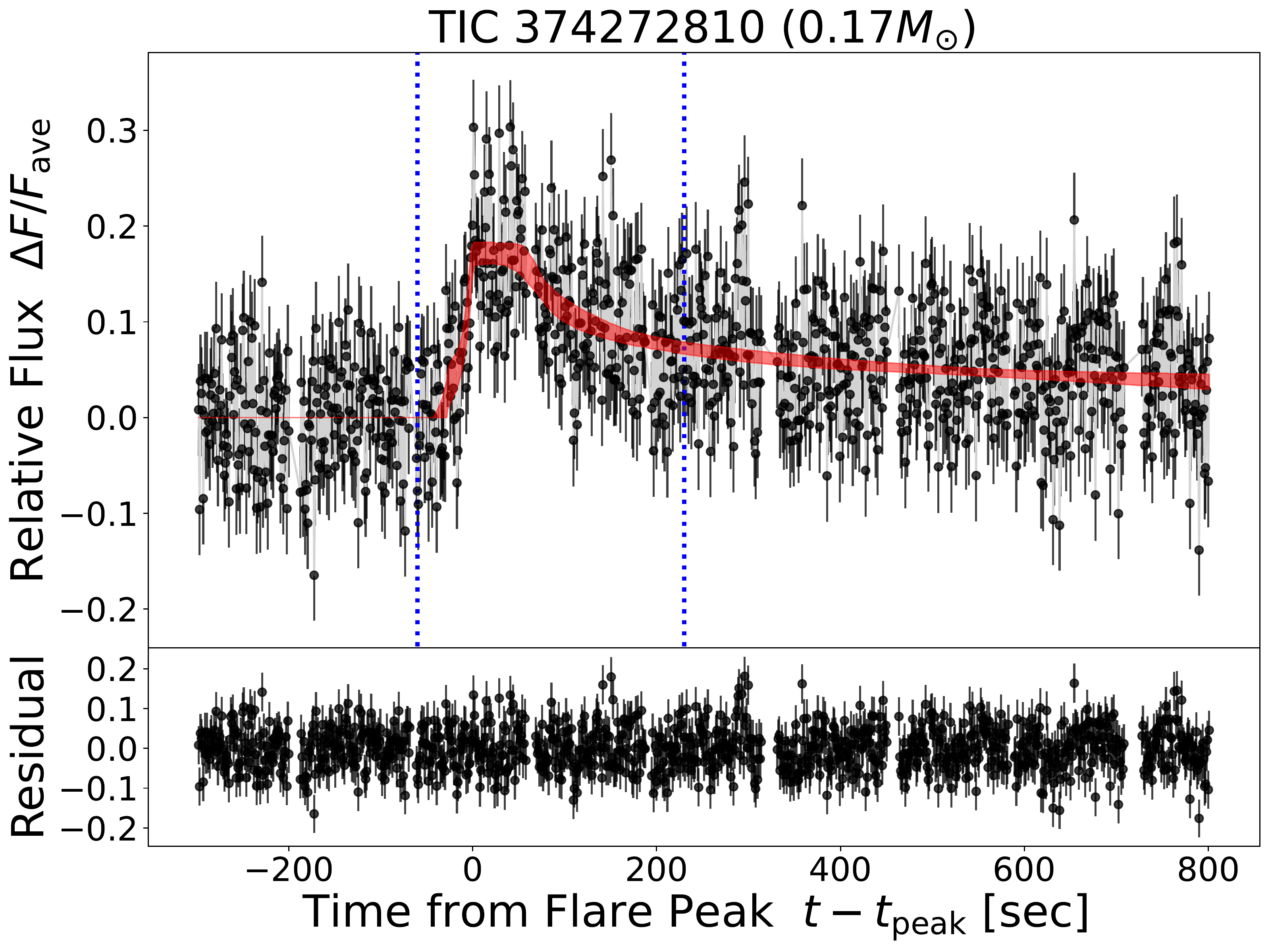}\hfill 
\includegraphics[width=0.49 \linewidth]{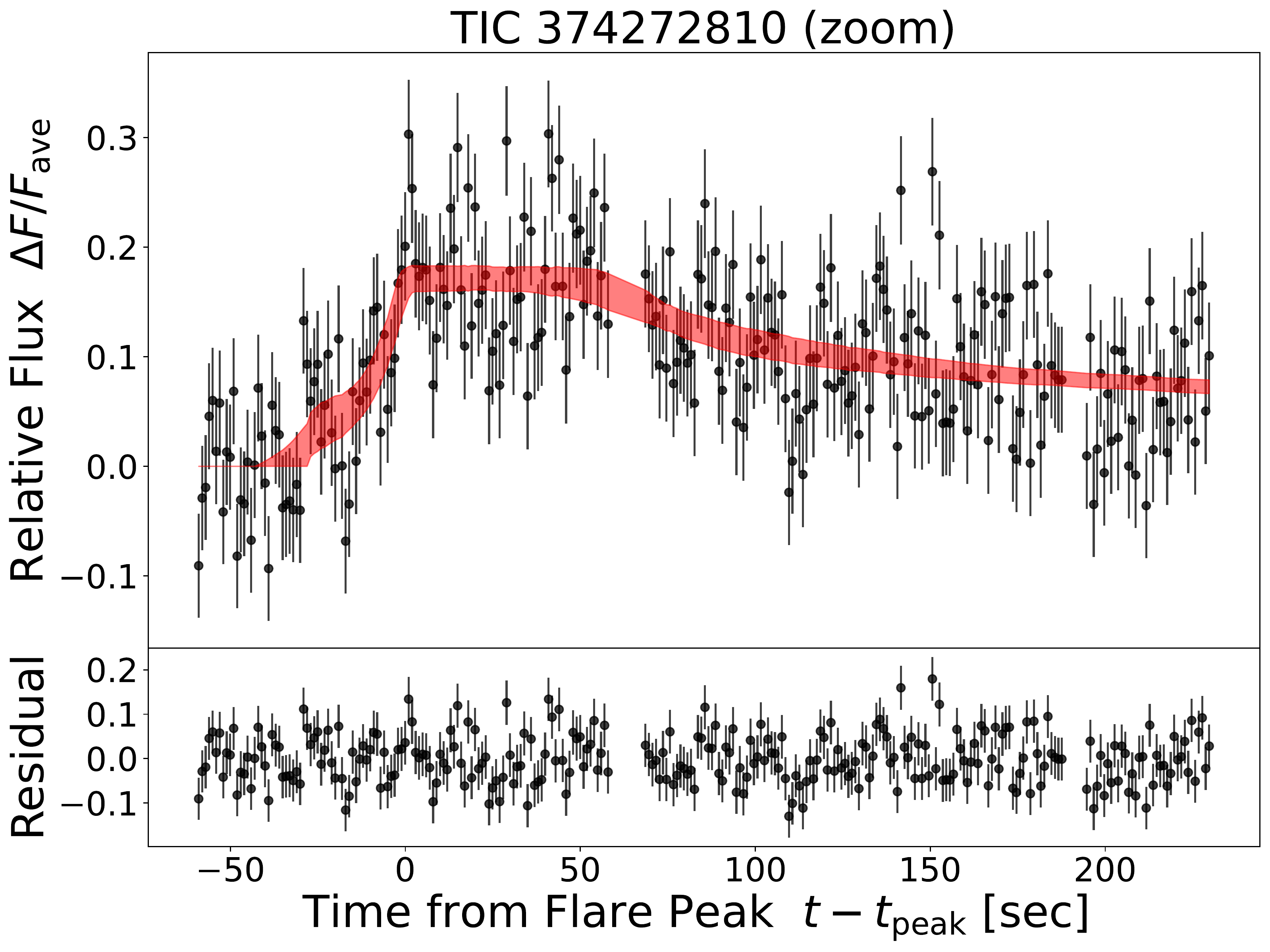}\hfill 
\includegraphics[width=0.49 \linewidth]{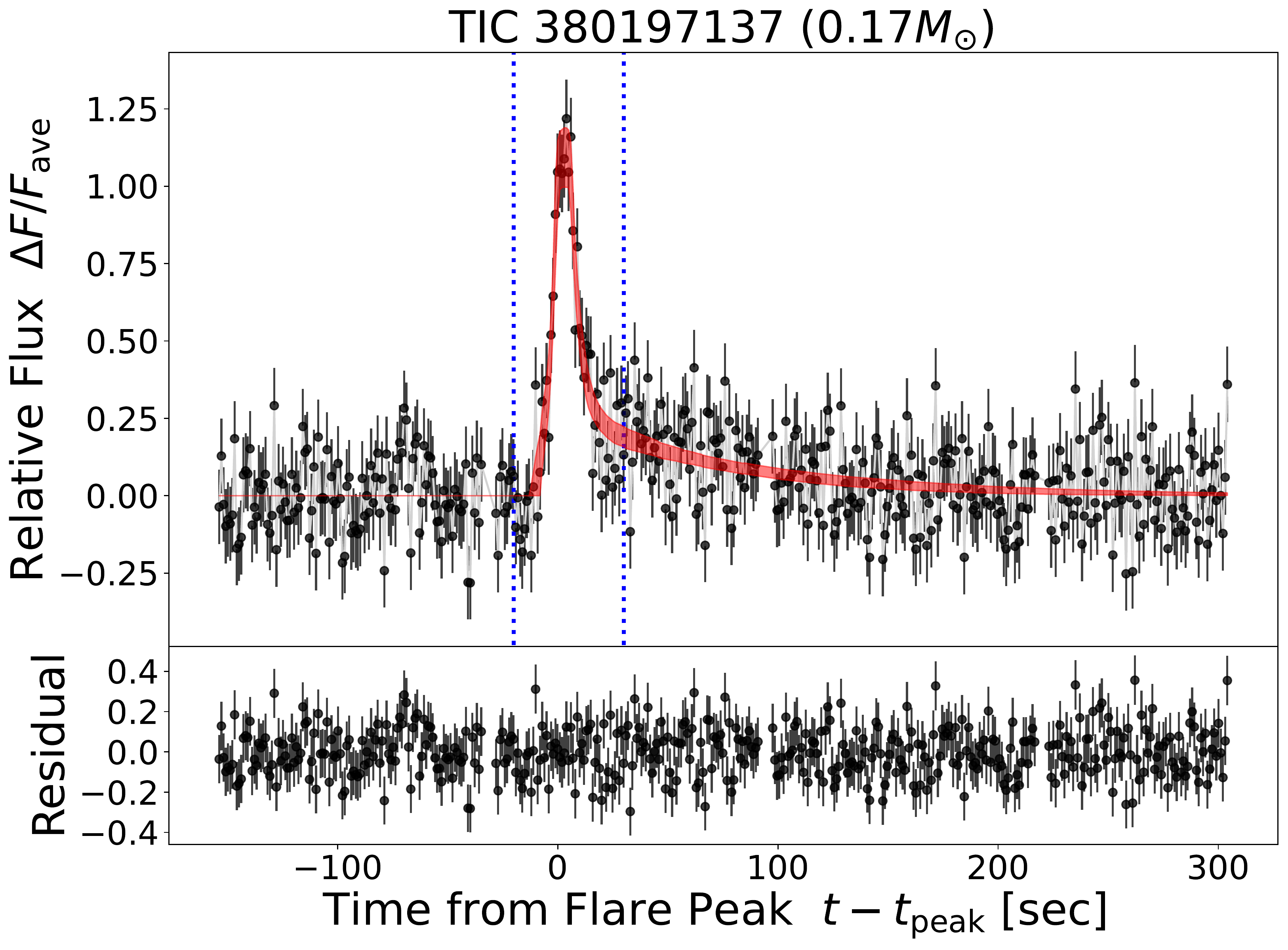}\hfill 
\includegraphics[width=0.49 \linewidth]{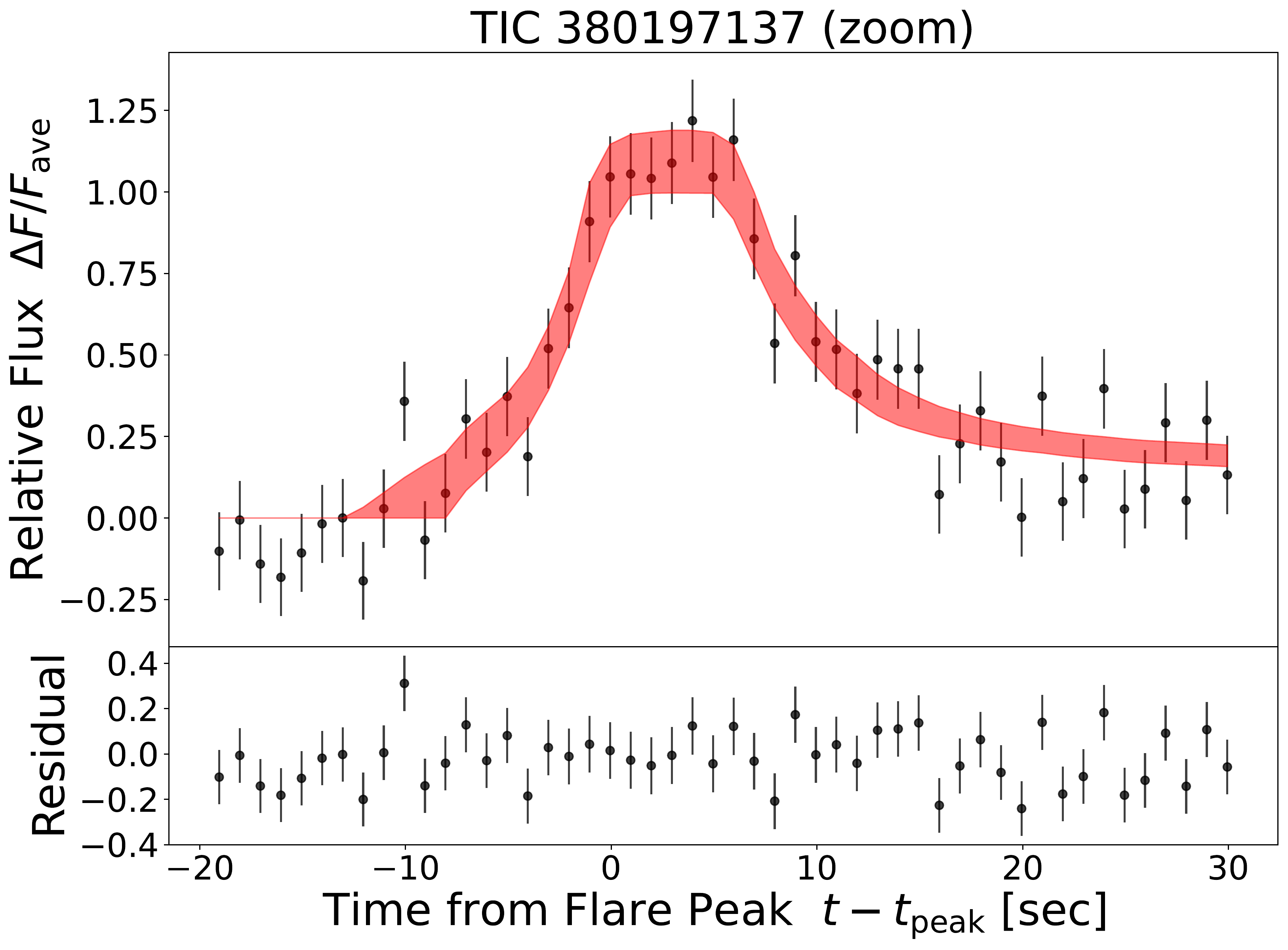}\hfill 
\includegraphics[width=0.49 \linewidth]{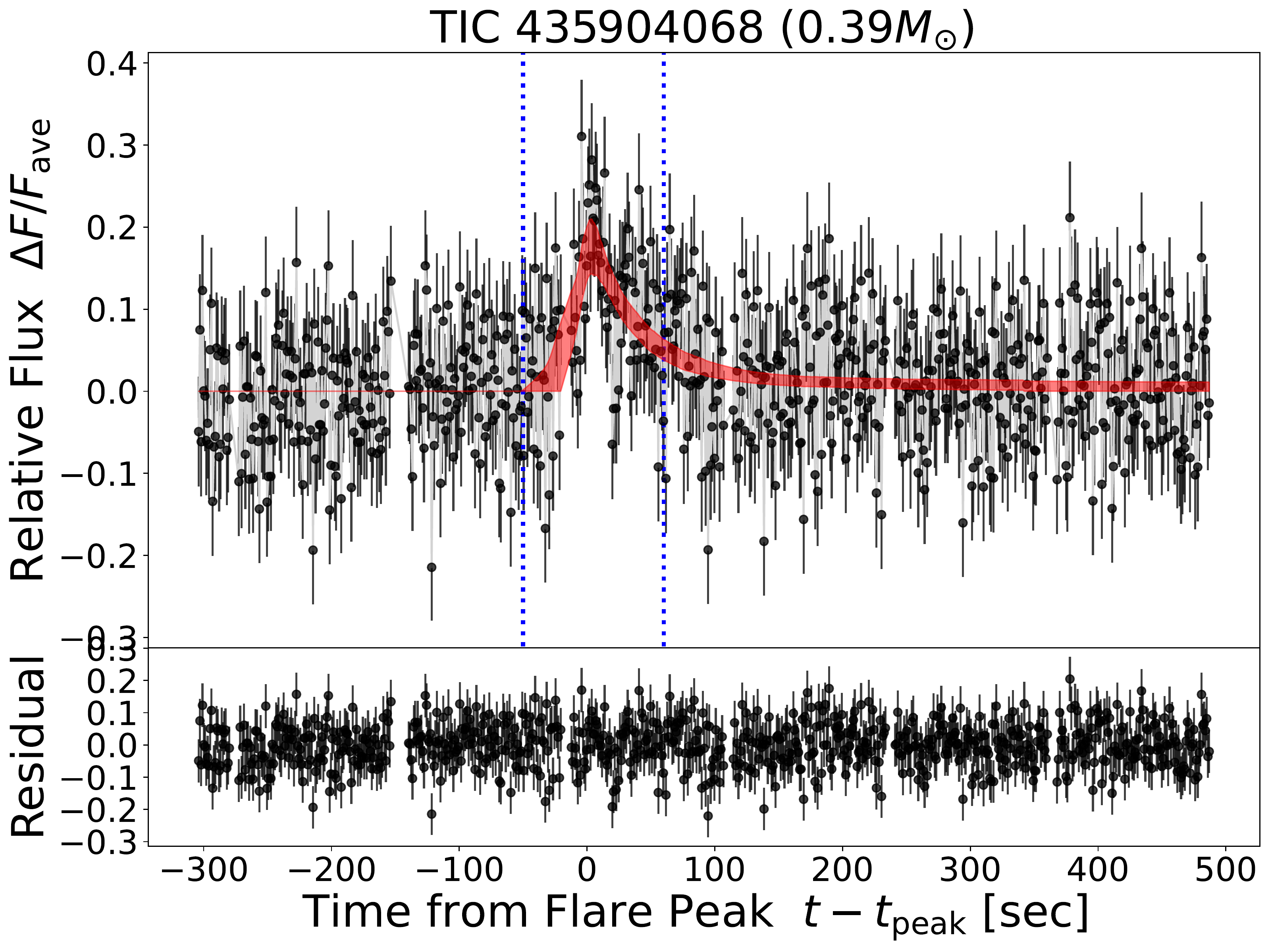}\hfill 
\includegraphics[width=0.49 \linewidth]{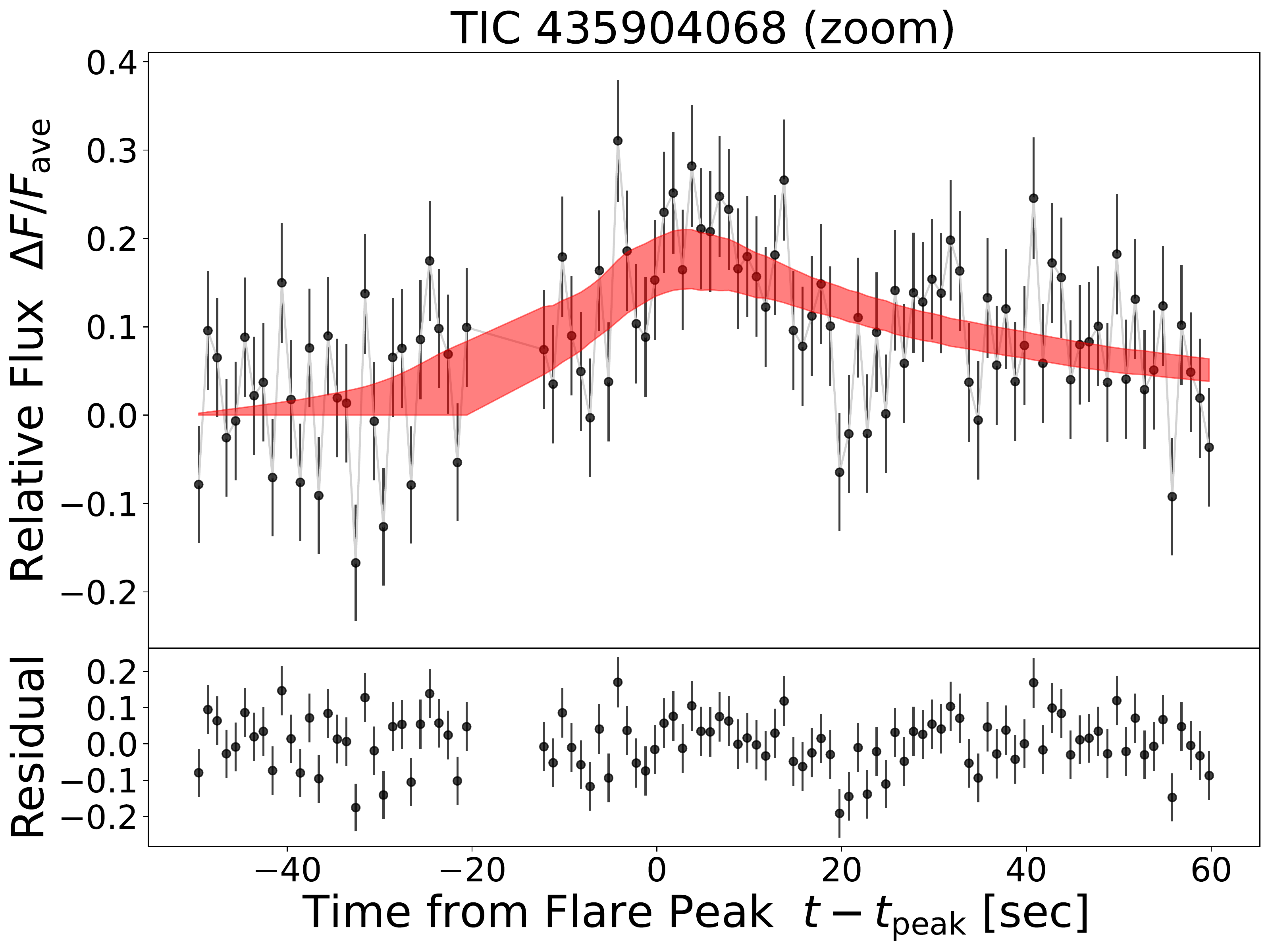}\hfill 
\caption{(Continue) }  \end{center}\end{figure*}  
 \addtocounter{figure}{-1}\begin{figure*}[htbp]\begin{center} 
 
 \caption{ (Continue) }  \end{center}\end{figure*}  
 